 \documentclass[preprint2,10pt]{aastex}

\usepackage[scriptsize]{subfigure}
\usepackage{array}
\usepackage{amssymb,amsmath}

\shorttitle{SDSS/SEGUE Spectral Feature Analysis For Stellar \\Atmospheric Parameter Estimation}
\shortauthors{Li et al.}

\begin{document}


\title{SDSS/SEGUE Spectral Feature Analysis For Stellar \\Atmospheric Parameter Estimation}


%
%
%

\author{Xiangru Li\altaffilmark{1,2}, Q.M. Jonathan Wu\altaffilmark{2}, Ali Luo\altaffilmark{3,4}, Yongheng Zhao\altaffilmark{3,4}, Yu Lu\altaffilmark{1}, Fang Zuo\altaffilmark{3,4}, Tan Yang\altaffilmark{1}, Yongjun Wang\altaffilmark{1}}
\affil{\altaffilmark{1}School of Mathematical Sciences, South China Normal University, 510631, China;xiangru.li@gmail.com}
\affil{\altaffilmark{2}Department of Electrical and Computer Engineering, University of Windsor, Canada}
\affil{\altaffilmark{3}National Astronomical Observatories, Chinese Academy of Sciences, Beijing 100012, China}
\affil{\altaffilmark{4}Key Laboratory of Optical Astronomy, NAOC, Beijing 100012, China}

%
%
%




\begin{abstract}
Large-scale and deep sky survey missions are rapidly collecting a large amount of stellar spectra, which necessitate the estimation of atmospheric parameters directly from spectra and makes it feasible to statistically investigate latent principles in a large dataset. We present a technique for estimating parameters $T_\texttt{eff}$, log$~g$ and [Fe/H] from stellar spectra. With this technique, we first extract features from stellar spectra using the LASSO algorithm; then, the parameters are estimated from the extracted features using the SVR. On a subsample of 20~000 stellar spectra from SDSS with reference parameters provided by SDSS/SEGUE Pipeline SSPP, estimation consistency are 0.007458 dex for log$~T_\texttt{eff}$ (101.609921 K for $T_\texttt{eff}$), 0.189557 dex for log$~g$ and 0.182060 for [Fe/H], where the consistency is evaluated by mean absolute error. Prominent characteristics of the proposed scheme are sparseness, locality, and physical interpretability. In this work, every spectrum consists of 3821 fluxes, and 10, 19, and 14 typical wavelength positions are detected respectively for estimating $T_\texttt{eff}$, log$~g$ and [Fe/H]. It is shown that the positions are related to typical lines of stellar spectra. This characteristic is important in investigating physical indications from analysis results. Then, stellar spectra can be described by the individual fluxes on the detected positions (PD) or local integration of fluxes near them (LI). The abovementioned consistency is the result based on features described by LI. If features are described by PD, consistency are 0.009092 dex for log$~T_\texttt{eff}$ (124.545075 K for $T_\texttt{eff}$), 0.198928 dex for log$~g$, and 0.206814 dex for [Fe/H].
\end{abstract}



\section{Introduction}\label{Sec:Introduction}

Large-scale and deep sky survey missions, such as the Sloan Digital Sky Survey \citep[SDSS;][]{Journal:York:2000,Journal:Ahn:2012}, the Large Sky Area Multi-Object Fiber Spectroscopic Telescope \citep[LAMOST/ Guoshoujing Telescope;][]{Journal:Zhao:2006,Journal:Cui:2012}, and the Global Astrometric Interferometer for Astrophysics \citep[GAIA;][]{Journal:Perryman:2001,Journal:Lobel:2011}, are collecting and will obtain a large number of stellar spectra. To achieve scientific goals and make full use of the potential values of the observations, it is necessary to estimate the atmospheric parameters (e.g. $T_\texttt{eff}$, log$~g$ and [Fe/H]) directly from the spectrum and statistically investigate latent principles in the large spectral dataset.

This paper investigates the representation problem of stellar spectra for physical parameter estimation, which is a vital procedure in the aforementioned tasks and usually called feature extraction in data mining, machine learning, and pattern recognition. For example, in physical parameter estimation, a spectrum can be represented by the observed spectrum \citep{Journal:Bailer-Jones:2000,Journal:Shkedy:2007}, corrected spectrum \citep{Journal:Prieto:2006}, description of some typical lines \citep{Journal:Muirhead:2012,Journal:Mishenina:2006}, statistical description \citep{Journal:Fiorentin:2007}, etc. Feature extraction determines the applicable range of a data analysis system, accuracy, efficiency, physical interpretability, and robustness to noise and distortion from calibration error.

We propose a feature extraction scheme based on the LASSO (least absolute shrinkage and selection operator) algorithm \citep{Journal:Tibshirani:1996} for stellar spectra. The fundamental idea of this proposed scheme is to statistically detect typical wavelength positions statistically that are significant/necessary for discriminating stellar spectra with different atmospheric physical parameters. In this study, the proposed scheme successfully detects 10, 19, and 14 typical wavelength positions from 3~821 sample points \footnote{By `3~821 sample points', we mean that every spectrum is described by 3821 fluxes in this study.} respectively for estimating atmospheric parameter $T_\texttt{eff}$, log$~g$ and [Fe/H]. In other words, a spectrum can be described by 10, 19, or 14 of the 3~821 observed fluxes at the detected positions, or the local integrations of fluxes around the specific positions. It is shown that the detected positions are closely related with some spectral lines. In contrast, the global method Principal Component Analysis (PCA) \citep{Journal:Li:2012}, which computes every feature from nearly all observed fluxes, locality makes the proposed scheme immune or robust to the aggregated influence of noise and calibration distortion. Therefore, prominent characteristics of the proposed scheme are sparseness and locality, based on that it is easier to backtrack the specific effective factors in estimating an atmospheric parameter than with global methods.

To evaluate the effectiveness of the detected features, we investigate the atmospheric parameter estimation problem based on the Support Vector Regression (SVR) method  \citep{Journal:Smola:2004,Journal:Schokopf:2002} and the detected features. Experimental results show excellent consistency between the estimates of our proposed scheme and that provided by SDSS/SEGUE Spectroscopic Parameter Pipeline \citep[SSPP;][]{Journal:Beers:2006,Journal:Lee:2008:a,Journal:Lee:2008:b,Journal:Prieto:2008,Journal:Smolinski:2011,Journal:Lee:2011} on a subsample of 20$~$000 stellar spectra from SDSS. The SSPP of SLOAN estimates the fundamental stellar parameters based on both stellar spectra and ugriz photometry by multiple techniques \citep{Journal:Lee:2008:a} and a robust decision tree scheme. Performance of the SSPP were also investigated from multiple aspects\citep{Journal:Prieto:2008,Journal:Lee:2008:b,Journal:Smolinski:2011}.

The proposed scheme is also evaluated on synthetic stellar spectra with ground-truth parameters. The synthetic spectra are computed based on the ¡°New Grids of ATLAS9 Model Atmospheres¡±  \citep{Journal:Castelli:2003}. On the synthetic spectra, the accuracy of the proposed scheme are 0.000801 dex for log$~T_\texttt{eff}$ , 0.017881 dex for log$~g$ and 0.013142 for [Fe/H], where the accuracy is evaluated by mean absolute error (MAE).


The rest of this paper is organized as follows. We describe the stellar spectra used in this study and the previously estimated physical parameters for reference in section \ref{Sec:Data}. In section \ref{Sec:Feature_extraction}, we introduce our proposed feature extracting scheme and analyze the extracted features. The parameterization model of stellar spectra and evaluation methods for accuracy/consistency are introduced in section \ref{Sec:Regression_model}. In section \ref{Sec:Feature_Description}, we propose our feature description schemes and present the parameterizing results. In section \ref{Sec:Feature:eva}, compactness of the detected features are evaluated. In section \ref{Sec:Config_evaOnSyn}, we evaluated the proposed scheme on synthetic spectra and discussed the configuration problem of the scheme. To highlight the characteristics of our proposed scheme, related research is reviewed and analyzed in section \ref{Sec:RelatedResearches}. Finally, we summarize this work in section \ref{Sec:Conclusion}.

\section{Data}\label{Sec:Data}
 \begin{figure*}
  \centering
  \subfigure[$T_{\texttt{eff}}$ and log~$g$]{
    \label{Fig:para:distrubution:Teff_logg} 
    \includegraphics[width =1.55in]{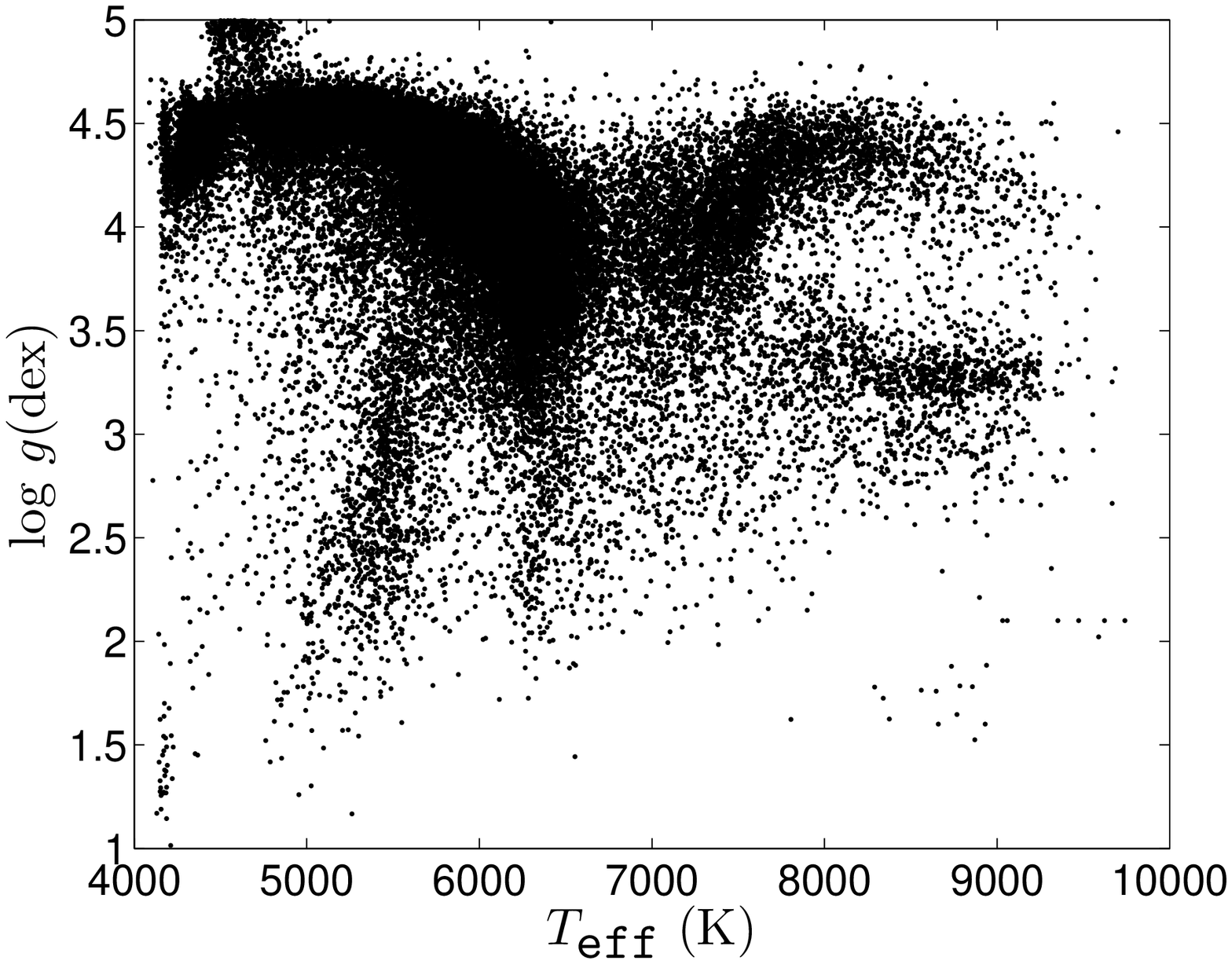}}
\hspace{-0.17in}
  \subfigure[$T_{\texttt{eff}}$ and $\texttt{[}$Fe/H$\texttt{]}$]{
    \label{Fig:para:distrubution:Teff_FeH} 
    \includegraphics[width =1.55in]{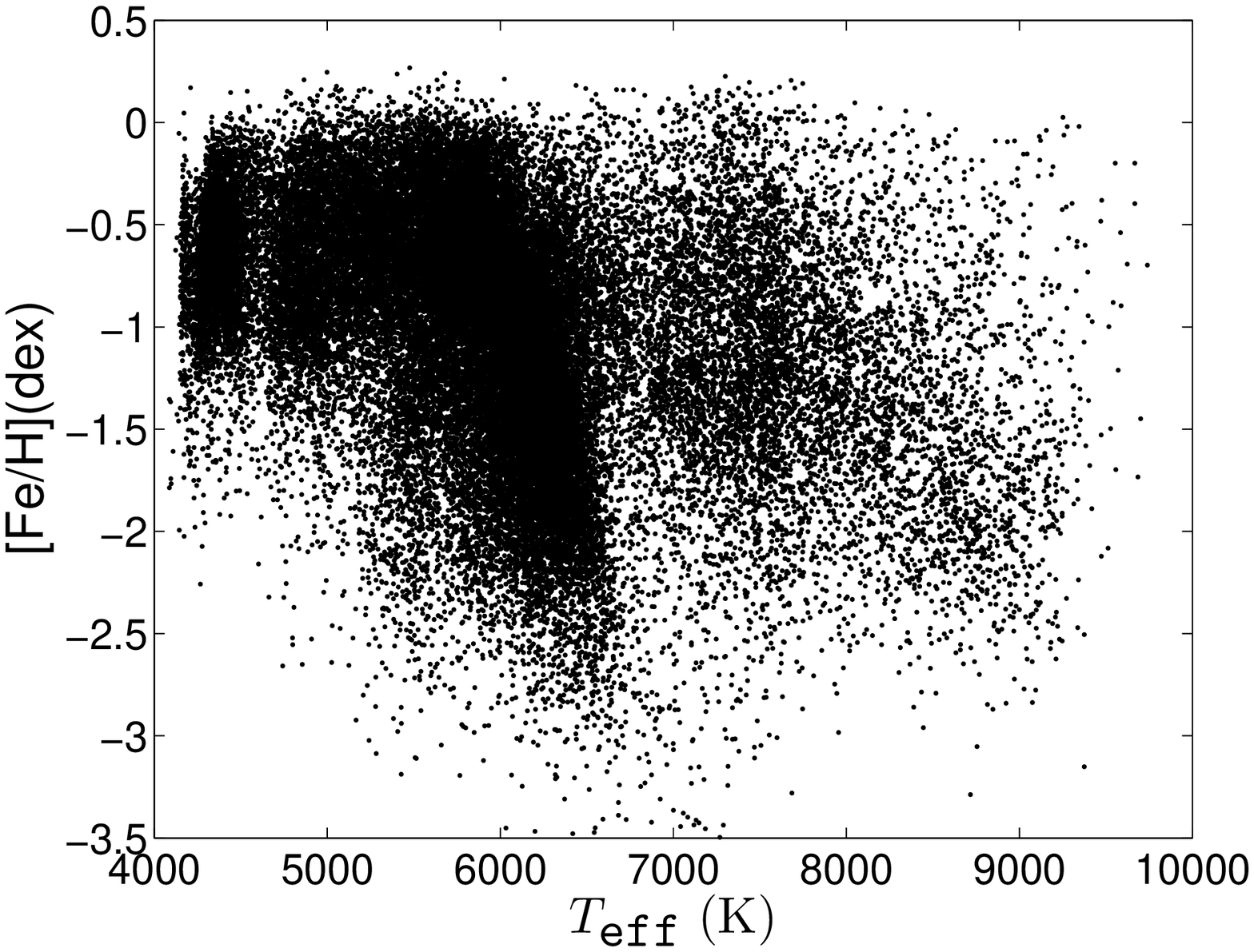}}
  \hspace{-0.17in}
  \subfigure[log$~g$ and $\texttt{[}$Fe/H$\texttt{]}$]{
    \label{Fig:para:distrubution:logg_FeH} 
    \includegraphics[width =1.55in]{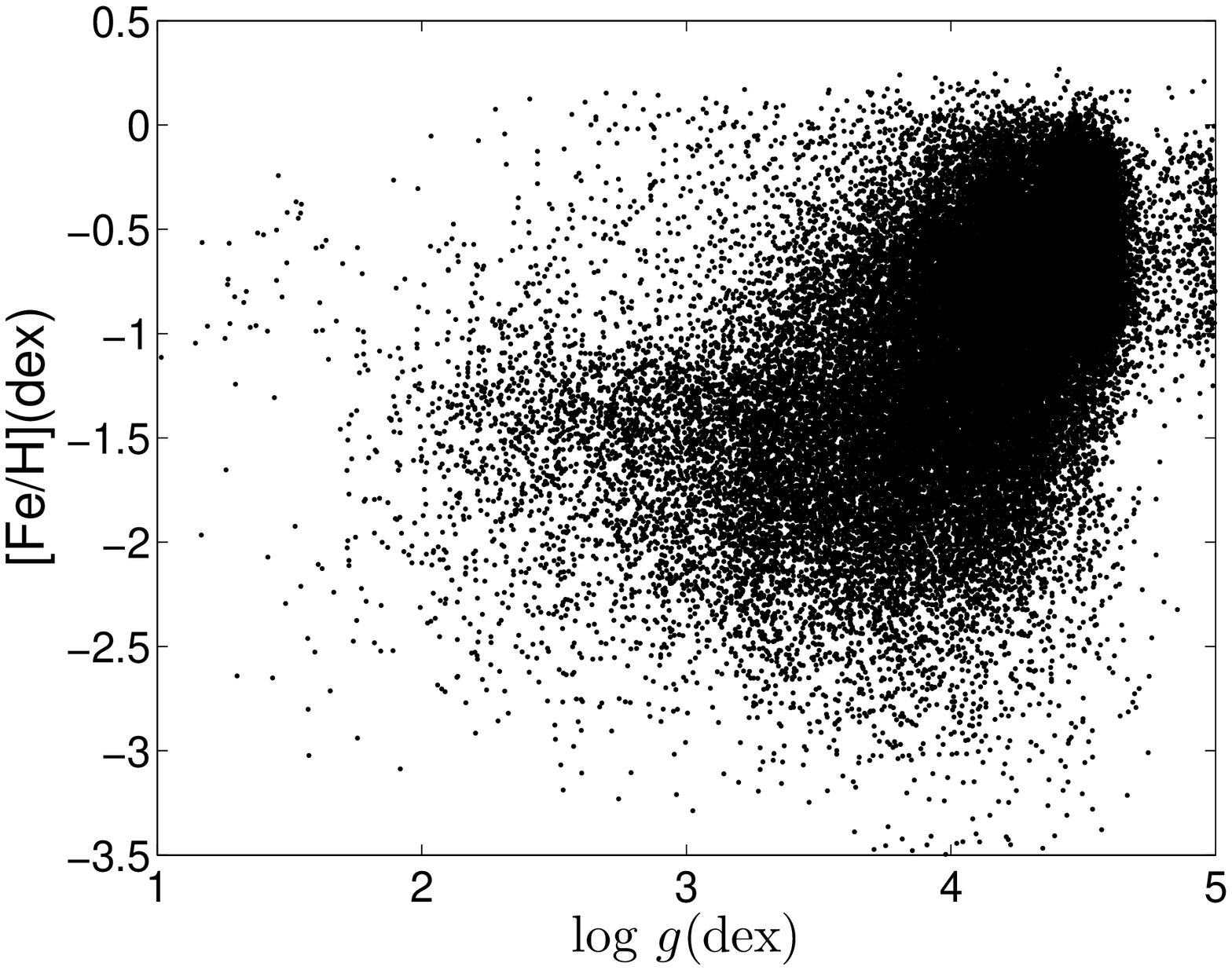}}
    \setlength{\abovecaptionskip}{-10pt}
  \caption{Scatter diagram of the atmospheric parameters of the selected spectra.}
  \label{Fig:para:distribution} 
\end{figure*}

   \begin{figure*}
  \centering
  \subfigure[log$~T_{\texttt{eff}}$]{
    \label{Fig:para:distrubution:Teff} 
    \includegraphics[width =1.5in]{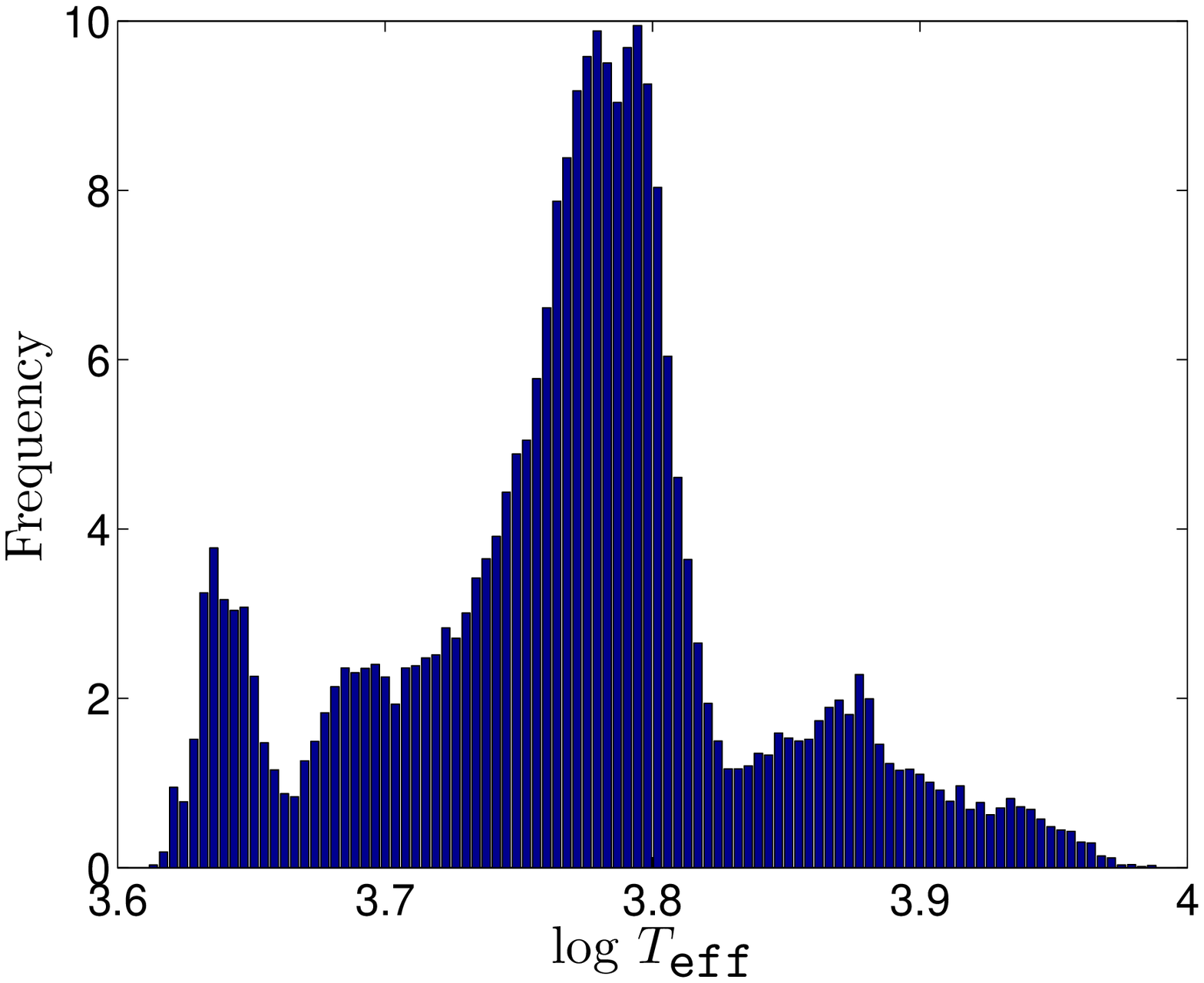}}
\hspace{-0.15in}
  \subfigure[log~$g$]{
    \label{Fig:para:distrubution:log} 
    \includegraphics[width =1.5in]{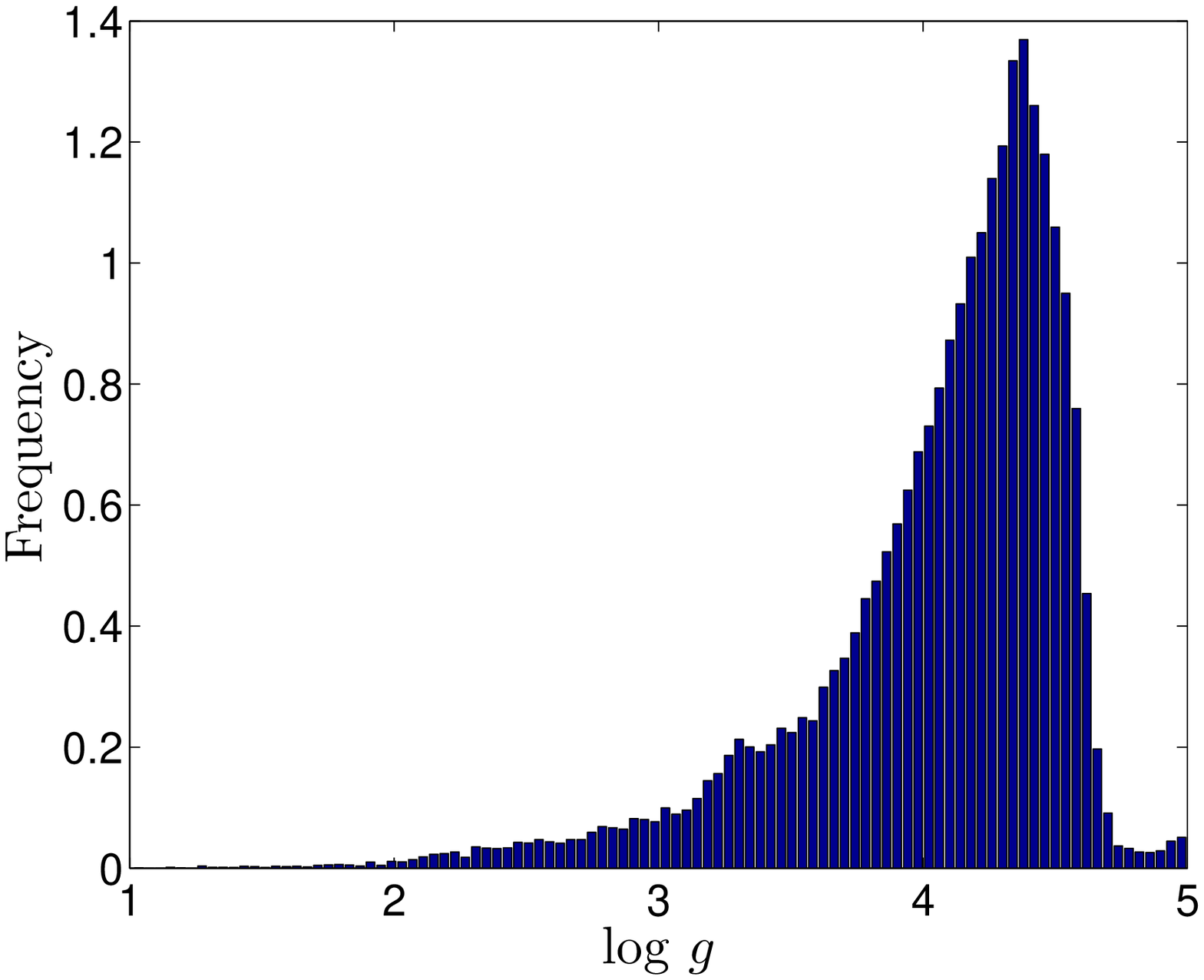}}
  \hspace{-0.15in}
  \subfigure[$\texttt{[}$Fe/H$\texttt{]}$]{
    \label{Fig:para:distrubution:FeH} 
    \includegraphics[width =1.5in]{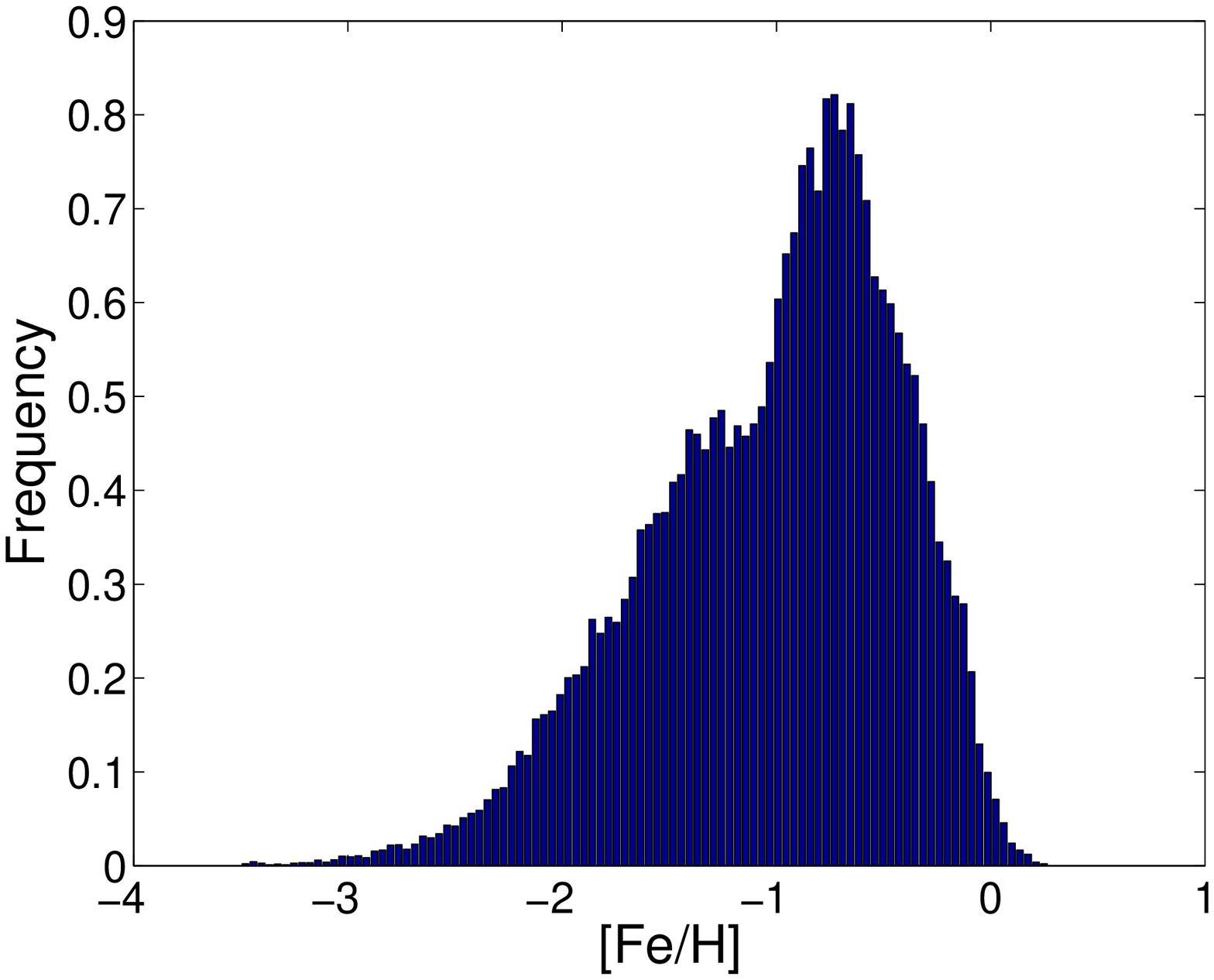}}
    \setlength{\abovecaptionskip}{-2pt}
  \caption{Distribution of the atmospheric parameters of the selected spectra.}
  \label{Fig:para:hist_AtmPar} 
\end{figure*}

In this work, we use 50~000 stellar spectra of SDSS/SEGUE observation \citep{Journal:Yanny:2009} and their previously computed physical parameters from the Seventh Sloan Data Release \citep{Journal:Abazajian:2009}. The selected spectra span the ranges [4088, 9740] K in effective temperature $T_{\texttt{eff}}$, [1.015000, 4.998000] dex in surface gravity log$~g$, and [-3.497000, 0.268000] dex in metallicity [Fe/H]; additional statistical information on the selected spectra is presented in Fig. \ref{Fig:para:distribution} and Fig. \ref{Fig:para:hist_AtmPar}. All of the stellar spectra are shifted to their rest frames (zero radial velocity) based on the previously estimated radial velocity provided by the SSPP and rebinned to a maximal common log(wavelength) range [3.581862, 3.963961]\footnote{Approximately, the common wavelength range is [3818.23, 9203.67]${\AA}$.} with a sampling step 0.0001.

Our proposed scheme belongs to the statistical learning method. The fundamental idea of this scheme is to discover the potentially predictive relationship based on empirical stellar spectra and corresponding atmospheric parameters, which are called training data. At the same time, performance of the discovered predictive relationships should also be evaluated objectively. Therefore, a separate set of stellar spectra is needed for evaluation, usually called a test set in pattern recognition. On the other hand, most learning methods tend to overfit the empirical data. That is to say, the statistical learning methods can discover some alleged relationships from the training data that do not hold in general. In order to avoid overfitting, we need some independent spectra for optimizing the parameters that need to be adjusted objectively in investigating the potential relationships, and these independent spectra and their reference parameters constitute a validation set. Therefore, the selected stellar spectra are partitioned into three subsets: training set, validation set, and test set. Sizes of the three subsets are 20~000, 20~000 and 10~000 respectively. The roles of the three subsets are presented in Table \ref{Tab:DataSets:roles}.

In the training and evaluation process based on SDSS spectra, we take the previously estimated atmospheric parameters provided by SDSS/SEGUE Spectroscopic Parameter Pipeline \citep[SSPP;][]{Journal:Beers:2006,Journal:Lee:2008:a,Journal:Lee:2008:b,Journal:Prieto:2008,Journal:Smolinski:2011,Journal:Lee:2011} as a reference. The SSPP of SLOAN estimates the fundamental stellar atmospheric parameters based on both stellar spectra and ugriz photometry by multiple techniques \citep{Journal:Lee:2008:a}, for example, spectral fitting with k24 \citep{Journal:Allende:2006} and ki13 Girds, extended WBG method (\citep{Journal:Wilhelm:1999,Journal:Lee:2008:a}) based on theoretical ugr colors and line parameters from synthetic spectra, nonlinear neural network models trained by real SDSS spectra or synthetic spectra (\citep{Journal:Fiorentin:2007}), the $\chi^2$ minimization technique based on synthetic spectral libraries NGS1 and NGS2 girds, Sensitive wavelength window selection methods G8(CaI1) and M8(CaIIK1) based on the synthetic NGS1 gird, CaII~K and autocorrelation function methods (CaIIK2, CaIIK3, and ACF \citep{Journal:Beers:1999}), M12 method based on Ca~II Triplet lines, CaI2 and MgH methods based on the CaI (4227 \AA), MgIb and MgH features, etc. The SSPP make the final decision by adaptively evaluating the reliabilities of the multiple estimates of every atmospheric parameter from a stellar spectrum and computing the weighted average of the reliable estimates. By doing this, limitations of a specific technique can be alleviated to a certain degree, for example, the restricted applicability from the coverage of the grids of utilized synthetic spectra, the methods used for spectral matching, and their sensitivity to the signal-noise ratio of a spectrum, the applicable range in parameter space, etc. The SSPP were validated by comparing its estimates with the sets of parameters obtained from the high-resolution spectra from SDSS-I/SEGUE stars \citep{Journal:Prieto:2008}, and with the available information from the literature for stars in Galactic open and globular clusters \citep{Journal:Lee:2008:b,Journal:Smolinski:2011}. Therefore, consistency between estimates of a proposed method and the SSPP results can reflects the performance of a method to a certain extent.

The proposed scheme were also evaluated on synthetic spectra with ground-truth parameters. The synthetic spectra and the experiments are introduced in section \ref{Sec_sub:eva_Syn}.

\begin{table}\scriptsize
\centering
\caption{Roles of three data sets.}
\begin{tabular}{ p{2cm}<{\centering} p{5cm}  }
  \hline \hline
Data sets         &\qquad \qquad \qquad \qquad  Roles                 \\ \hline
Training set      & Be used in
                    \begin{enumerate} \setlength{\itemsep}{0pt} \setlength{\parsep}{0pt} \setlength{\parskip}{0pt}
                    \item[1)] detecting features (Section \ref{Sec:Feature_extraction});
                    \item[2)] estimating preprocessing parameters $\{ \hat{\mu_j}\}$ (equation (\ref{Equ:training_sp_normalized_muj_tilde})), $\{ \hat{\sigma_j}\}$ (equation  (\ref{Equ:training_sp_normalized_sigmaj_tilde}));
                    \item[3)] parameterizing model (Section \ref{Sec:Regression_model}).
                    \end{enumerate}\\
Validation set    & Be used in
 \begin{enumerate} \setlength{\itemsep}{0pt} \setlength{\parsep}{0pt} \setlength{\parskip}{0pt}
                    \item[1)] estimating feature description parameter $k$ in equations (\ref{Equ:featuredescription:inte:Teff}), (\ref{Equ:featuredescription:inte:Logg}) and (\ref{Equ:featuredescription:inte:FeH});
                    \item[2)] feature evaluation \& refinement (Section \ref{Sec:Feature:eva}).
                    \end{enumerate}\\                                \\
Testing set          & Be used in performance evaluation (Section \ref{Sec:Regression_model:Evaluation}).                                  \\ \hline
\end{tabular}\label{Tab:DataSets:roles}
\end{table}

\begin{table}\scriptsize
\centering
\caption{Detected typical positions for estimating $T_{\texttt{eff}}$ from SDSS stellar spectra. TPW $\lambda^w$: Typical position in wavelength, TPL $\lambda^l$: Typical position in log(wavelength), TP: typical position.}
\begin{tabular}{ c c  c c   }
  \hline \hline
label         &TPW $\lambda^w$ ({\AA})       &TPL $\lambda^l$               & lines near TP \\ \hline
  T1             &   3840.2721    &3.5844         &   Fe~I                           \\
  T2             &   3936.0626    &3.5951         &   KP,Ca~IIK                       \\
  T3             &   3936.9690    &3.5952         &   KP,Ca~IIK                       \\
  T4             &   3969.7394    &3.5988         &   Ca~IIHKp,Heps                   \\
  T5             &   4341.7219    &3.6377         &   $H_{\gamma}$                   \\
  T6             &   4680.1740    &3.6703         &   CC12                           \\
  T7             &   5182.7708    &3.7146         &   MgH+MgI                        \\
  T8             &   6569.9490    &3.8176         &   $H_{\alpha}$,CaH               \\
  T9             &   9148.7551    &3.9614         &   Fe~I,O~I                       \\
  T10            &   9150.8619    &3.9615         &   Fe~I,O~I                       \\  \hline
\end{tabular}\label{Tab:LASSO:Features:Teff}
\end{table}

\section{Feature Extraction}\label{Sec:Feature_extraction}

We investigate the feature extraction problem by using the LASSO algorithm\citep{Journal:Tibshirani:1996,Journal:Efron:2004} for automatically estimating atmospheric parameters from stellar spectra. Suppose the training set is represented by
\begin{equation}\label{Equ:training_sp}
S_{tr} = \{ (x^i, y_i), i = 1, 2, \cdots, N \},
\end{equation}
where $x^i = (x^i_{1}, \cdots, x^i_{p})^T$ is an observed spectra and $y_i$ is the corresponding atmospheric parameter\footnote{In this paper, $y_i$ can be effective temperature, surface gravity, or metallicity. The stellar spectra are analyzed three times respectively for the three parameters.}, $x^i_{j}$ is a specific observed flux, and $N$ is the size of training data set (in this study, $N = 20~000$). Let $(x, y)$ represents a general stellar spectrum and its corresponding atmospheric parameter in consideration, where
\begin{equation}\label{Equ:spectrum}
  x = (x_{1}, \cdots, x_{p})^T.
\end{equation}
The validation set and testing set can be represented similarily by $S_{val}$ and  $S_{te}$.

\subsection{Preprocessing}\label{Sec:Feature_extraction:Preprocessing}
In feature analyzing, we conduct the following preprocessing procedures:
\begin{itemize}
\item Replace $T_{\texttt{eff}}$ with log $T_{\texttt{eff}}$ to reduce the dynamical range and to better represent the uncertainties of spectral data\citep{Journal:Fiorentin:2007}.
\item Normalize the features by setting every variable with zero mean and unit variance, which helps to put all of the variables on an equal footing. That is to say, the spectrum in equation (\ref{Equ:spectrum}) is transformed into
      \begin{equation}\label{Equ:spectrum_feature_Nor}
            \tilde{x} = (\tilde{x}_{1}, \cdots, \tilde{x}_{p})^T
      \end{equation}
      and the training set in equation (\ref{Equ:training_sp}) is transformed into
      \begin{equation}\label{Equ:training_sp_normalized}
           \tilde{S}_{tr}^{F} = \{ (\tilde{x}^i, y_i), i = 1, 2, \cdots, N \},
      \end{equation}
      where
         \begin{equation}\label{Equ:training_sp_normalized_xi_tilde}
            \tilde{x}^i = (\tilde{x}^{i}_{1}, \cdots, \tilde{x}^{i}_{p})^T,
         \end{equation}
         \begin{equation}\label{Equ:training_sp_normalized_xj_tilde}
            \tilde{x}_{j} = \frac{x_{j} - \hat{\mu}_{j}} {\hat{\sigma}_{j}},
         \end{equation}
         \begin{equation}\label{Equ:training_sp_normalized_xij_tilde}
            \tilde{x}^{i}_{j} =  \frac{x^{i}_{j} - \hat{\mu}_{j}} {\hat{\sigma}_{j}},
         \end{equation}
         \begin{equation}\label{Equ:training_sp_normalized_muj_tilde}
            \hat{\mu}_{j} = \frac{\sum_{i=1}^{N}{x^{i}_{j}} }{N},
         \end{equation}
         \begin{equation}\label{Equ:training_sp_normalized_sigmaj_tilde}
             \hat{\sigma}_{j} = \sqrt{\frac{\sum_{i=1}^{N}({x^{i}_{j} - \hat{\mu}_{j}})^2 } {N}},
         \end{equation}
         $$j = 1, \cdots, p,$$
         $$i = 1, \cdots, N.$$
\end{itemize}
The validation set and testing set are preprocessed similarily by equation (\ref{Equ:training_sp_normalized_xij_tilde}) based on the parameters $\hat{\mu}_{j}$ in equation (\ref{Equ:training_sp_normalized_muj_tilde}) and $\hat{\sigma}_{j}$ in equation (\ref{Equ:training_sp_normalized_sigmaj_tilde}), and converted into $\tilde{S}_{val}$ and $\tilde{S}_{te}$.

\begin{figure*}
\begin{center}
\includegraphics[ width =1.6in, angle=90]{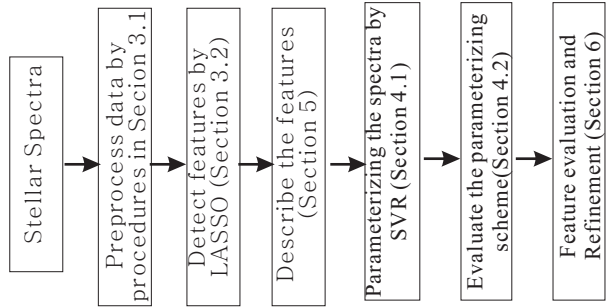}
\end{center}
\setlength{\abovecaptionskip}{1pt} \caption{A flowchart to show the order that the statistical procedures are used in analyzing.}
\label{Fig:Flowchart}
\end{figure*}

There are multiple statistical procedures that will be performed in this paper. To be readable, a flowchart is presented in Fig. \ref{Fig:Flowchart} to demonstrate the end-to-end flow in the analysis.

\subsection{Detect Features}\label{Sec:Feature_extraction:Detect_Features}

In the LASSO scheme, features are identified by the following model
\begin{equation}\label{Equ:LASSO}
   (\hat{\alpha}, \hat{\beta}) = arg\min\limits_{(\alpha, \beta)} \{\Sigma_{i=1}^{N}( y_i - \alpha - \Sigma_{j=1}^{p}{\beta_j \tilde{x}^{i}_{j}} )^2\}
\end{equation}
subject to
\begin{equation} \label{Equ:LASSO_constraint}
  \Sigma_{j=1}^{p}{|\beta_j |} \le t,
\end{equation}
where $t> 0$ is a preset parameter, $\alpha$ and $\beta= (\beta_1, \cdots, \beta_p)^T$ are parameters to be optimized.  In the model (\ref{Equ:LASSO}), only a few $\hat{\beta}_j$ will be nonzero and the wavelength positions of the corresponding $\tilde{x}_{j}$ or $\tilde{x}^i_j$ with nonzero $\hat{\beta}_{j}$ are exactly the detected positions of spectral features. In this model, the parameter $t$ controls the sparsity of the solution. The sparsity refers to number of detected features. In this work, the parameter $t$ is estimated by 10-fold cross validation \citep{Journal:Tibshirani:1996,Software:Sjostrand:2005}.

\begin{figure*}
  \centering
  \subfigure[Features for estimating $T_{\texttt{eff}}$]{
    \label{Fig:Lasso:feature:Teff} 
    \includegraphics[width =2in]{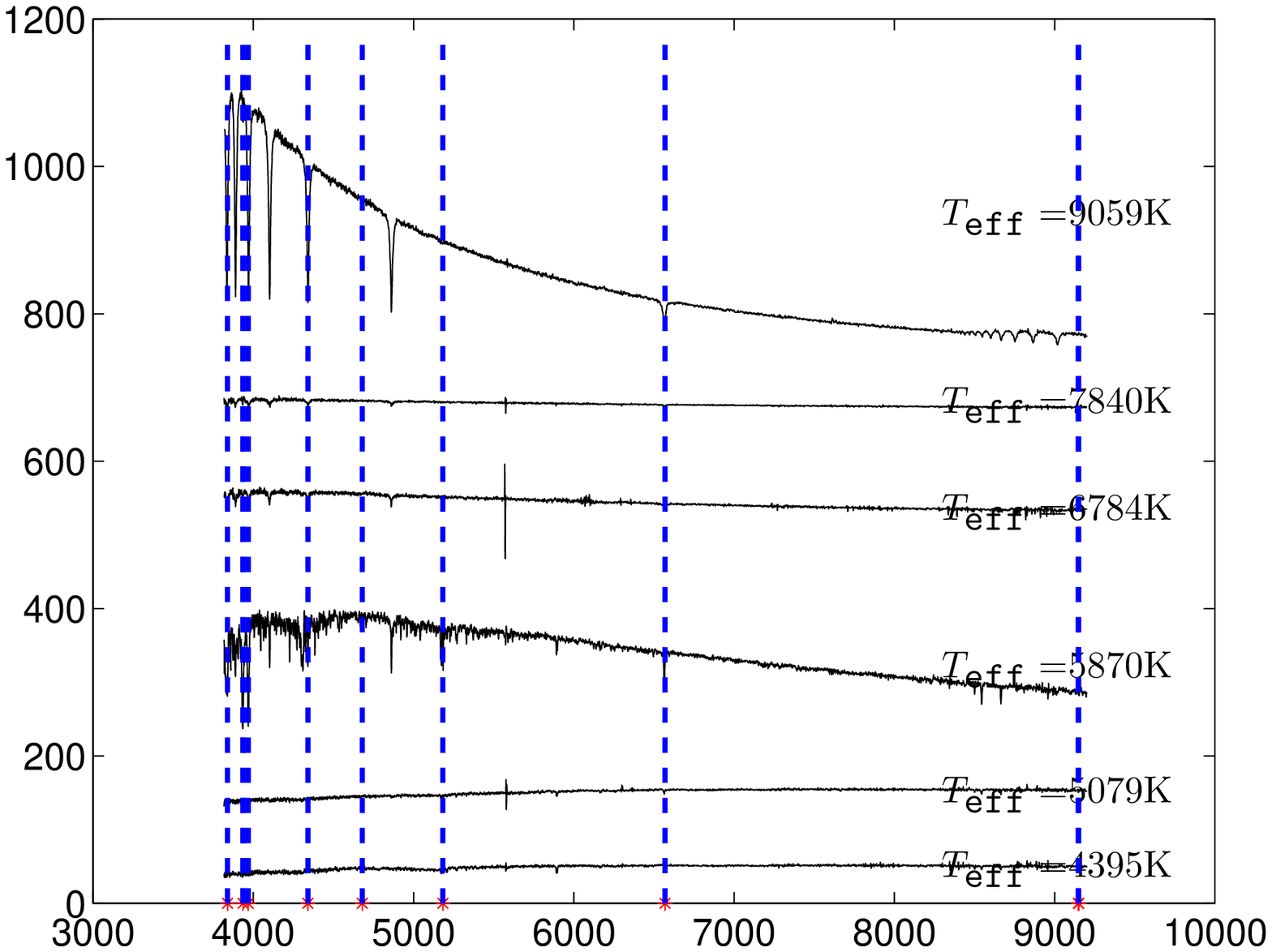}}
\hspace{-0.15in}
  \subfigure[Features for estimating log$~g$]{
    \label{Fig:Lasso:feature:logg} 
    \includegraphics[width =2in]{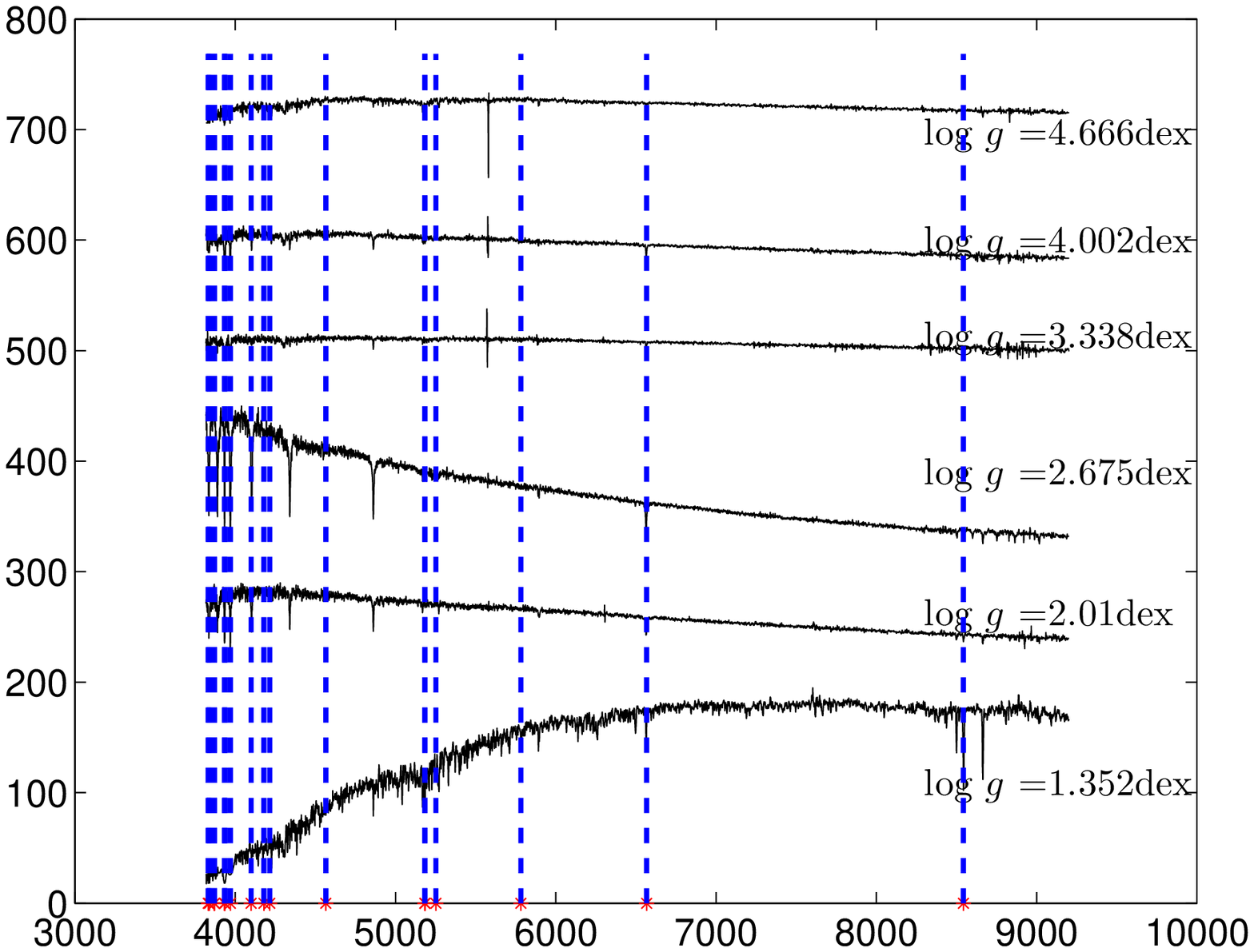}}
  \hspace{-0.15in}
  \subfigure[Features for estimating metallicity $\texttt{[}$Fe/H$\texttt{]}$]{
    \label{Fig:Lasso:feature:FeH} 
    \includegraphics[width =2in]{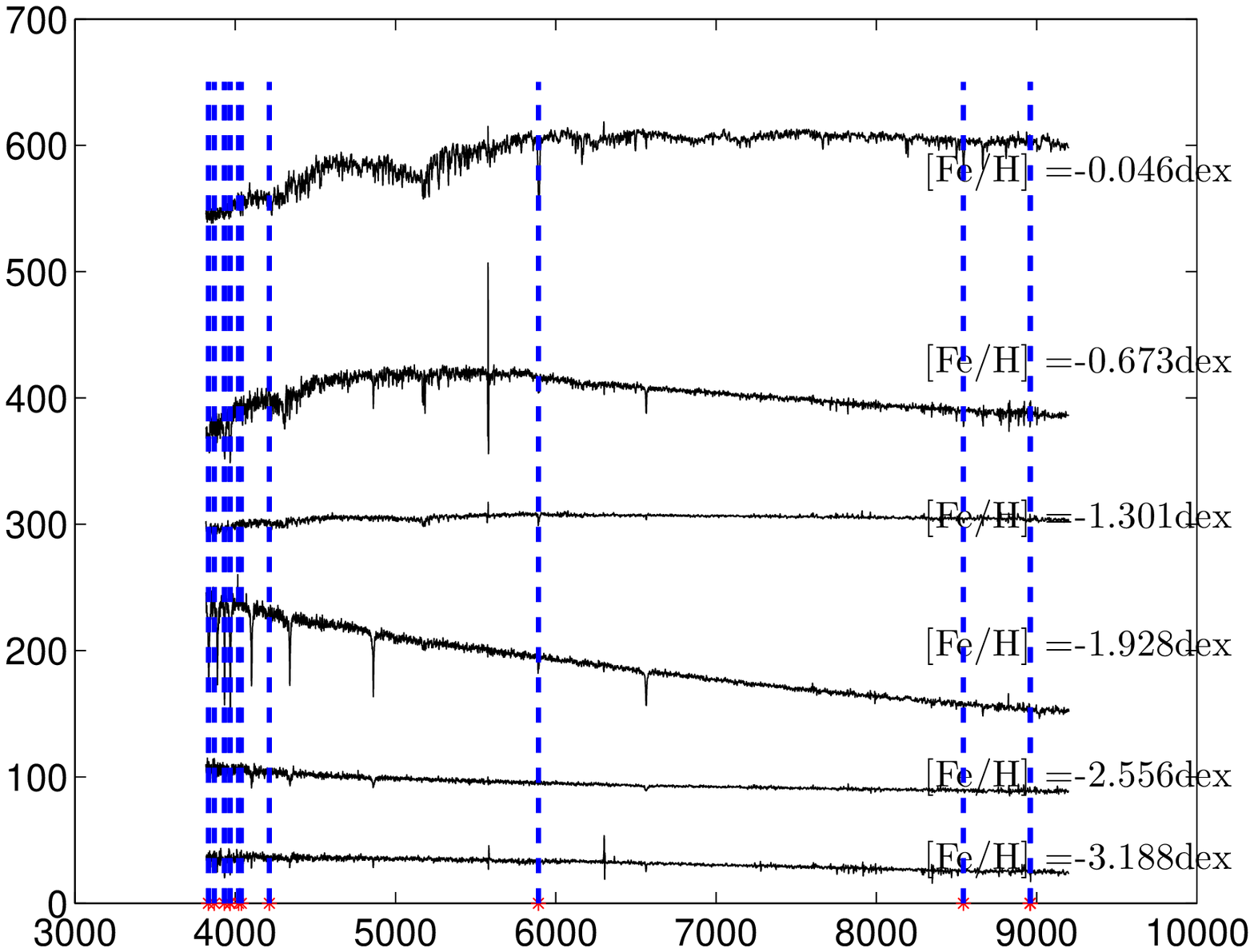}}
    \setlength{\abovecaptionskip}{0pt}
  \caption{Detected features for estimating the atmospheric parameters from SDSS stellar spectra. Black curves are stellar spectra with different parameters, red stars mark the positions of the detected features, and vertical dashed lines are to help us observe the representativeness of the detected features. The horizontal axis and vertical axis represent wavelength ($\AA$) and flux respectively.}
  \label{Fig:Lasso:feature:overall} 
\end{figure*}

\begin{figure*}
  \centering
  \subfigure[~]{
    \label{Fig:Lasso:feature:Teff:1} 
    \includegraphics[width =2in]{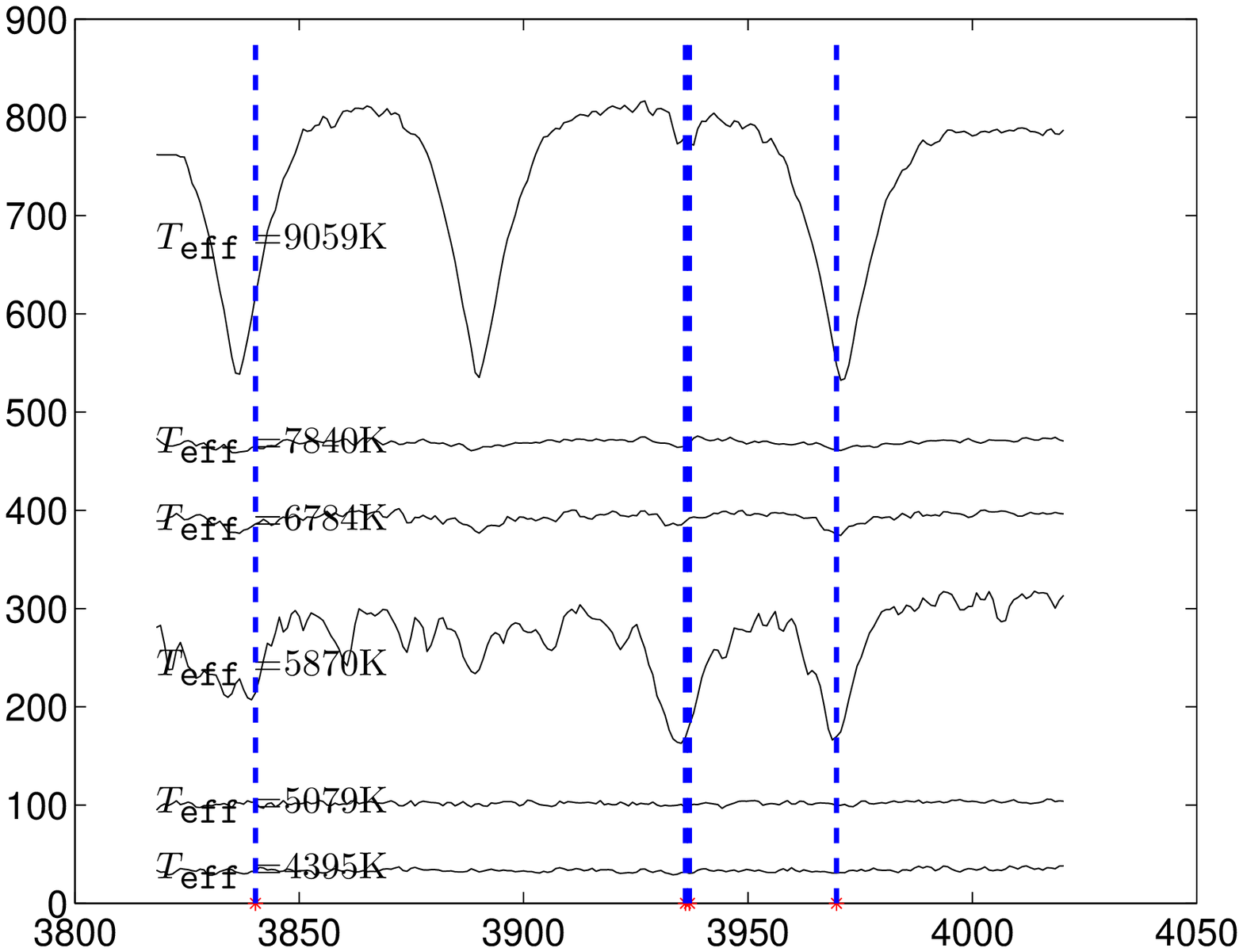}}
\hspace{-0.15in}
  \subfigure[~]{
    \label{Fig:Lasso:feature:Teff:2} 
    \includegraphics[width =2in]{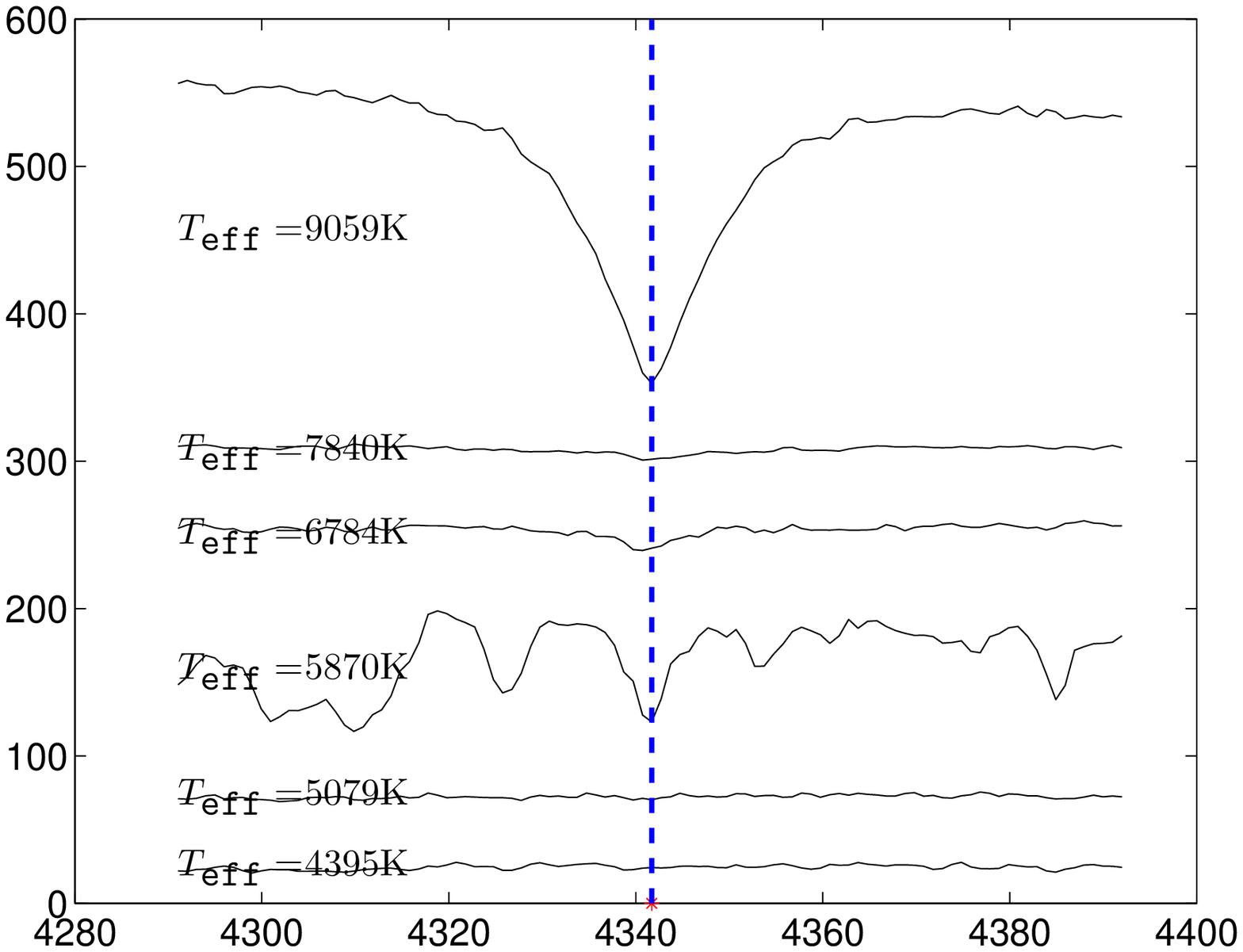}}
  \hspace{-0.15in}
  \subfigure[~]{
    \label{Fig:Lasso:feature:Teff:3} 
    \includegraphics[width =2in]{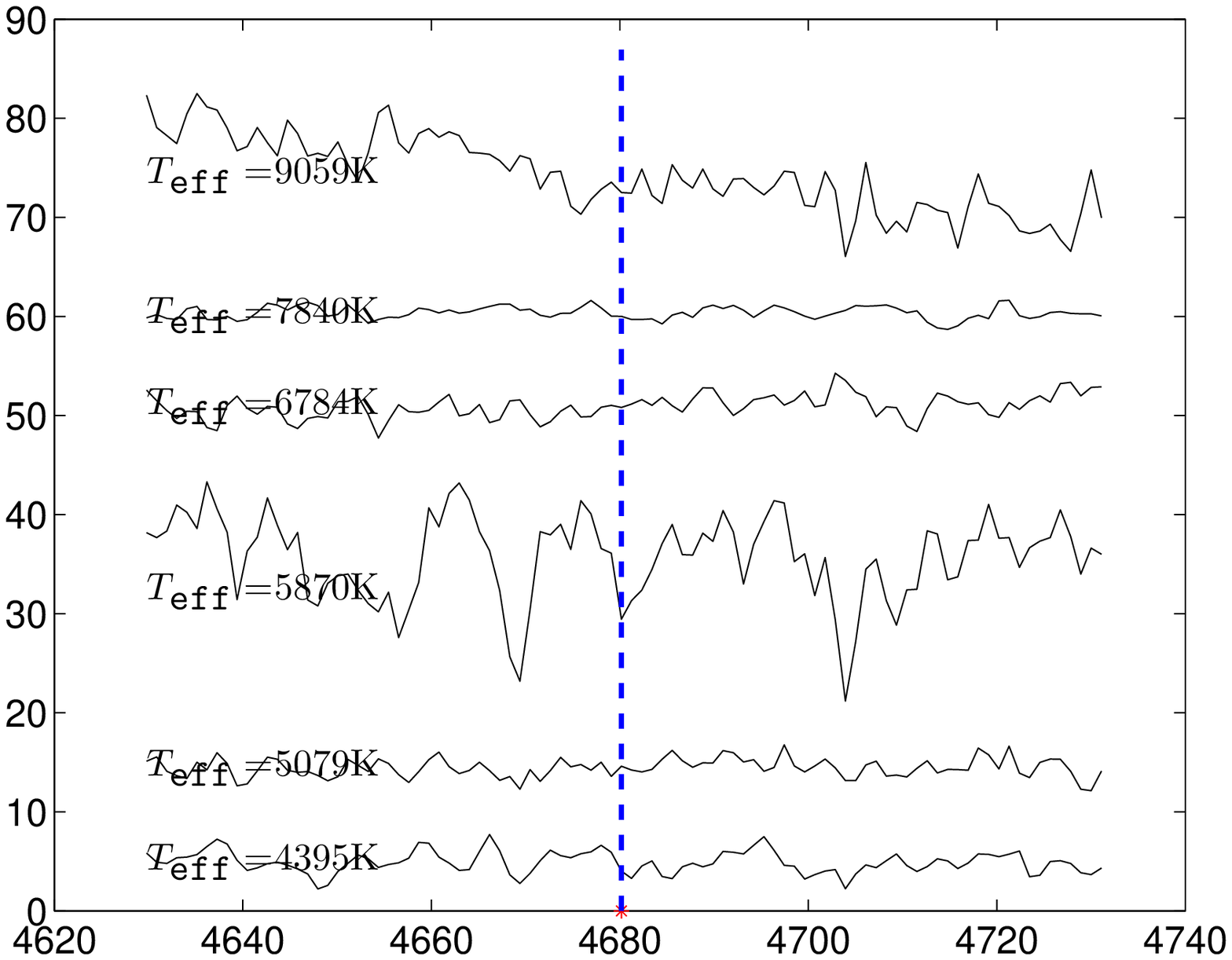}}
  \hspace{-0.15in}
    \subfigure[~]{
    \label{Fig:Lasso:feature:Teff:4} 
    \includegraphics[width =2in]{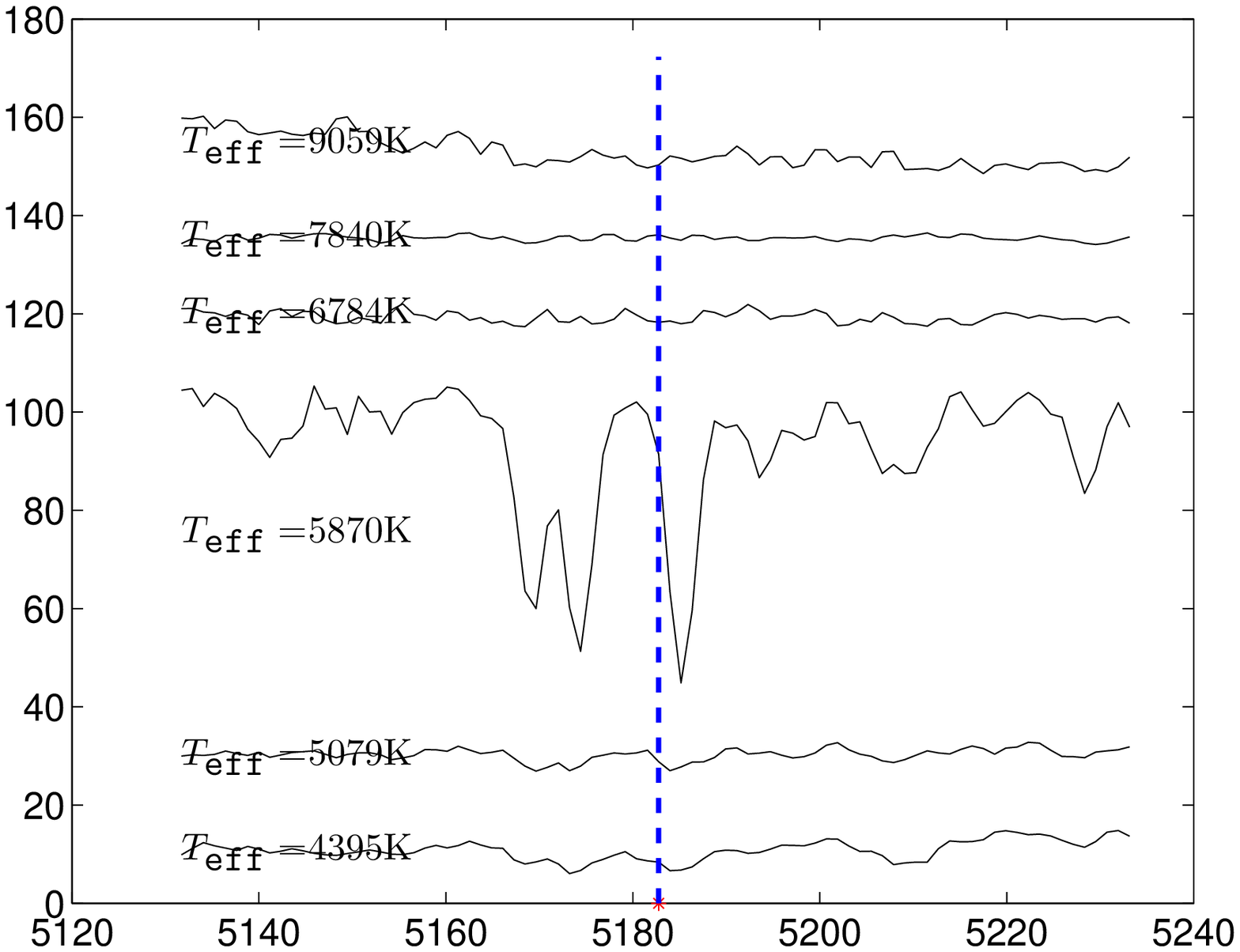}}
\hspace{-0.15in}
  \subfigure[~]{
    \label{Fig:Lasso:feature:Teff:5} 
    \includegraphics[width =2in]{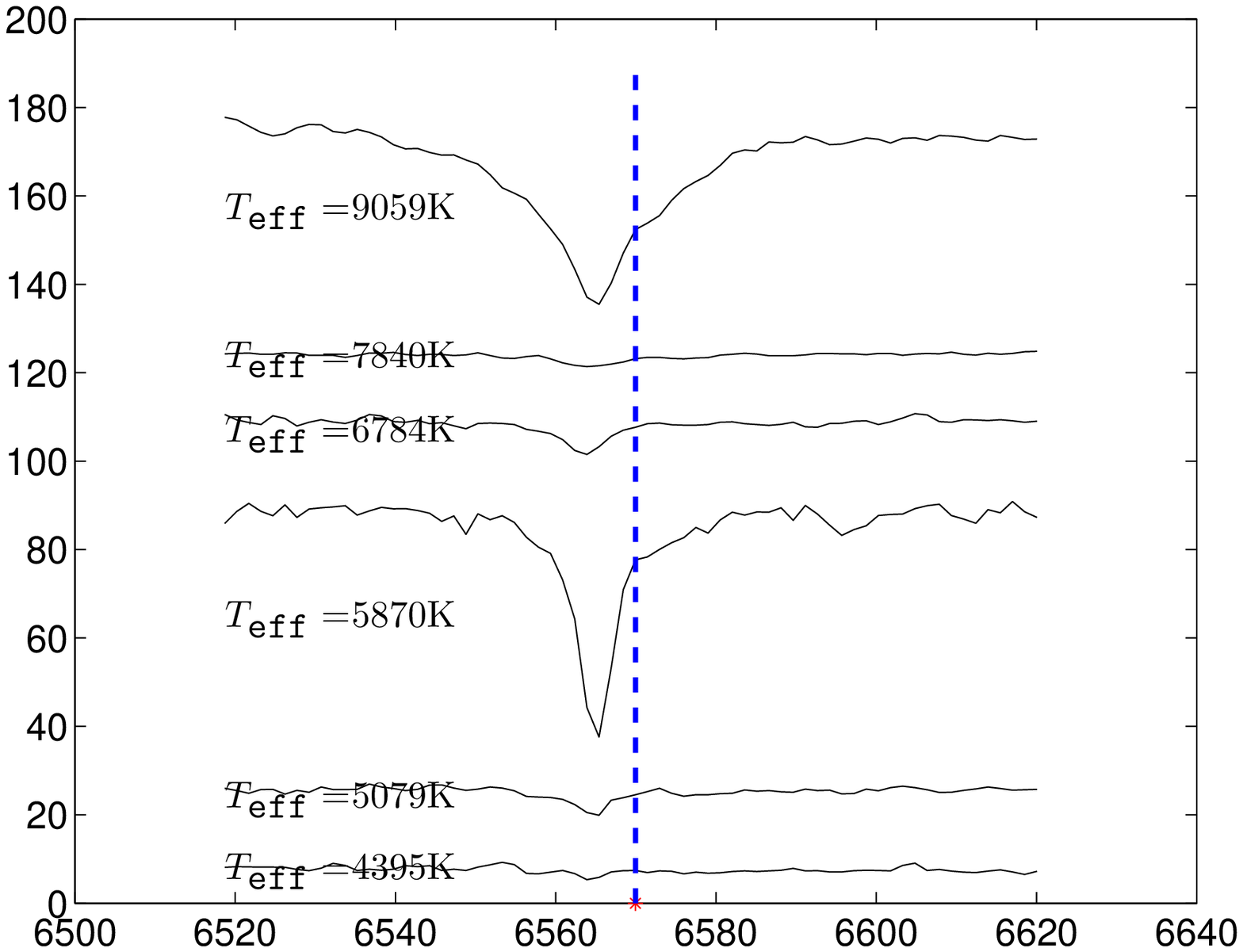}}
  \hspace{-0.15in}
  \subfigure[~]{
    \label{Fig:Lasso:feature:Teff:6} 
    \includegraphics[width =2in]{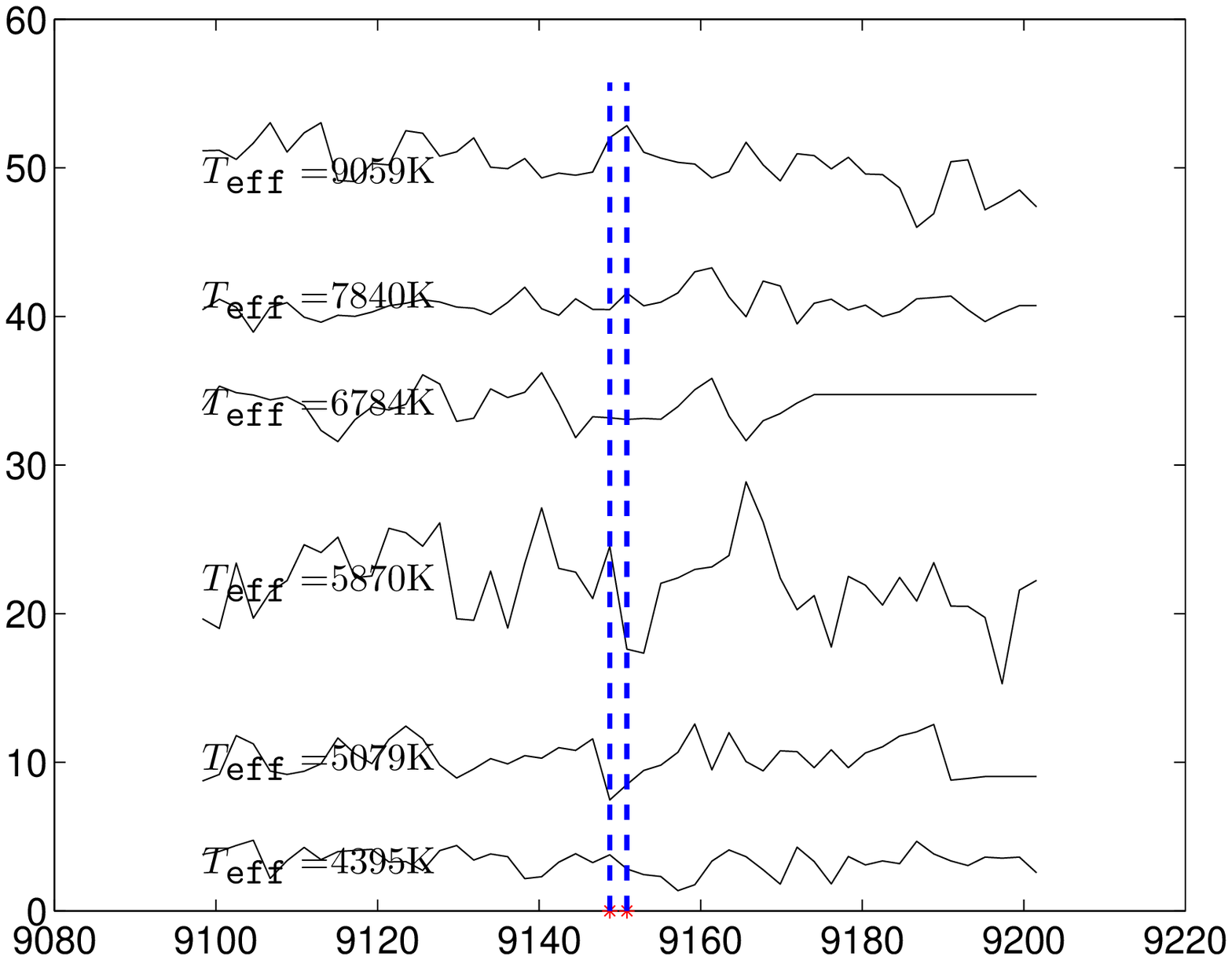}}
    \setlength{\abovecaptionskip}{-2pt}
  \caption{Close-range observations of the detected features for estimating $T_{\texttt{eff}}$.}
  \label{Fig:Lasso:feature:Teff:zoom} 
\end{figure*}

\begin{figure*}
  \centering
  \subfigure[~]{
    \label{Fig:Lasso:feature:logg:1} 
    \includegraphics[width =1.5in]{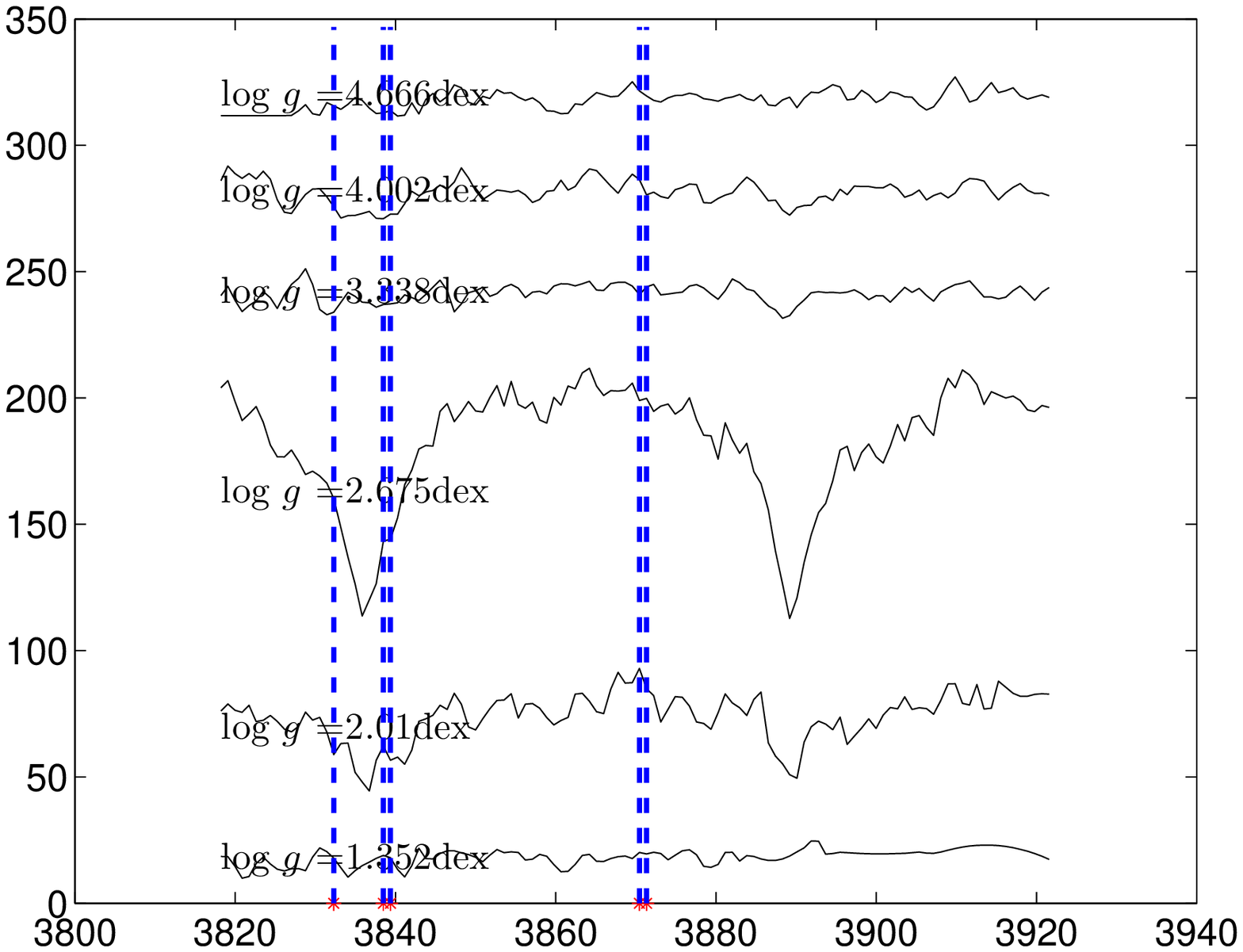}}
\hspace{-0.15in}
  \subfigure[~]{
    \label{Fig:Lasso:feature:logg:2} 
    \includegraphics[width =1.5in]{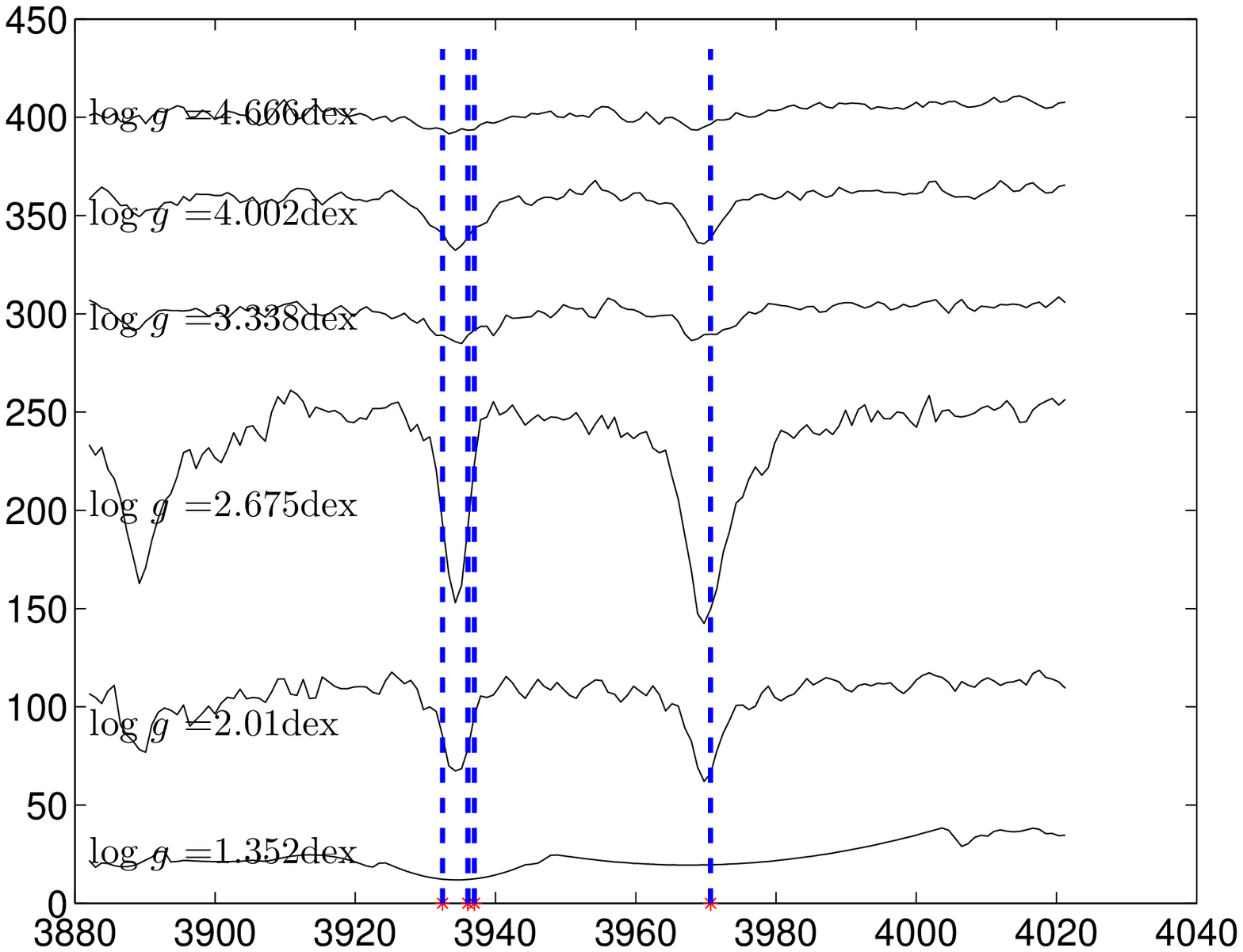}}
  \hspace{-0.15in}
  \subfigure[~]{
    \label{Fig:Lasso:feature:logg:3} 
    \includegraphics[width =1.5in]{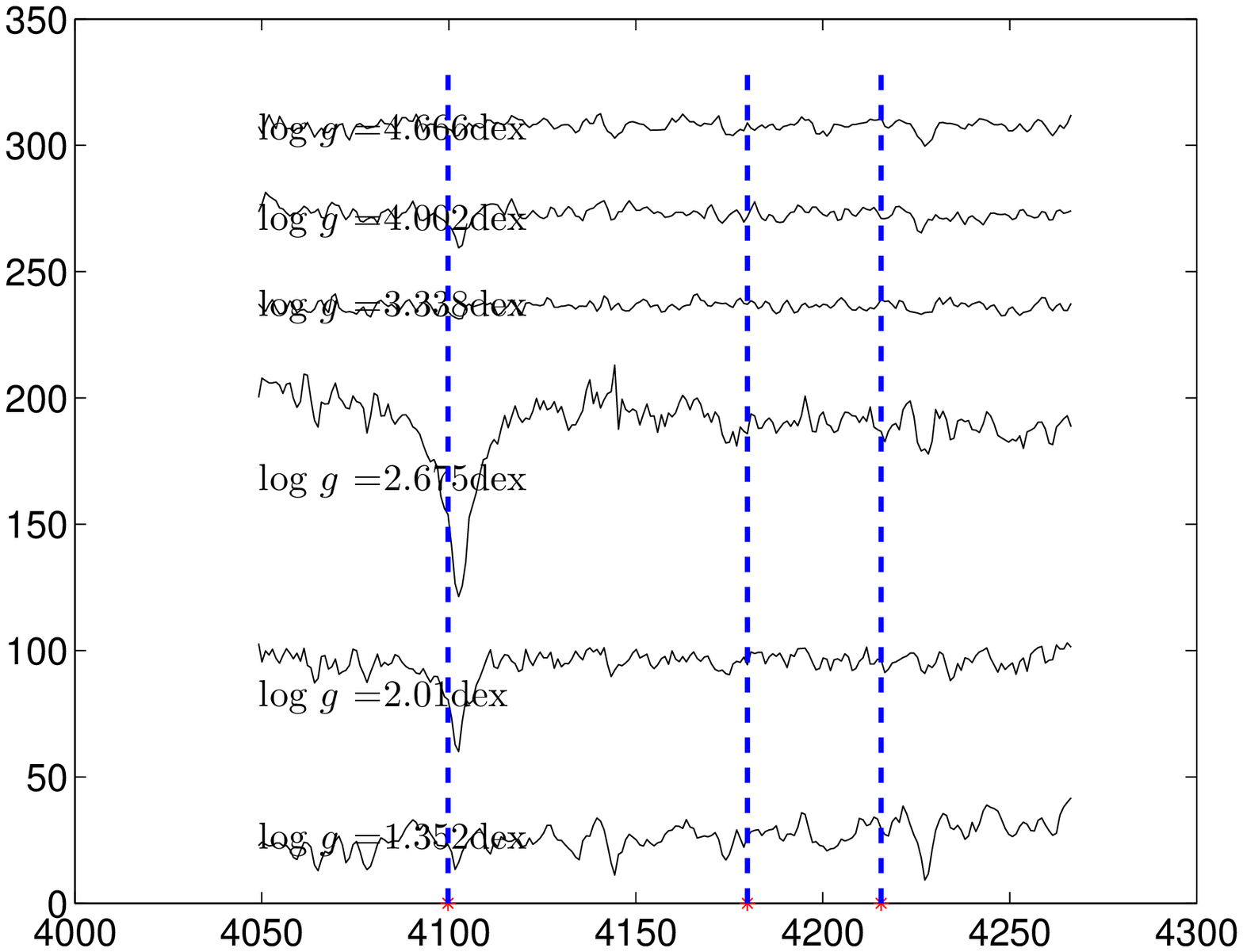}}
  \hspace{-0.15in}
    \subfigure[~]{
    \label{Fig:Lasso:feature:logg:4} 
    \includegraphics[width =1.5in]{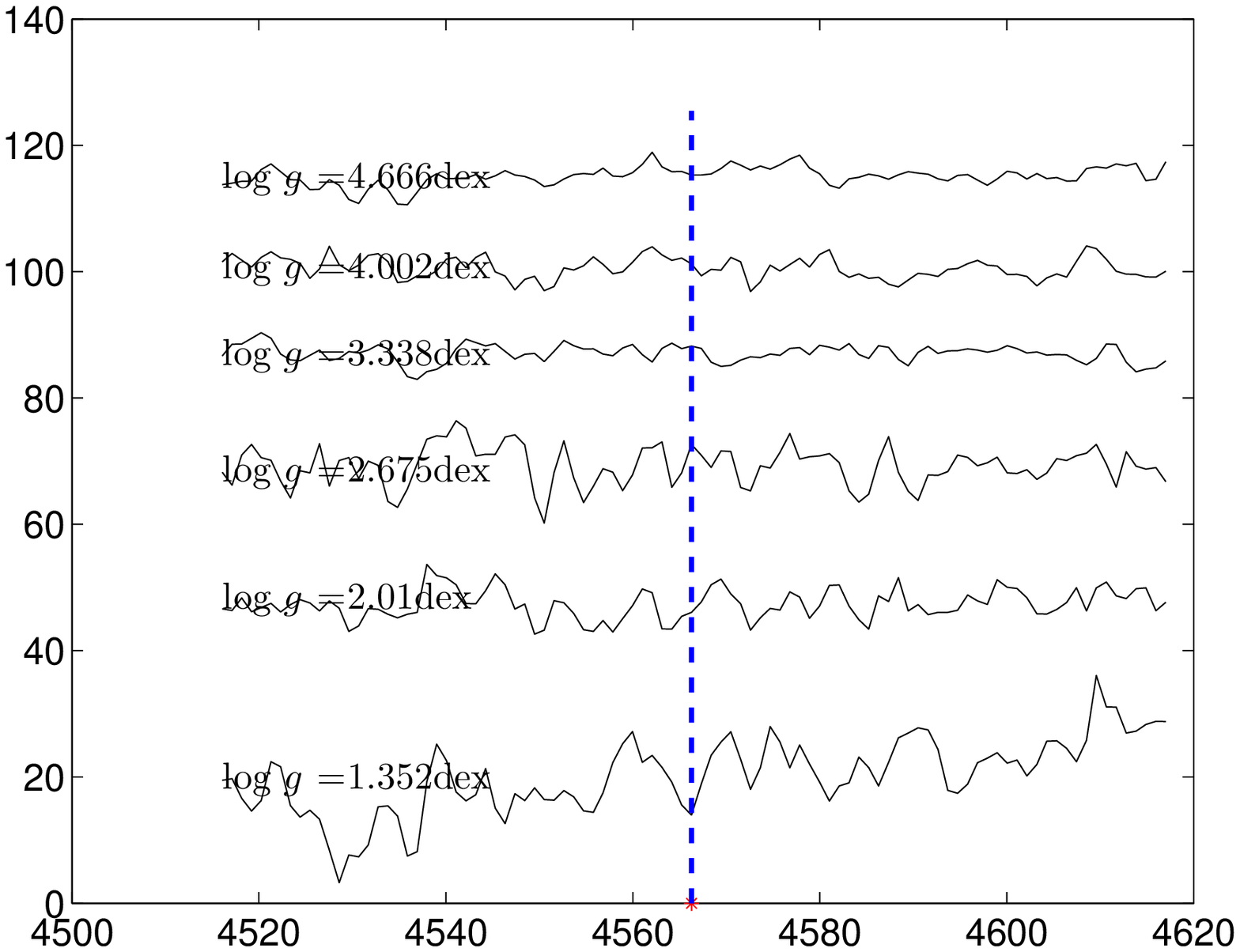}}
\hspace{-0.15in}
  \subfigure[~]{
    \label{Fig:Lasso:feature:logg:5} 
    \includegraphics[width =1.5in]{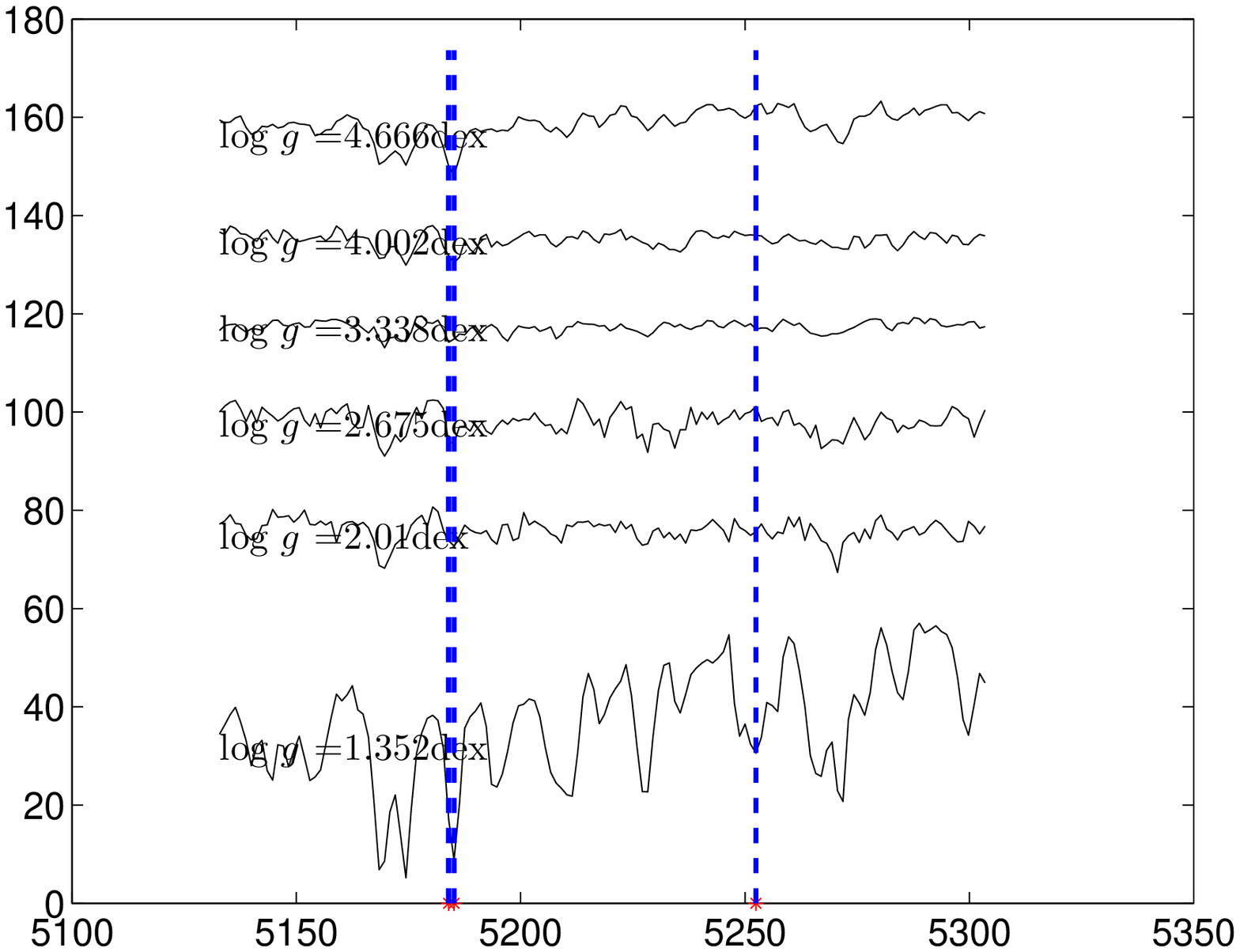}}
  \hspace{-0.15in}
  \subfigure[~]{
    \label{Fig:Lasso:feature:logg:6} 
    \includegraphics[width =1.5in]{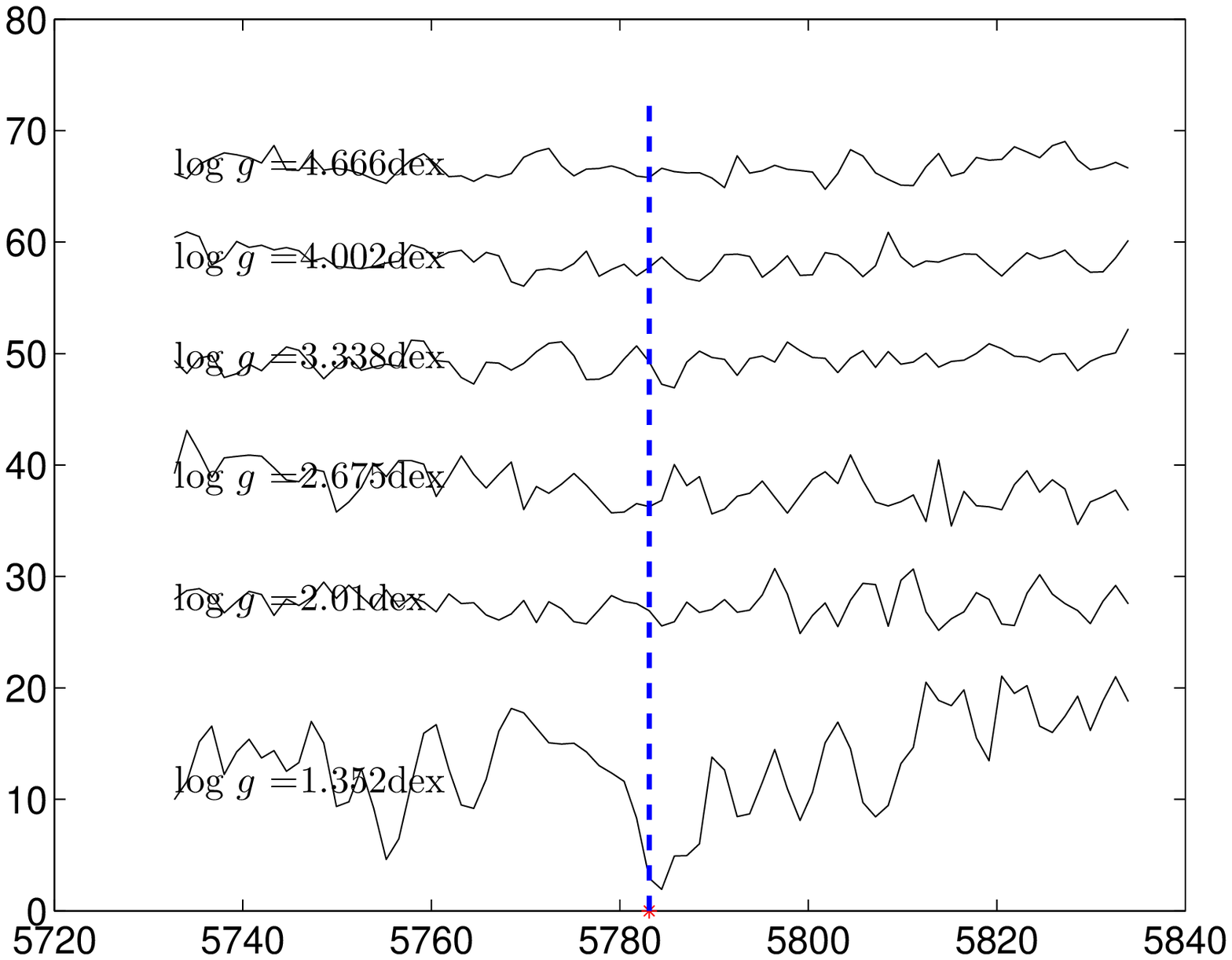}}
    \hspace{-0.15in}
  \subfigure[~]{
    \label{Fig:Lasso:feature:logg:7} 
    \includegraphics[width =1.5in]{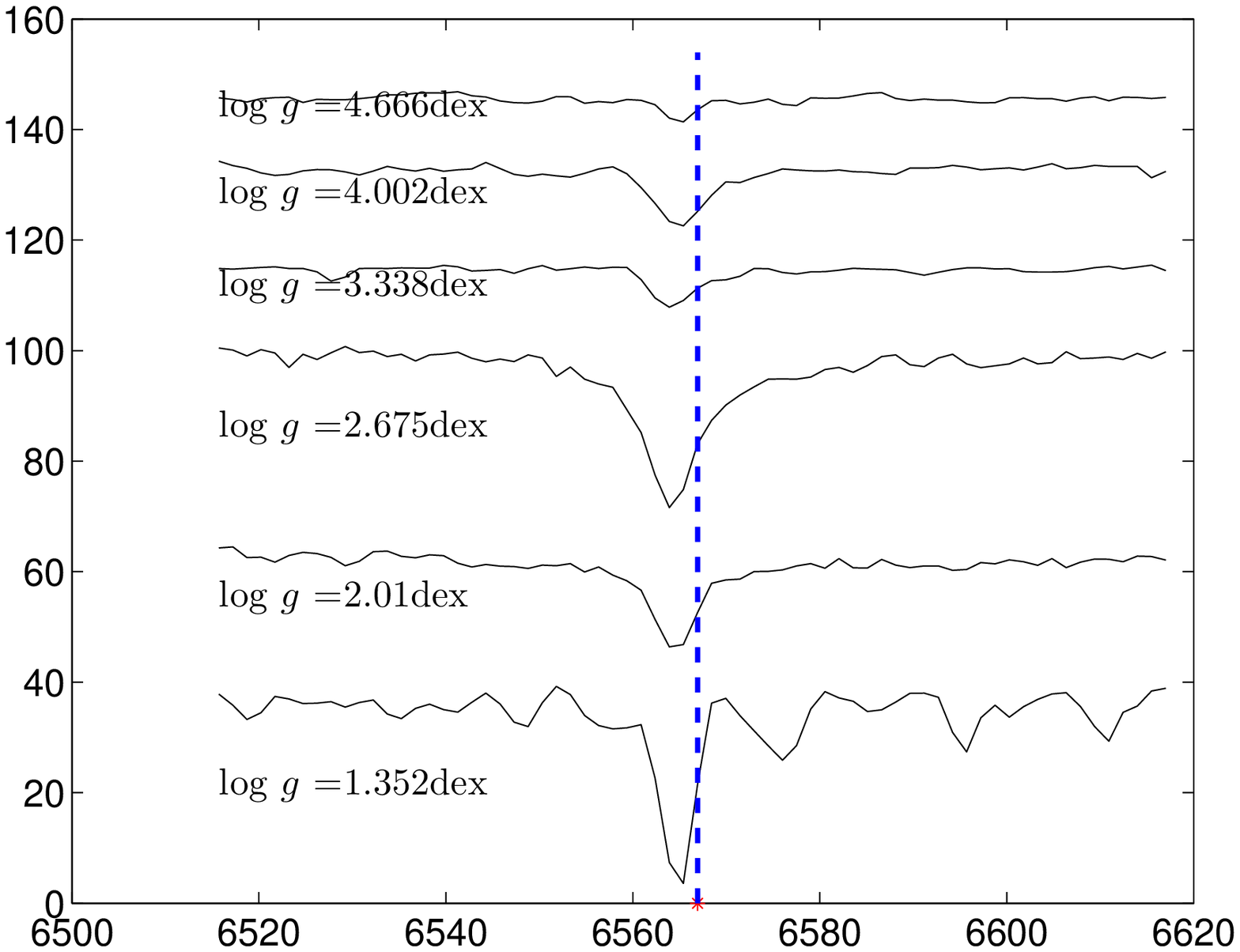}}
  \hspace{-0.15in}
  \subfigure[~]{
    \label{Fig:Lasso:feature:logg:8} 
    \includegraphics[width =1.5in]{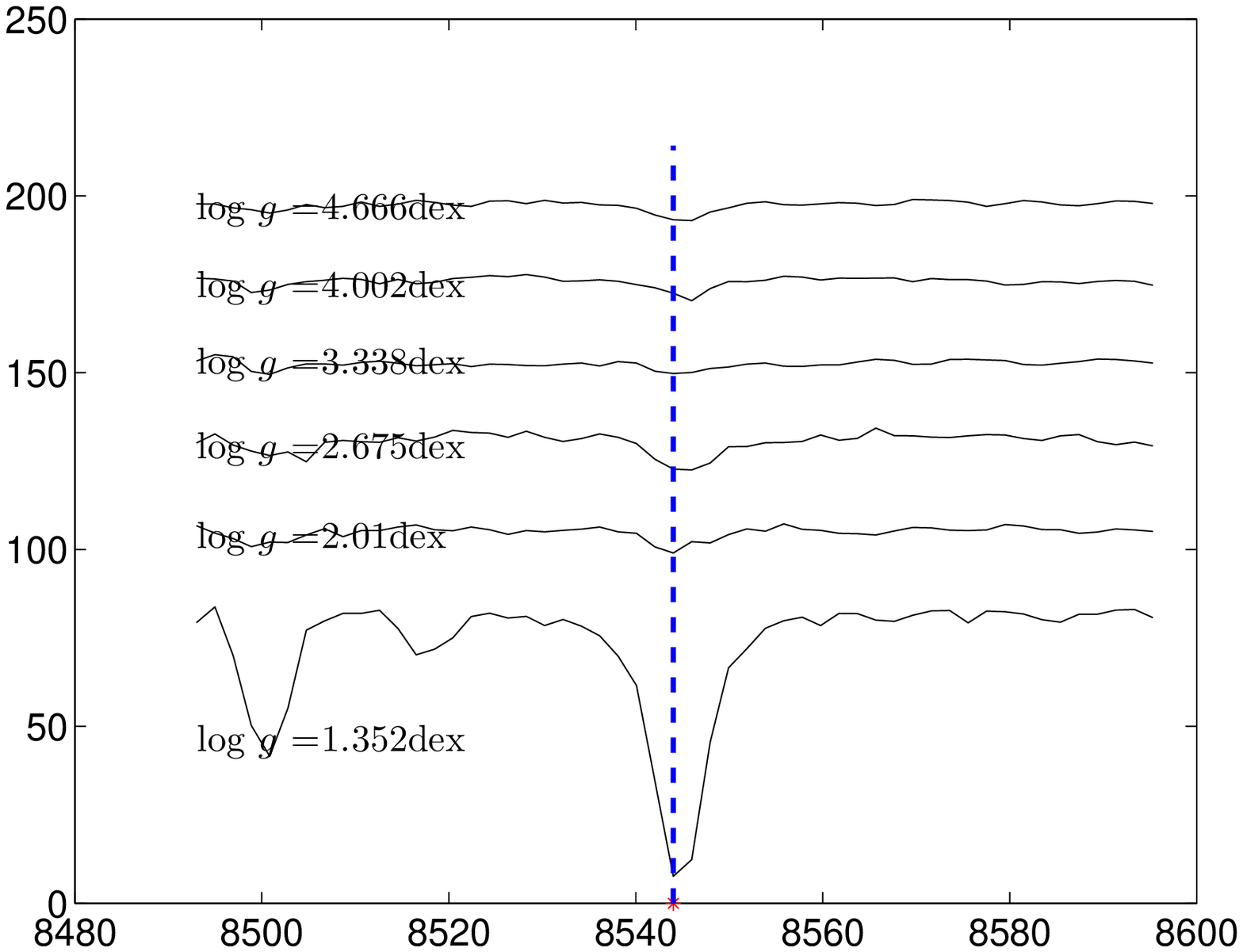}}
    \setlength{\abovecaptionskip}{-2pt}
  \caption{Close-range observations of the detected features for estimating log$~g$.}
  \label{Fig:Lasso:feature:logg:zoom} 
\end{figure*}

\begin{figure*}
  \centering
  \subfigure[~]{
    \label{Fig:Lasso:feature:FeH:1} 
    \includegraphics[width =2in]{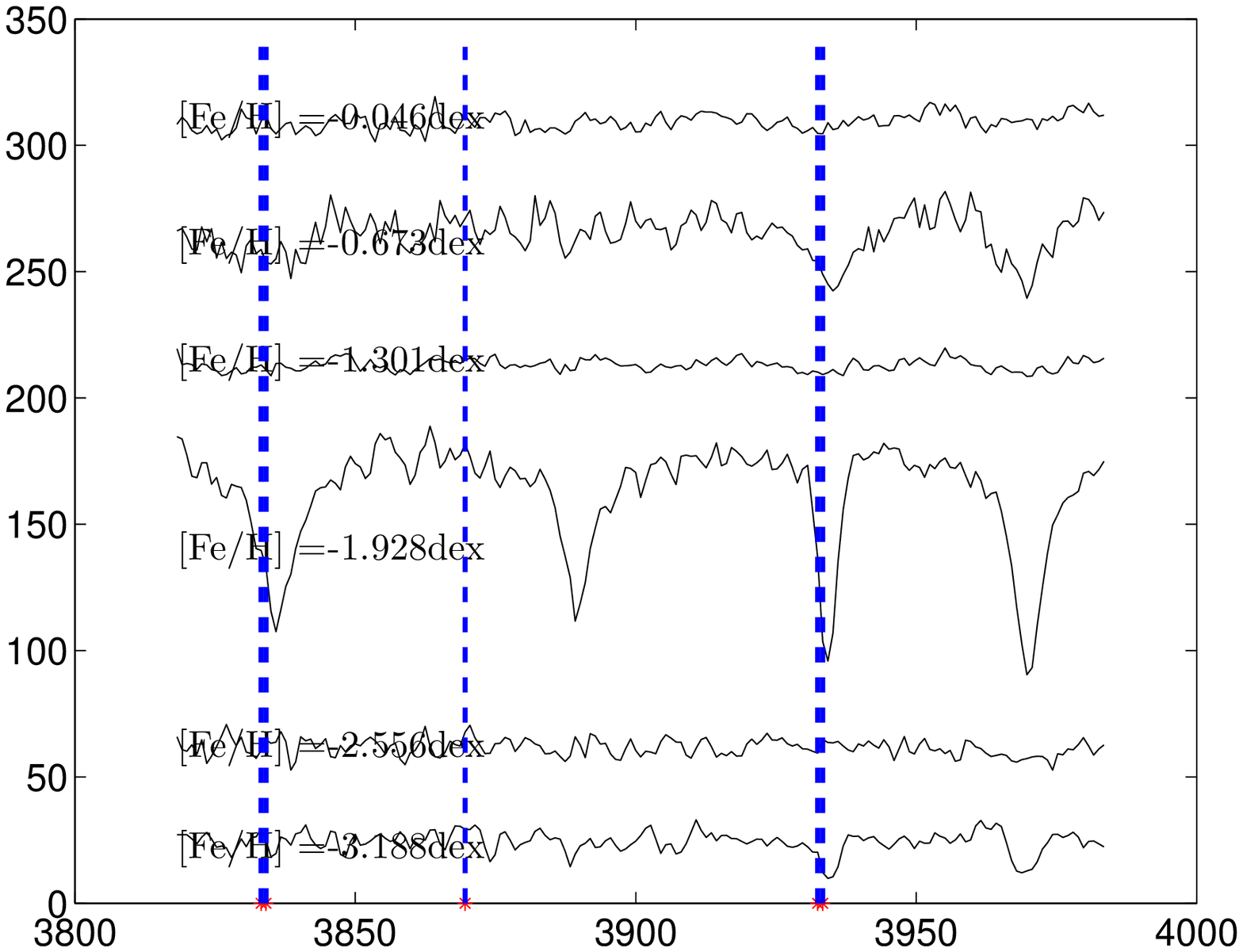}}
\hspace{-0.15in}
  \subfigure[~]{
    \label{Fig:Lasso:feature:FeH:2} 
    \includegraphics[width =2in]{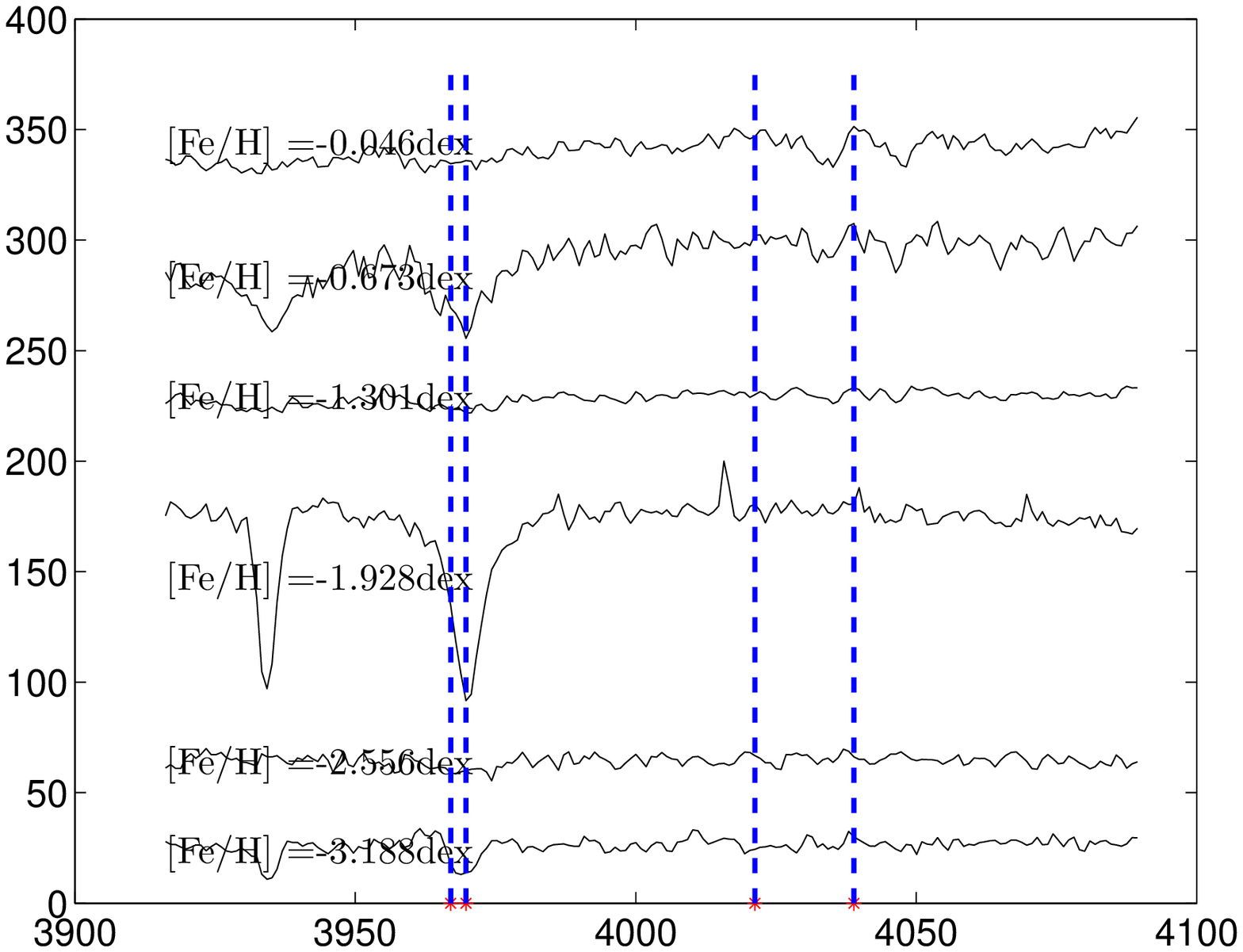}}
  \hspace{-0.15in}
  \subfigure[~]{
    \label{Fig:Lasso:feature:FeH:3} 
    \includegraphics[width =2in]{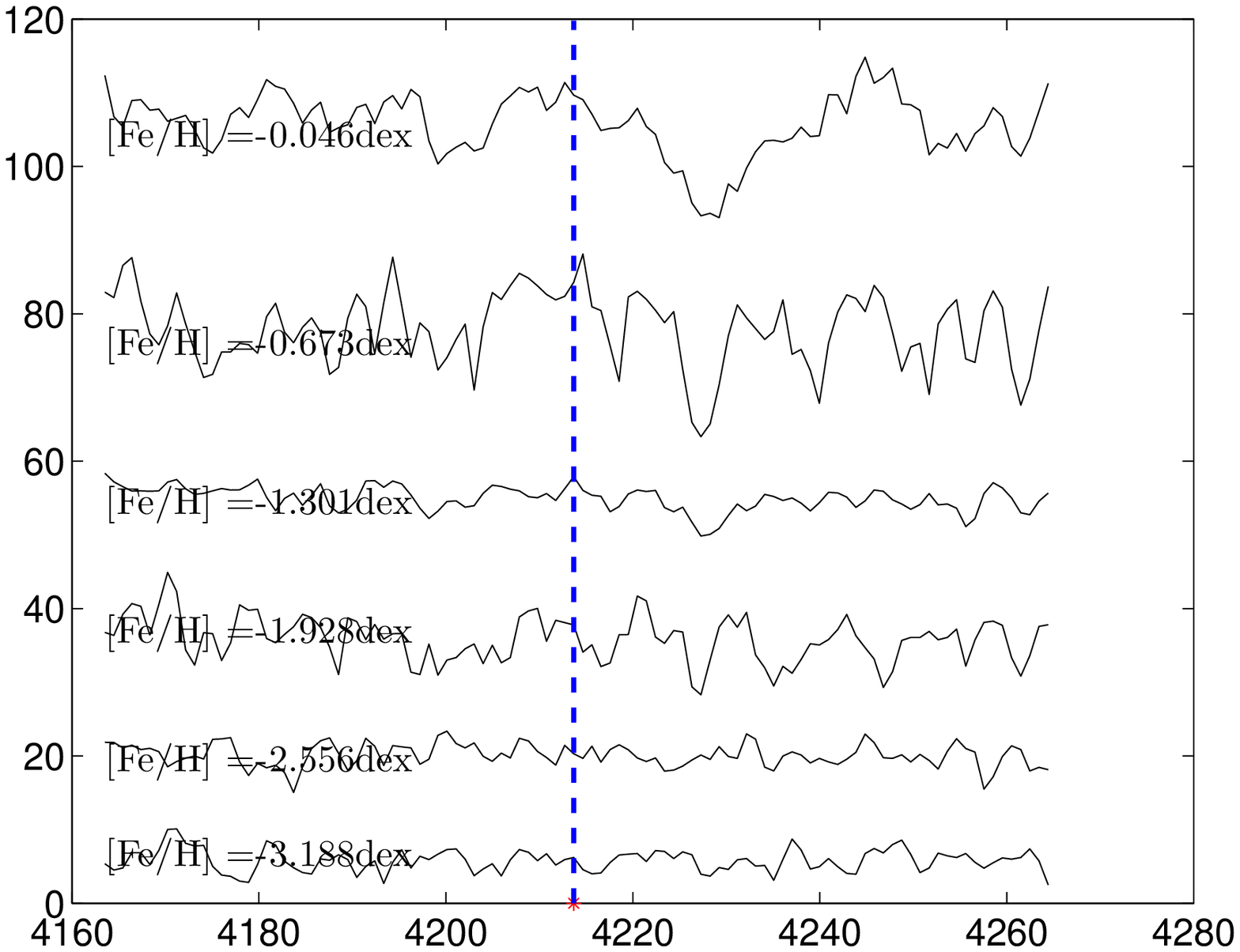}}
  \hspace{-0.15in}
    \subfigure[~]{
    \label{Fig:Lasso:feature:FeH:4} 
    \includegraphics[width =2in]{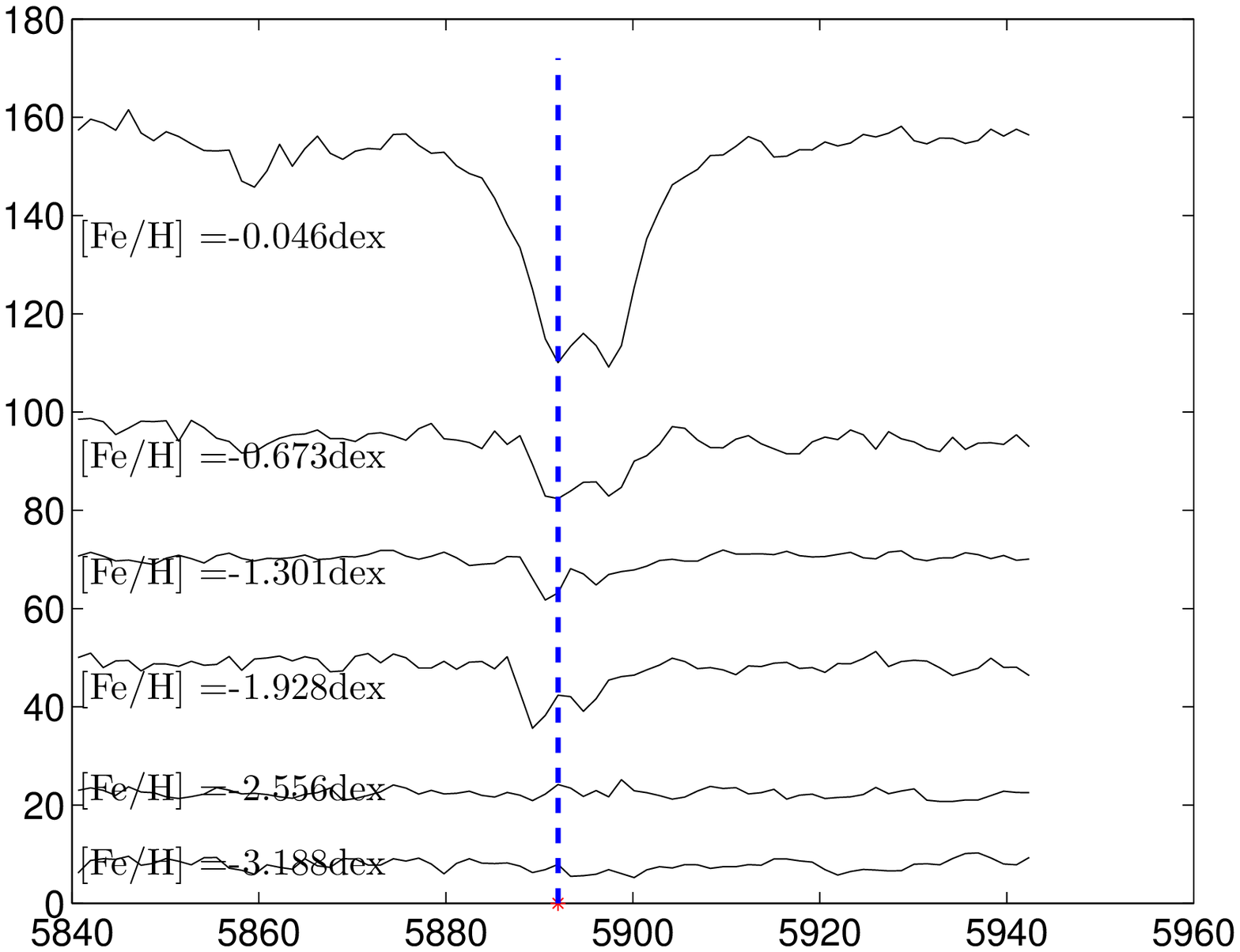}}
\hspace{-0.15in}
  \subfigure[~]{
    \label{Fig:Lasso:feature:FeH:5} 
    \includegraphics[width =2in]{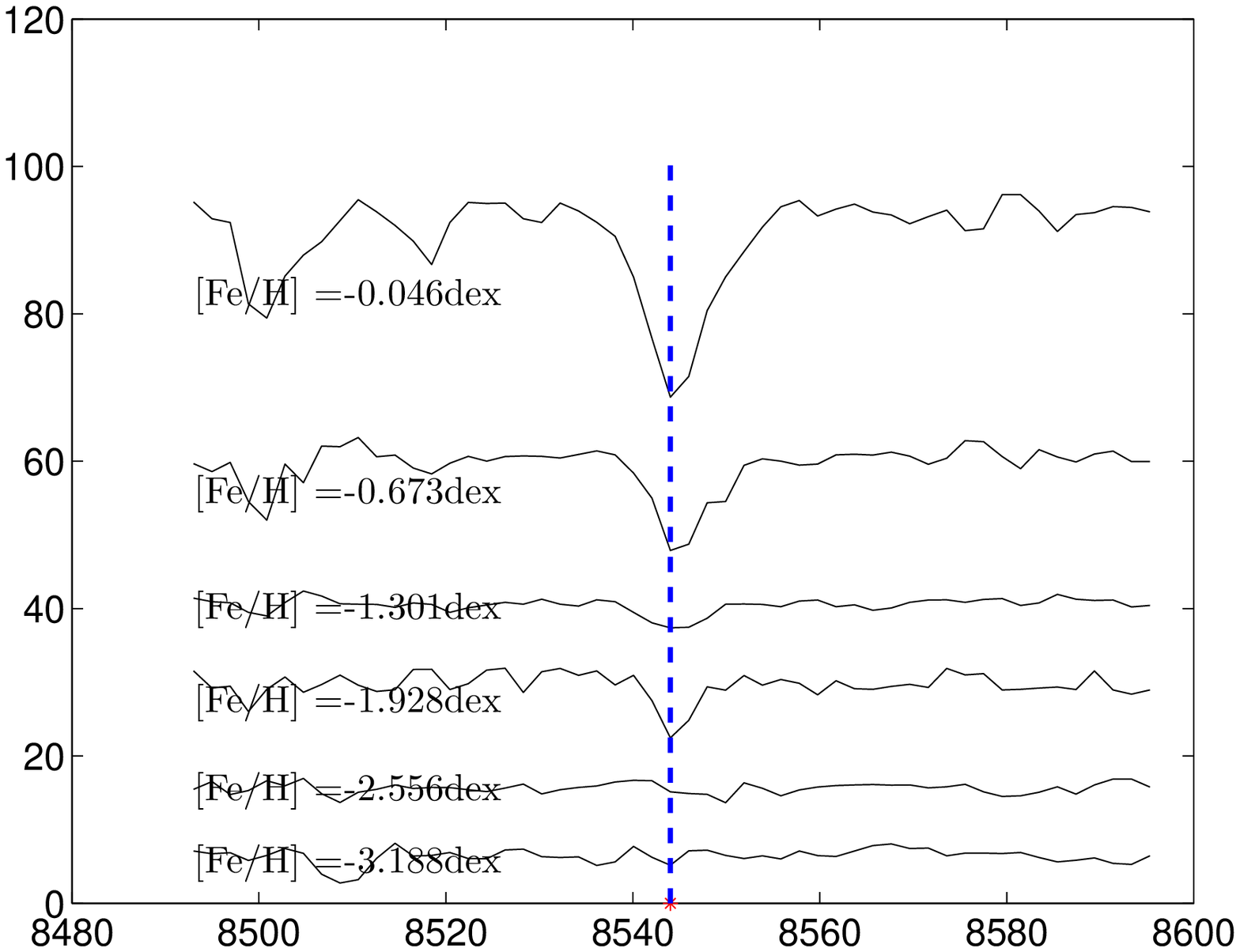}}
  \hspace{-0.15in}
  \subfigure[~]{
    \label{Fig:Lasso:feature:FeH:6} 
    \includegraphics[width =2in]{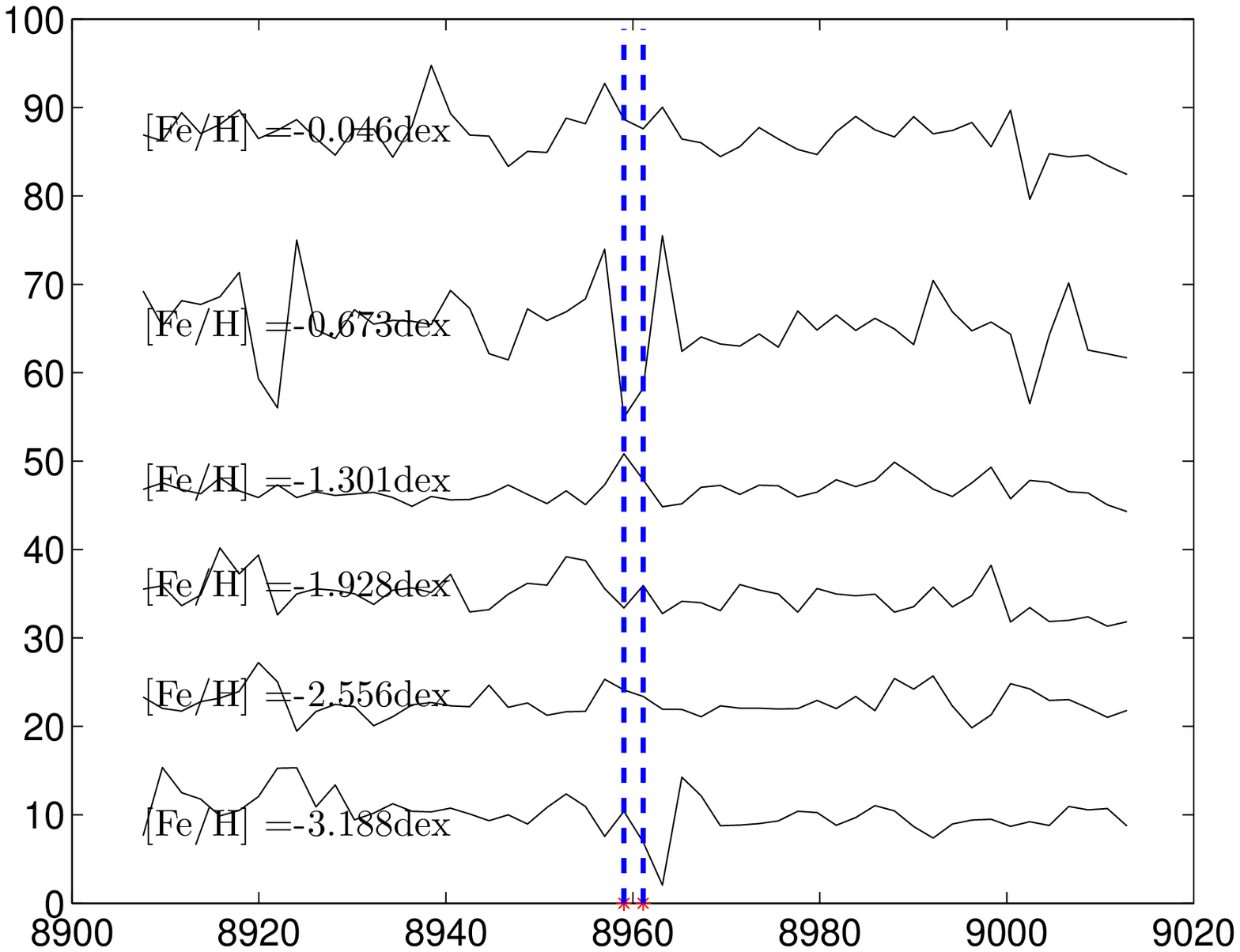}}
    \setlength{\abovecaptionskip}{-2pt}
  \caption{Close-range observations of the detected features for estimating [Fe/H].}
  \label{Fig:Lasso:feature:FeH:zoom} 
\end{figure*}

Detected features are presented in Fig. \ref{Fig:Lasso:feature:overall} visually, and their specific wavelength positions are listed in Table \ref{Tab:LASSO:Features:Teff}, Table \ref{Tab:LASSO:Features:Logg}, and Table \ref{Tab:LASSO:Features:FeH}. A specific feature can be referred to by its label, position in wavelength, or in log(wavelength). For example, $\lambda^w_{T2}$ and $\lambda^l_{T2}$ refer to the position of the second feature of $T_{\texttt{eff}}$ in wavelength and in log(wavelength) respectively (Table \ref{Tab:LASSO:Features:Teff}), $\lambda^w_{L3}$ and $\lambda^l_{L3}$ represent the position of the third feature of log$~g$ in wavelength and log(wavelength) respectively (Table \ref{Tab:LASSO:Features:Logg}):
 \begin{equation}
 \lambda^w_{T2} = 3936.0626~{\AA},
 \end{equation}
 \begin{equation}
 \lambda^l_{T2} = 3.5951~dex,
 \end{equation}
 \begin{equation}
 \lambda^w_{L3} = 3839.3880~{\AA},
 \end{equation}
 \begin{equation}
 \lambda^l_{L3} = 3.5843~dex.
 \end{equation}
To facilitate finding the characteristics of the detected features, we also show the features by some close-range views in Fig. \ref{Fig:Lasso:feature:Teff:zoom}, Fig. \ref{Fig:Lasso:feature:logg:zoom}, and Fig. \ref{Fig:Lasso:feature:FeH:zoom}.

 In this work, spectral features are extracted by the following two procedures: 1) detect the positions of spectral features where the spectral fluxes have some variance with the parameter in theory; 2) describe the features based on one or several fluxes near the detected positions (Section \ref{Sec:Feature_Description}). To highlight the variance of fluxes at one specific detected position, we sometimes use term `feature' instead of `position' or `descriptor' (Fig. \ref{Fig:Lasso:feature:overall}, Fig. \ref{Fig:Lasso:feature:Teff:zoom}, Fig. \ref{Fig:Lasso:feature:logg:zoom}, and Fig. \ref{Fig:Lasso:feature:FeH:zoom}). The variance is closely related to the discriminability of a spectrum, and is essential for a good feature.

The proposed feature extracting technique has the following advantages:
\begin{itemize}
\item{\textbf{Interpretability}~~The detected features all have specific wavelength positions, based on which we can backtrack the specific effective factors (Fig.~\ref{Fig:Lasso:feature:Teff:zoom}, Fig.~\ref{Fig:Lasso:feature:logg:zoom} and Fig.~\ref{Fig:Lasso:feature:FeH:zoom}) and evaluate their contributions to estimating the atmospheric parameters from stellar spectra (Section \ref{Sec:Feature:eva}). For example,
    $H_\gamma$ is a sensitive line to surface temperature (T5 in Table \ref{Tab:LASSO:Features:Teff} and Fig.~\ref{Fig:Lasso:feature:Teff:2});
    Ca~II, H~I, $H_\delta$, and Ca~I are sensitive to surface gravity (L9 \& L10 in Table \ref{Tab:LASSO:Features:Logg}, Fig.~\ref{Fig:Lasso:feature:logg:2} and Fig.~\ref{Fig:Lasso:feature:logg:3});
      H$_\alpha$ and Ca~H are sensitive to both surface temperature and gravity (T8 in table \ref{Tab:LASSO:Features:Teff} and Fig.~\ref{Fig:Lasso:feature:Teff:5}, L18 in table \ref{Tab:LASSO:Features:Logg} and Fig.~\ref{Fig:Lasso:feature:logg:7});
      Ca~II line (L19 in table \ref{Tab:LASSO:Features:Logg} and Fig.~\ref{Fig:Lasso:feature:logg:8}, F12 in table \ref{Tab:LASSO:Features:FeH} and Fig.~\ref{Fig:Lasso:feature:FeH:5}) is an effective factor for both surface gravity and stellar metal abundance \citep{Journal:Cenarro:2001}.
    }
\item{\textbf{Efficiency}~~Very few features are detected for every parameter estimation problem, and every feature can be described by, at most, 17 fluxes near the detected wavelength position (Section \ref{Sec:Feature_Description}). Therefore, it is very efficient to compute the features and estimate the atmospheric parameters from stellar spectra based on this scheme. For example, only 10 features need to be computed to estimate $T_{\texttt{eff}}$, 19 features to estimate log~$g$, and 14 features to estimate [Fe/H]. More analysis on efficiency is presented in section \ref{Sec:Conclusion}.}
\item{\textbf{Good generalization}~~In this study, every spectrum is described by 3~821 fluxes. LASSO can identify 10 local features to estimate  $T_{\texttt{eff}}$, 19 features  to estimate log$~g$, 14 features to estimate [Fe/H], and the parameterization results are excellent comparing with the similar studies in literatures (Section \ref{Sec:Feature_Description} and \citep{Journal:Fiorentin:2007}). Therefore, the proposed scheme enhances the generalization performance by rejecting redundancy, which usually cannot improve the performance of the estimating system except introducing disturbances and overfitting. }
\item{\textbf{High robustness}~~The commonly used method PCA is of a global scheme. In PCA, every feature is computed from nearly all or most of the observed fluxes. This contributes to accumulation of the negative influence from noise, observation error, and calibration distortion. Our proposed method can determine the specific positions of effective features and obtain their descriptions only from one or several observed fluxes near the detected positions (Section \ref{Sec:Feature_Description}). Therefore, this scheme is more robust or immune to the aforementioned undue influences in theory, and this also is validated by the excellent performance on SDSS/SEGUE spectra.}
\end{itemize}

\begin{table}\scriptsize
\centering
\caption{Detected typical positions for estimating log$~g$ from stellar spectra. TPW $\lambda^w$: Typical position in wavelength, TPL $\lambda^l$: Typical position in log(wavelength), TP: typical position.}
\begin{tabular}{ c c  c c   }
  \hline \hline
label         &TPW $\lambda^w$ ({\AA})       &TPL $\lambda^l$               & lines near TP \\ \hline
   L1            &      3832.3221 &3.5835         &   Mg~I,Fe~I, He~I,Na~I               \\
   L2            &      3838.5040 &3.5842         &   He~I, Mg~I,VI                    \\
   L3            &      3839.3880 &3.5843         &   Fe~I, Fe~V                       \\
   L4            &      3870.4548 &3.5878         &     H8                           \\
   L5            &      3871.3461 &3.5879         &     H8                           \\
   L6            &      3932.4390 &3.5947         &     KP,Ca~IIK                     \\
   L7            &      3936.0626 &3.5951         &     KP,Ca~IIK                     \\
   L8            &      3936.9690 &3.5952         &     KP,Ca~IIK                     \\
   L9            &      3970.6536 &3.5989         &     Ca~II,H~I                 \\
   L10           &      4099.7937 &3.6128         &     $H_{\delta}$,CaI             \\
   L11           &      4179.8625 &3.6212         &      VI                          \\
   L12           &      4215.6253 &3.6249         &      CaI                         \\
   L13           &      4566.2743 &3.6596         &     Ba                           \\
   L14           &      5183.9643 &3.7147         &     Mg~I,Mg~Ic                     \\
   L15           &      5185.1581 &3.7148         &     Mg~I,Mg~Ic                     \\
   L16           &      5252.4509 &3.7204         &     Fe~II                        \\
   L17           &      5783.1173 &3.7622         &     Fe~II, Fe~I, O~II,VI            \\
   L18           &      6566.9241 &3.8174         &     $H_{\alpha}$, Ca~H                 \\
   L19           &      8544.0150 &3.9317         &     Ca~II                         \\ \hline
\end{tabular}\label{Tab:LASSO:Features:Logg}
\end{table}

\begin{table}\scriptsize
\centering
\caption{Detected typical positions for estimating [Fe/H] from SDSS stellar spectra. TPW $\lambda^w$: Typical position in wavelength, TPL $\lambda^l$: Typical position in log(wavelength), TP: typical position.}
\begin{tabular}{ c c  c c   }
  \hline \hline
label         &TPW $\lambda^w$ ({\AA})       &TPL $\lambda^l$               & lines near TP \\ \hline
   F1            &       3833.2046&3.5836         &   O~II,FI,Ca~III,He~I                \\
   F2            &       3834.0873&3.5837         &   FeI, OVI, NI,FeII               \\
   F3            &       3869.5637&3.5877         &   Fe~I                             \\
   F4            &       3932.4390&3.5947         &   KP, Ca~IIK                       \\
   F5            &       3933.3446&3.5948         &   KP, Ca~IIK                       \\
   F6            &       3966.9982&3.5985         &   Ca~IIHKp,Heps                    \\
   F7            &       3969.7394&3.5988         &   Ca~IIHKp,Heps                    \\
   F8            &       4021.2586&3.6044         &   He~I                             \\
   F9            &       4038.8898&3.6063         &   He~I                             \\
   F10           &       4213.6844&3.6247         &   Ca~I                             \\
   F11           &       5891.9900&3.7703         &   Na~I,Na                          \\
   F12           &       8544.0150&3.9317         &   Ca~II,Ca~IIa                      \\
   F13           &       8959.0508&3.9523         &   Fe~I,Fe~II,Ne~II                   \\
   F14           &       8961.1140&3.9524         &   Fe~I                            \\ \hline
\end{tabular}\label{Tab:LASSO:Features:FeH}
\end{table}

\section{Non-linear Regression Model for Atmospheric Parameter Estimation and Evaluation Scheme}\label{Sec:Regression_model}
Let
\begin{equation}\label{Equ:spectrum_feature}
   \tilde{x}_F = (\tilde{x}^{F}_{(1)}, \cdots, \tilde{x}^{F}_{(q)})^T
\end{equation}
 represents a stellar spectrum, $x$, in equation (\ref{Equ:spectrum}) based on the features detected in section \ref{Sec:Feature_extraction}, where $q>0$ is the number of extracted features. Based on the spectral features, the training set in equation (\ref{Equ:training_sp}) can be denoted by
\begin{equation}\label{Equ:training_sp_fe}
      S_{tr}^{F} = \{ (\tilde{x}^i_{F}, y_i), i = 1, 2, \cdots, N \},
\end{equation}
where $x^i_{F} = (\tilde{x}_{i1}^{F}, \cdots, \tilde{x}_{iq}^{F})^T$. Similarily, based on the extracted features, the validation set and test set can be denoted by $S_{val}^F$ and $S_{te}^F$.

\subsection{Estimation model for atmospheric parameters}\label{Sec:Regression_model:SVR}
We utilize the Support Vector Regression (SVR) algorithm\footnote{Support Vector Machine (SVM) is a learning algorithm that can be used for classification and regression. To be unambiguous, it is denoted by Support Vector Classification (SVC) and Support Vector Regression (SVR) in scenarios of recognition and estimation respectively.} \citep{Journal:Smola:2004,Journal:Schokopf:2002} to estimate the mapping between stellar spectra and atmospheric parameters. The SVR estimation can be described by
\begin{equation}\label{Equ:SVR}
   f(\tilde{x}_F) = \sum_{m=1}^{l} {\alpha_m}k(\tilde{x}_F^{i_m}, \tilde{x}_F) + b
\end{equation}
where $k(\cdot, \cdot)$ is a kernel function\footnote{Gaussian kernel is used in our experiments unless otherwise stated.} and $\{\tilde{x}_F^{i_m}, m = 1, \cdots, l\}$  are some members of training spectra in equation (\ref{Equ:training_sp_fe}) (called support vectors in literature \citep{Journal:Vapnik:1995}). In SVR, the estimation model (\ref{Equ:SVR}) is learnt from the training set $\hat{S}_{tr}^{F}$ based on the structural risk minimization principle, which combines empirical error and model complexity evaluation. Extensive research shows that this model has excellent generalization capacity. A typical characteristic of SVR is that the set of support vectors usually consists of a small fraction of the training samples; therefore, the obtained model is very efficient, which is important for large data processing. In this work, we used the implementation of SVR in \citet{Tec:Chang:2001}.

\subsection{Evaluation methods}\label{Sec:Regression_model:Evaluation}
Suppose $\bar{S}_{te}^{F} = \{ (\tilde{x}^m, y_m), m = 1, 2, \cdots, M \}$ is a test set. In this work, the performance of the proposed scheme is evaluated by Mean Absolute Error (MAE), and Standard Deviation (SD). They are defined as follows:
\begin{equation}\label{Equ:MAE}
   MAE = \frac{1}{M}\sum_{m=1}^{M}|e_{m}|,
\end{equation}
\begin{equation}\label{Equ:SD}
   SD = \sqrt{\frac{1}{M}\sum_{m=1}^{M}(e_{m}-\bar{e})^2},
\end{equation}
where $e_m$ is the error/difference between the reference value of stellar parameter and its estimation
\begin{equation}\label{Equ:deviation}
   e_m = y_m - f(\tilde{x}^m), ~m = 1,~\cdots, M.
\end{equation}
and $\bar{e} = \frac{1}{M}\sum_{m=1}^{M}{e_{m}}$.

MAE and SD are all widely used in evaluating performance of an estimation scheme. Each of two evaluation schemes focuses on different aspects of an estimation method. MAE measures the average magnitude of the deviation by ignoring the sign/direction of error. SD shows how much variation exists in an estimation error, and reflects the stability/robustness of an estimation scheme. A low SD indicates that the performance of the proposed estimation scheme is very stable; a high SD indicates that its performance is sensitive to a specific spectrum to be processed.

\section{Feature Description and its Application in Atmospheric Parameter Estimation}\label{Sec:Feature_Description}

Suppose $x$ and $\tilde{x}$ are a spectrum in equation (\ref{Equ:spectrum}) and its preprocessed edition in equation (\ref{Equ:spectrum_feature_Nor}), $\lambda^l$ is a given position of a detected feature in log(wavelength). For ease of introduction to feature description, we assume $\tilde{x}(\lambda^l)$ represents the preprocessed flux of spectrum $x$ at log(wavelength) position $\lambda^l$.

In section \ref{Sec:Feature_extraction}, we detect the positions of features from stellar spectra. A direct description of the features is just to pick up observed fluxes at the detected positions:
\begin{equation}\label{Equ:featuredescription:inte:Teff:flux}
    \tilde{x}^F_{(m)} = \tilde{x}(\lambda^l_{Tm}),~ m =1,\cdots, 10
\end{equation}
for $T_\texttt{eff}$ (Table \ref{Tab:LASSO:Features:Teff}),
\begin{equation}\label{Equ:featuredescription:inte:Logg:flux}
    \tilde{x}^F_{(m)} = \tilde{x}(\lambda^l_{Lm}),~ m =1,\cdots, 19
\end{equation}
for log$~g$ (Table \ref{Tab:LASSO:Features:Logg}), and
\begin{equation}\label{Equ:featuredescription:inte:FeH:flux}
    \tilde{x}^F_{(m)} = \tilde{x}(\lambda^l_{Fm}),~ m =1,\cdots, 14
\end{equation}
for [Fe/H] (Table \ref{Tab:LASSO:Features:FeH}). The labels $\lambda^l_{Tm}$, $\lambda^l_{Lm}$ and $\lambda^l_{Fm}$ are defined in section \ref{Sec:Feature_extraction}. Experimental results based on this kind description are presented in Table \ref{Tab:Accuracy:R0}. In this scheme, only 10 observed fluxes are picked up directly and used for estimating $T_\texttt{eff}$, 19 observed fluxes for estimating log$~g$, and 14 observed fluxes for estimating [Fe/H]. Therefore, it is very efficient to extract features in application. The performance of the proposed scheme is also excellent compared with a similar study in literature (Table 2 in \citep{Journal:Fiorentin:2007}) in which 50 PCA features were used, every feature was computed from approximately 2~000 observed fluxes, and MAE is 0.0126 for log$~T_{\texttt{eff}}$, 0.3644 for log$~g$ and 0.1949 for [Fe/H]. More direct comparisions are presented in section \ref{Sec:Conclusion}.

\begin{table}\scriptsize
\centering
\caption{Consistency/Accuracy on test set with features described by the observed fluxes on the detected typical positions.}
\begin{tabular}{ c c  c  c }
   \hline \hline
   evaluation method & log $T_{\texttt{eff}}$     &  log~g     & [Fe/H]  \\ \hline
   MAE               & 0.009092                   & 0.198928   & 0.206814\\
   SD                & 0.012978                   & 0.282752   & 0.274245\\   \hline
\end{tabular}\label{Tab:Accuracy:R0}
\end{table}

However, real spectra are inevitably corrupted by noise, which usually degrades accuracy. Therefore, to further improve accuracy, we propose the following feature description method based on the local average of preprocessed spectral fluxes in a local area around the detected positions:
\begin{equation}\label{Equ:featuredescription:inte:Teff}
    \tilde{x}^F_{(m)} = \sum _{j = -k}^{j = k}\tilde{x}(\lambda^l_{Tm} + j \times \Delta_{\lambda}^{\l}),~ m =1,\cdots, 10
\end{equation}
for $T_\texttt{eff}$ (Table \ref{Tab:LASSO:Features:Teff}),
\begin{equation}\label{Equ:featuredescription:inte:Logg}
    \tilde{x}^F_{(m)} = \sum _{j = -k}^{j = k}\tilde{x}(\lambda^l_{Lm} + j \times \Delta_{\lambda}^{\l}),~ m =1,\cdots, 19
\end{equation}
for log$~g$ (Table \ref{Tab:LASSO:Features:Logg}), and
\begin{equation}\label{Equ:featuredescription:inte:FeH}
    \tilde{x}^F_{(m)} = \sum _{j = -k}^{j = k}\tilde{x}(\lambda^l_{Fm} + j \times \Delta_{\lambda}^{\l}),~ m =1,\cdots, 14
\end{equation}
for [Fe/H] (Table \ref{Tab:LASSO:Features:FeH}), where $k\geq 0$ is an integer representing radius of integration, and $\Delta_{\lambda}^{\l}$ is the sampling step of a spectrum whose value is 0.0001 in this work (Section \ref{Sec:Data}). For convenience, we name the two describing methods Point Description (PD) and Local Integration (LI) respectively.

\begin{figure*}
  \centering
  \subfigure[~]{
    \label{Fig:Eva_MAE:Teff_val} 
    \includegraphics[width =1.5in]{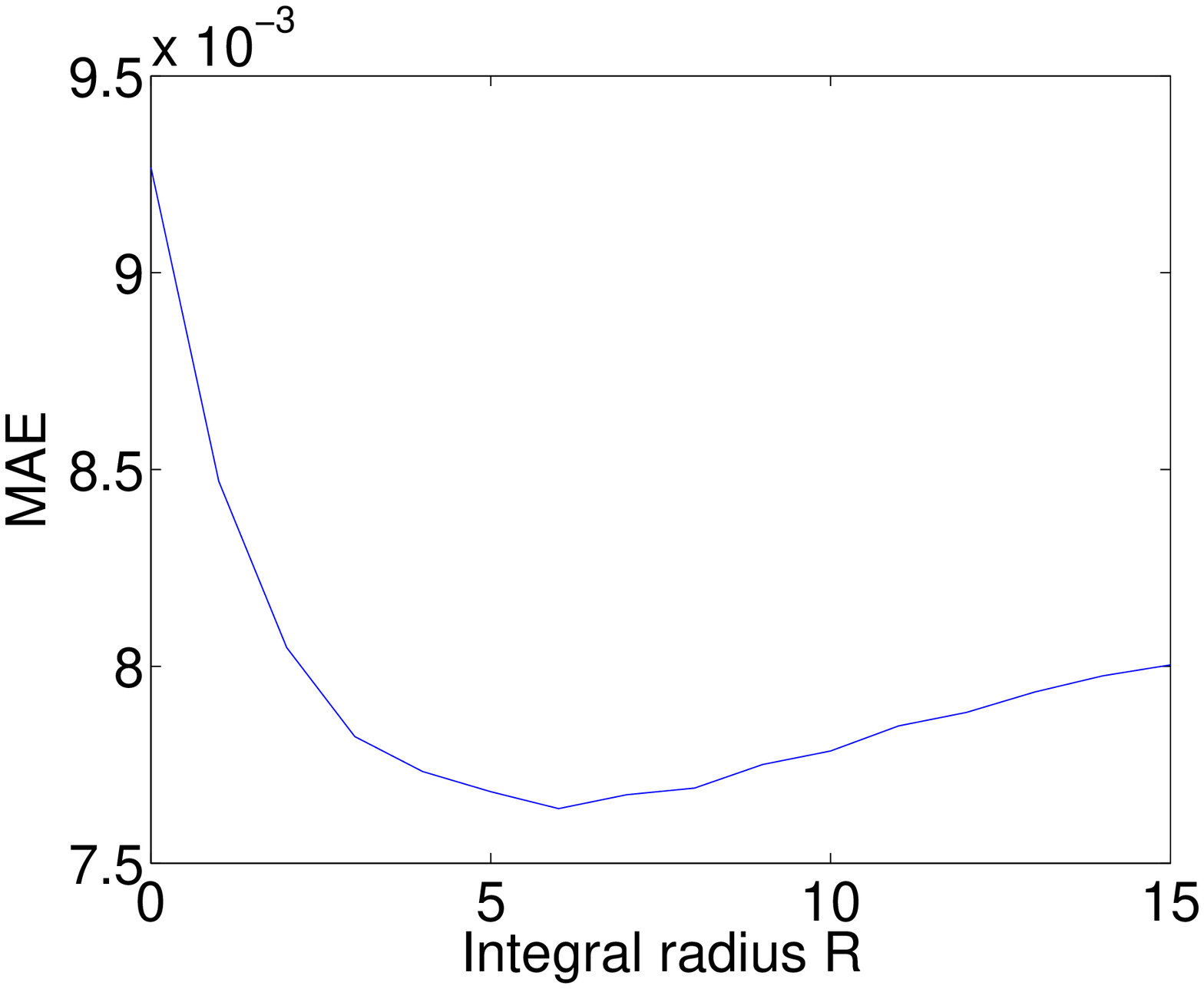}}
    \hspace{-0.15in}
    \subfigure[~]{
    \label{Fig:Eva_MAE:logg_tr} 
    \includegraphics[width =1.5in]{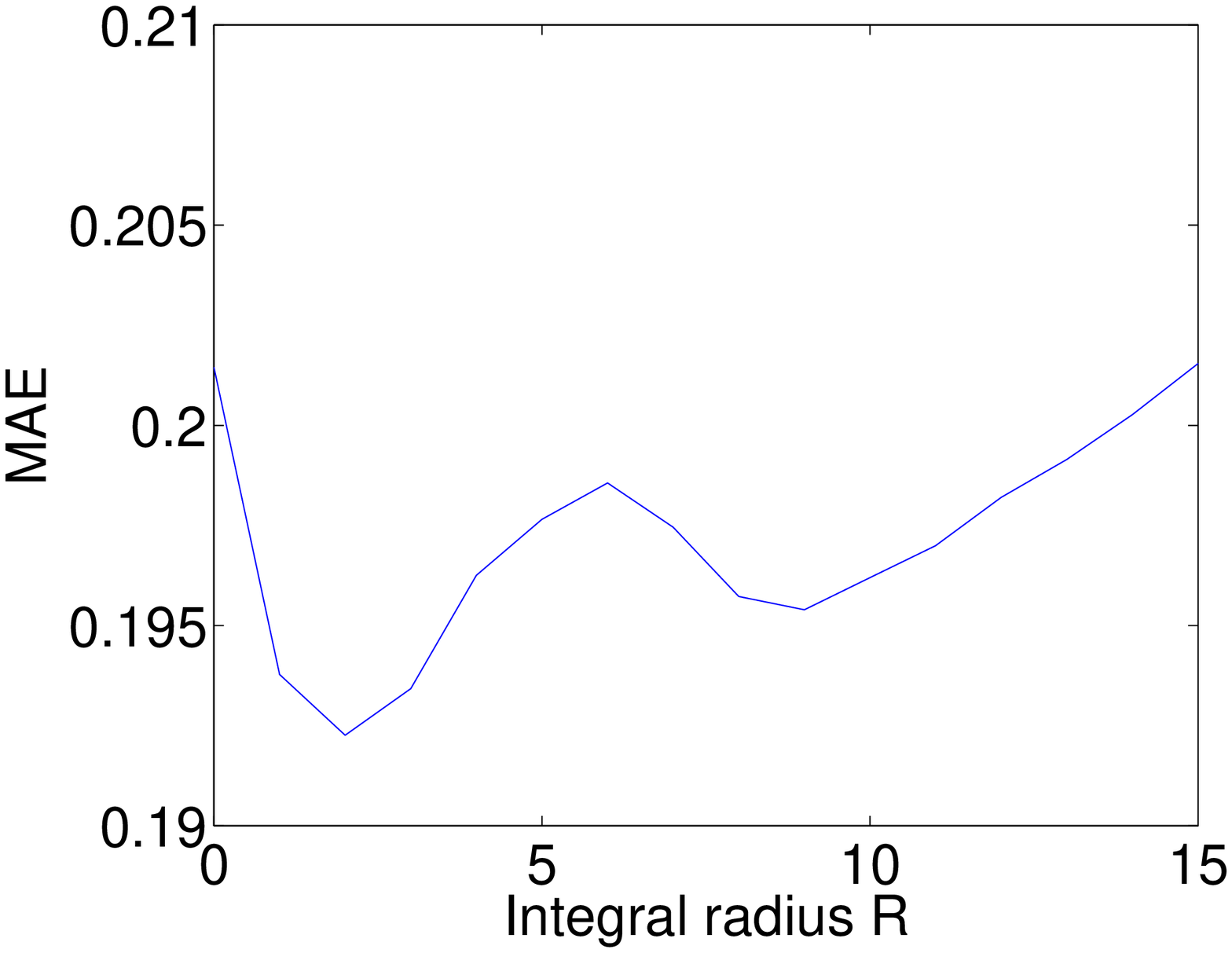}}
    \hspace{-0.15in}
    \subfigure[~]{
    \label{Fig:Eva_MAE:Feh_tr} 
    \includegraphics[width =1.5in]{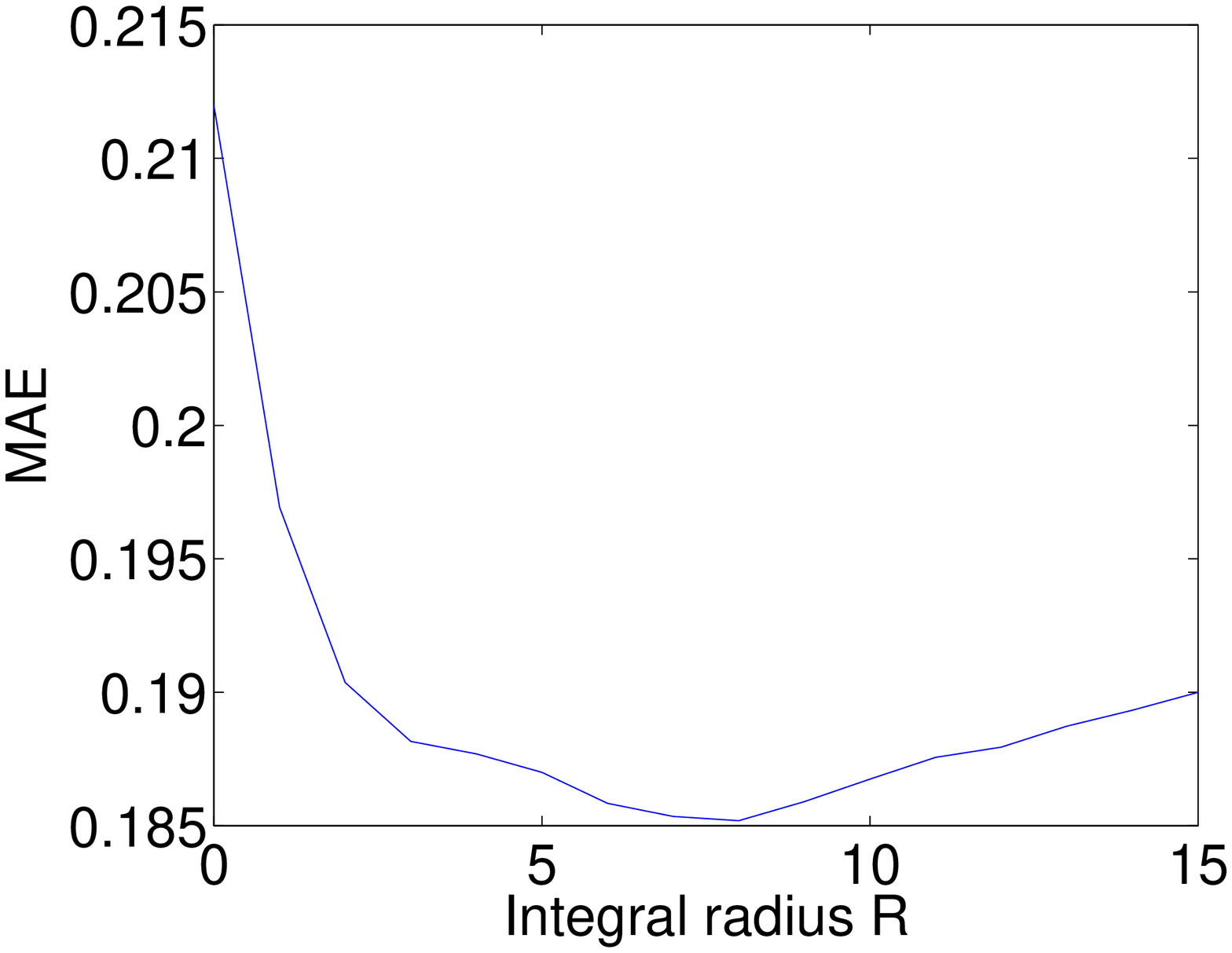}}
    \setlength{\abovecaptionskip}{-2pt}
  \caption{Variation of Mean Absolute Error (MAE) on validation set with integral radius $R$. Subfigures (a), (b), and (c) show MAEs of the estimations with different integration radii on validation set respectively for log $T_\texttt{eff}$, log$~g$ and [Fe/H].}
  \label{Fig:Eva_MAE:TR} 
\end{figure*}

\begin{figure*}
  \centering
  \subfigure[log$~T_{\texttt{eff}}$]{
    \label{Fig:consistency_AtmPar_1_im_2} 
    \includegraphics[width =2.0in]{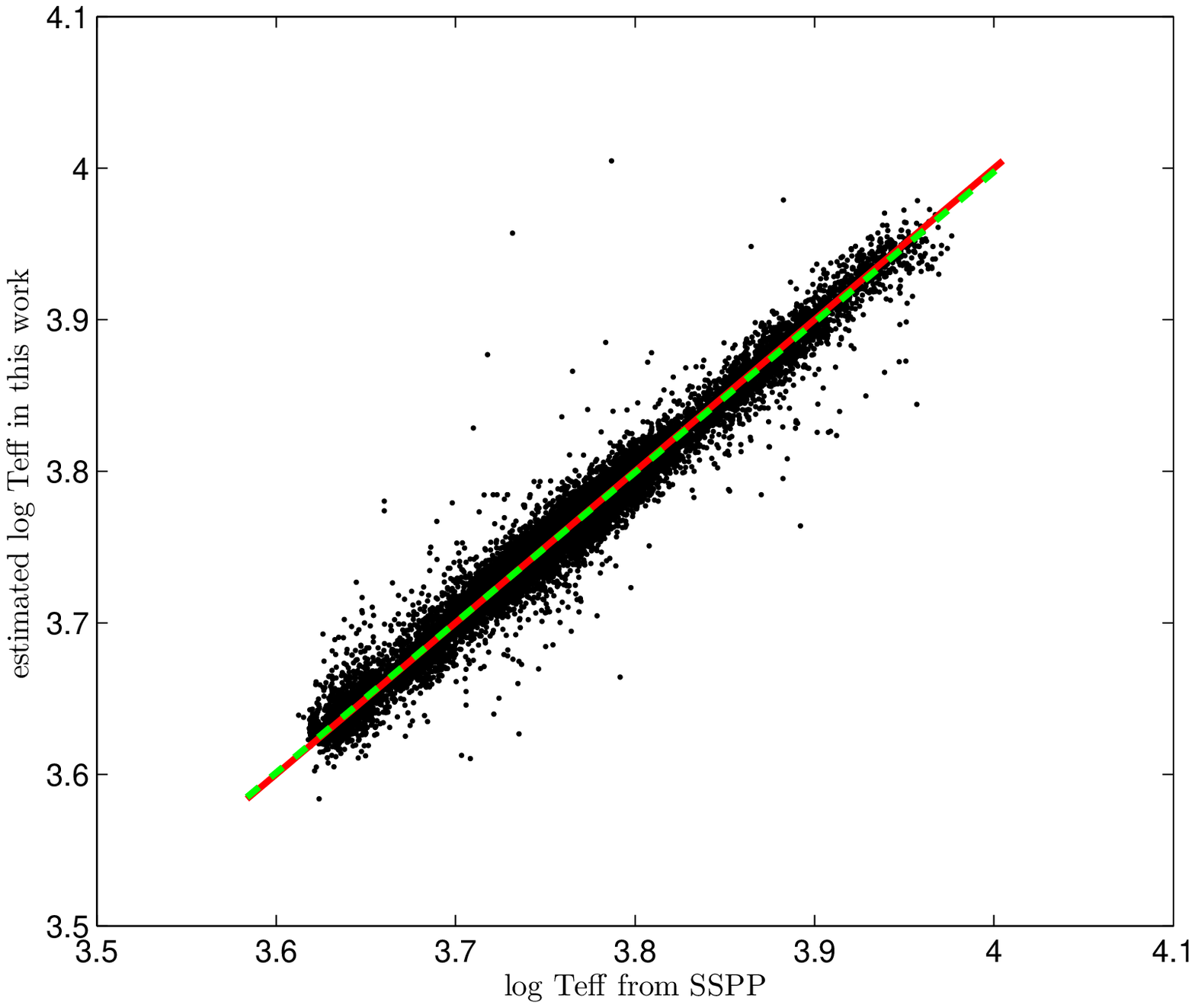}}
\hspace{-0.15in}
  \subfigure[log$~g$]{
    \label{Fig:consistency_AtmPar_2_im_3} 
    \includegraphics[width =2.0in]{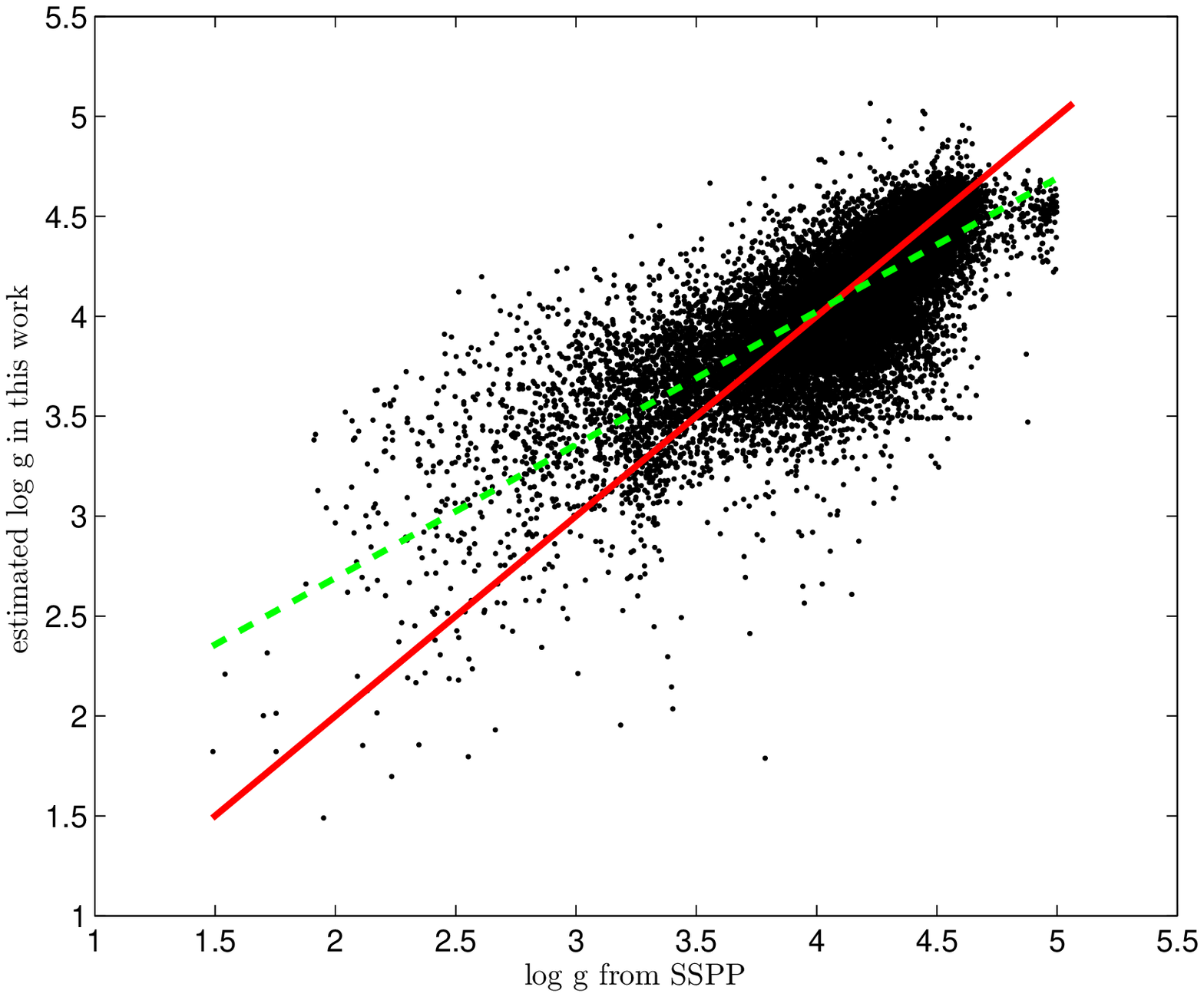}}
  \hspace{-0.15in}
  \subfigure[$\texttt{[}$Fe/H$\texttt{]}$]{
    \label{Fig:consistency_AtmPar_3_im_2} 
    \includegraphics[width =2.0in]{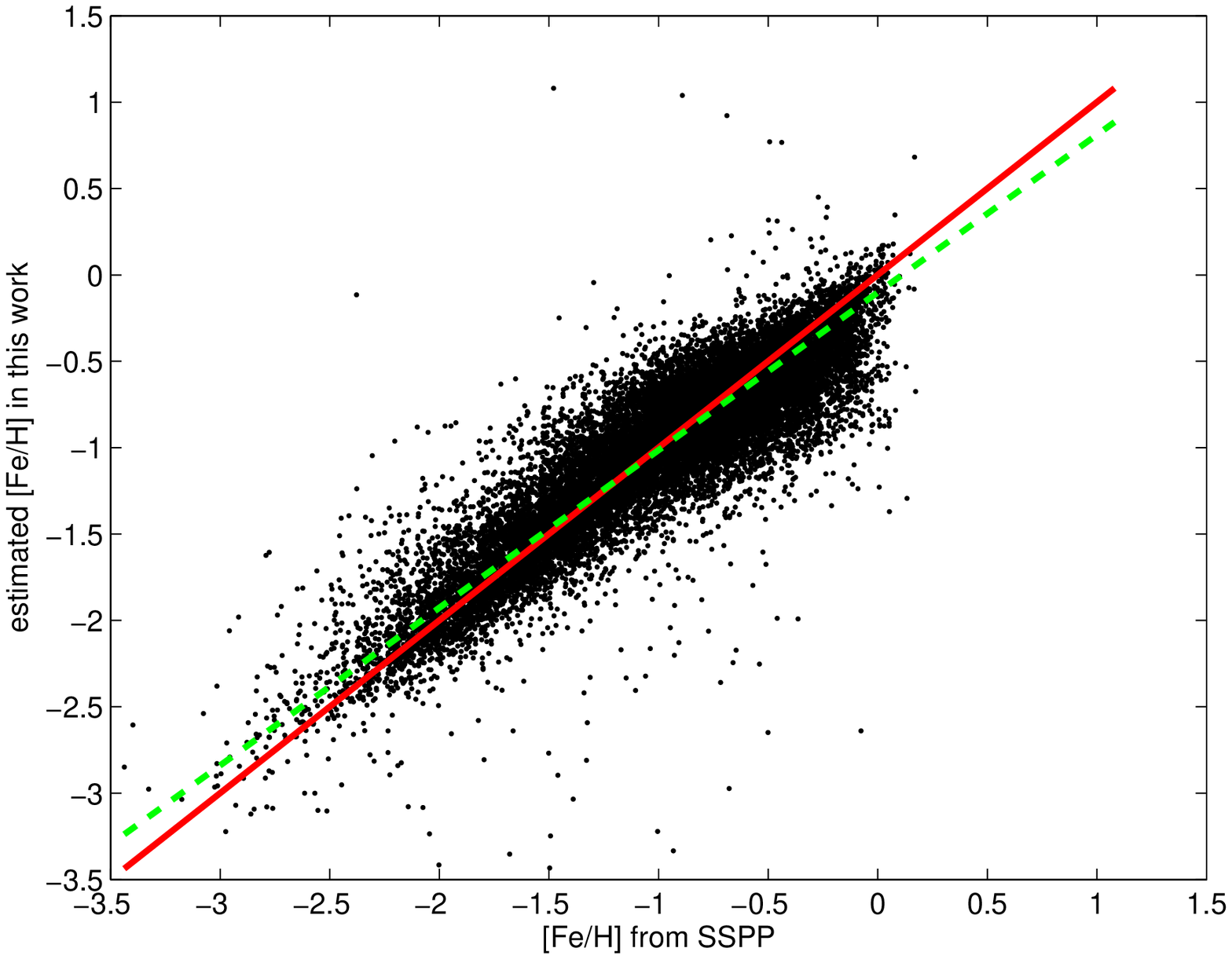}}
    \setlength{\abovecaptionskip}{-2pt}
  \caption{Consistency. We compare our estimation of log$~T_{\texttt{eff}}$, log$~g$ and [Fe/H] with the corresponding reference values provided by SSPP of SLOAN on the test set. The horizontal axis and vertical axis are the reference parameters provided by SSPP of SLOAN and the estimation of our proposed method. In this experiment, features are described by the LI method.}
  \label{Fig:consistency:im} 
\end{figure*}

\begin{figure*}
  \centering
  \subfigure[log$~T_{\texttt{eff}}$]{
    \label{Fig:Discripancy_AtmPar_1} 
    \includegraphics[width =2.0in]{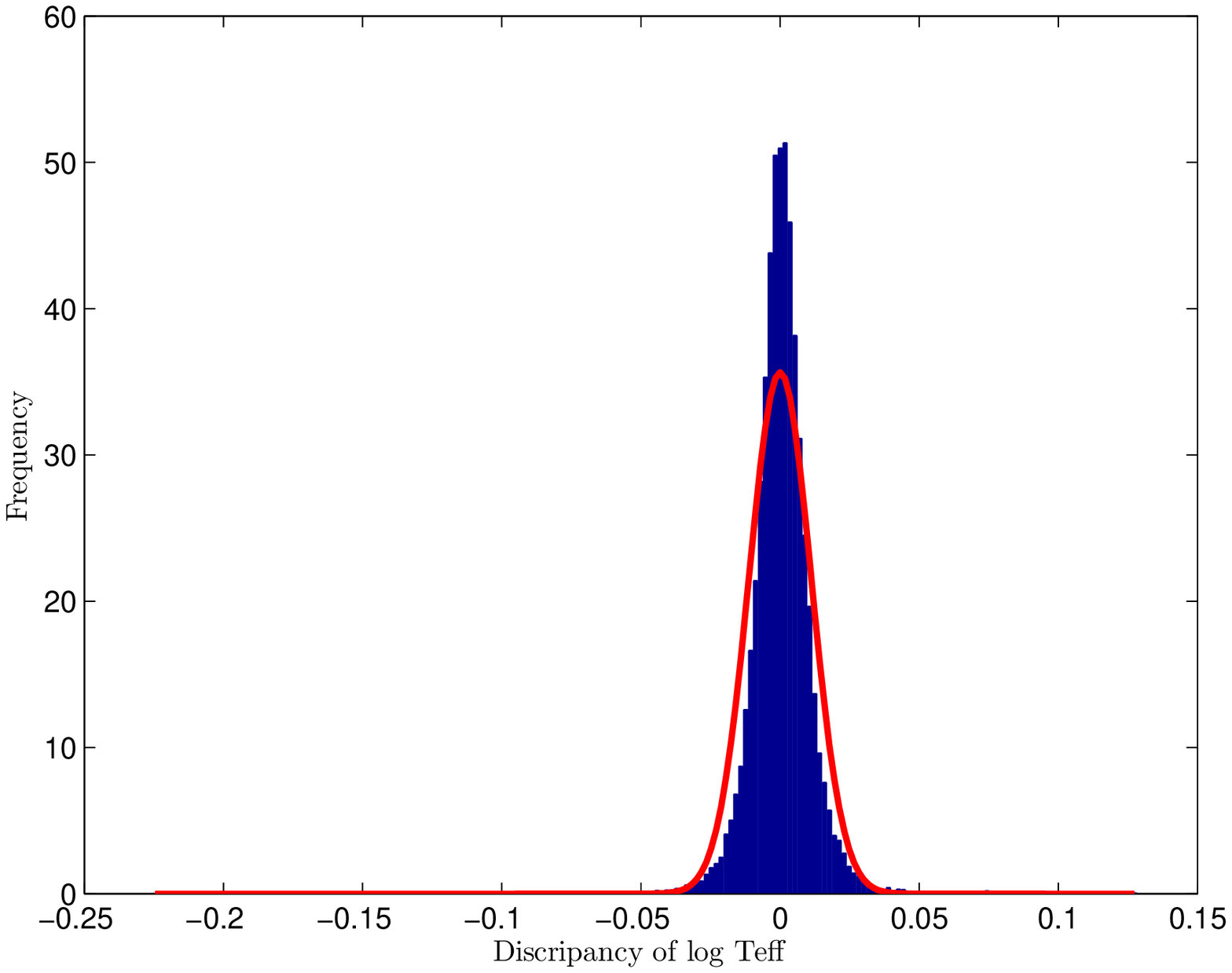}}
\hspace{-0.15in}
  \subfigure[log$~g$]{
    \label{Fig:Discripancy_AtmPar_2} 
    \includegraphics[width =2.0in]{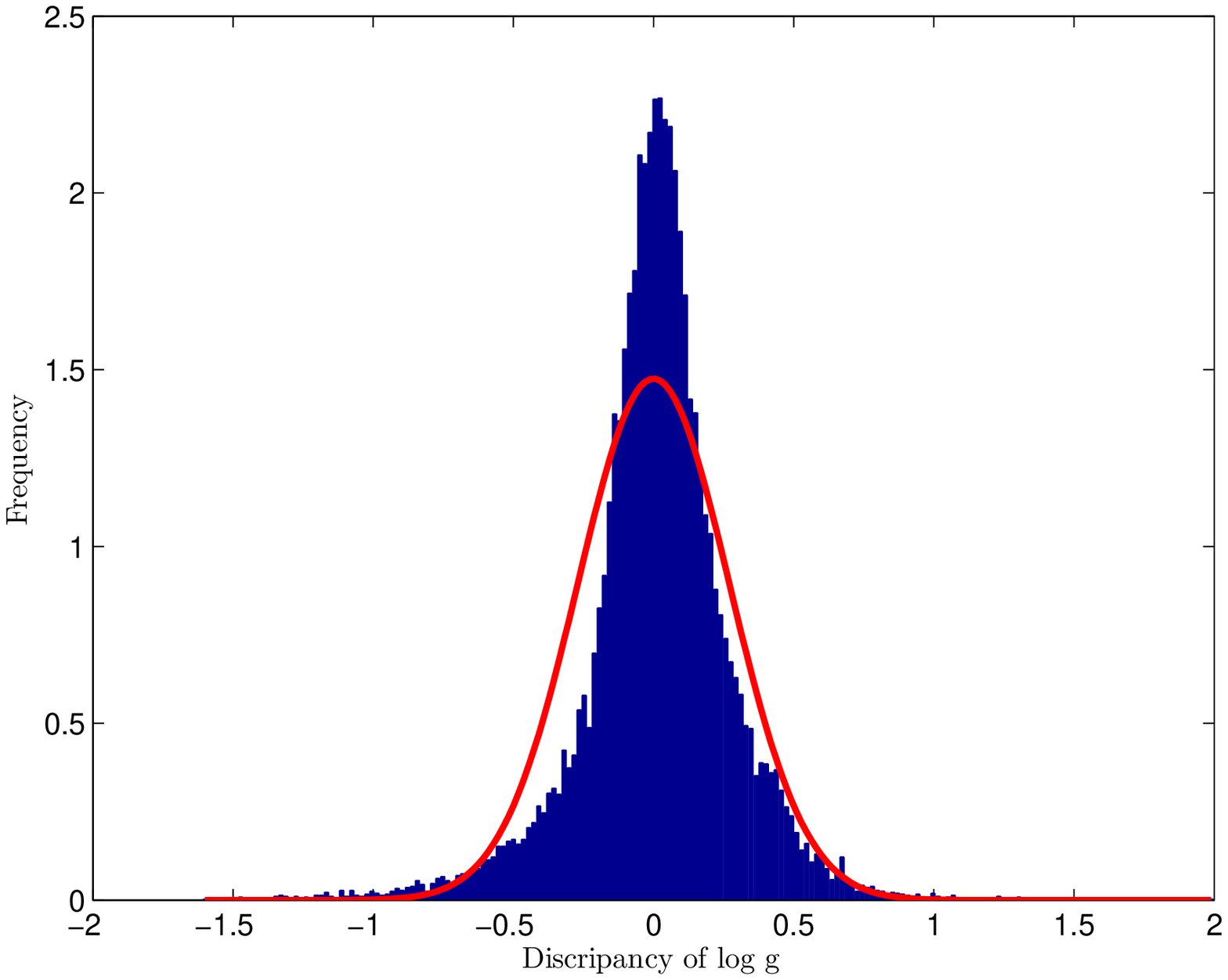}}
  \hspace{-0.15in}
  \subfigure[$\texttt{[}$Fe/H$\texttt{]}$]{
    \label{Fig:Discripancy_AtmPar_3} 
    \includegraphics[width =2.0in]{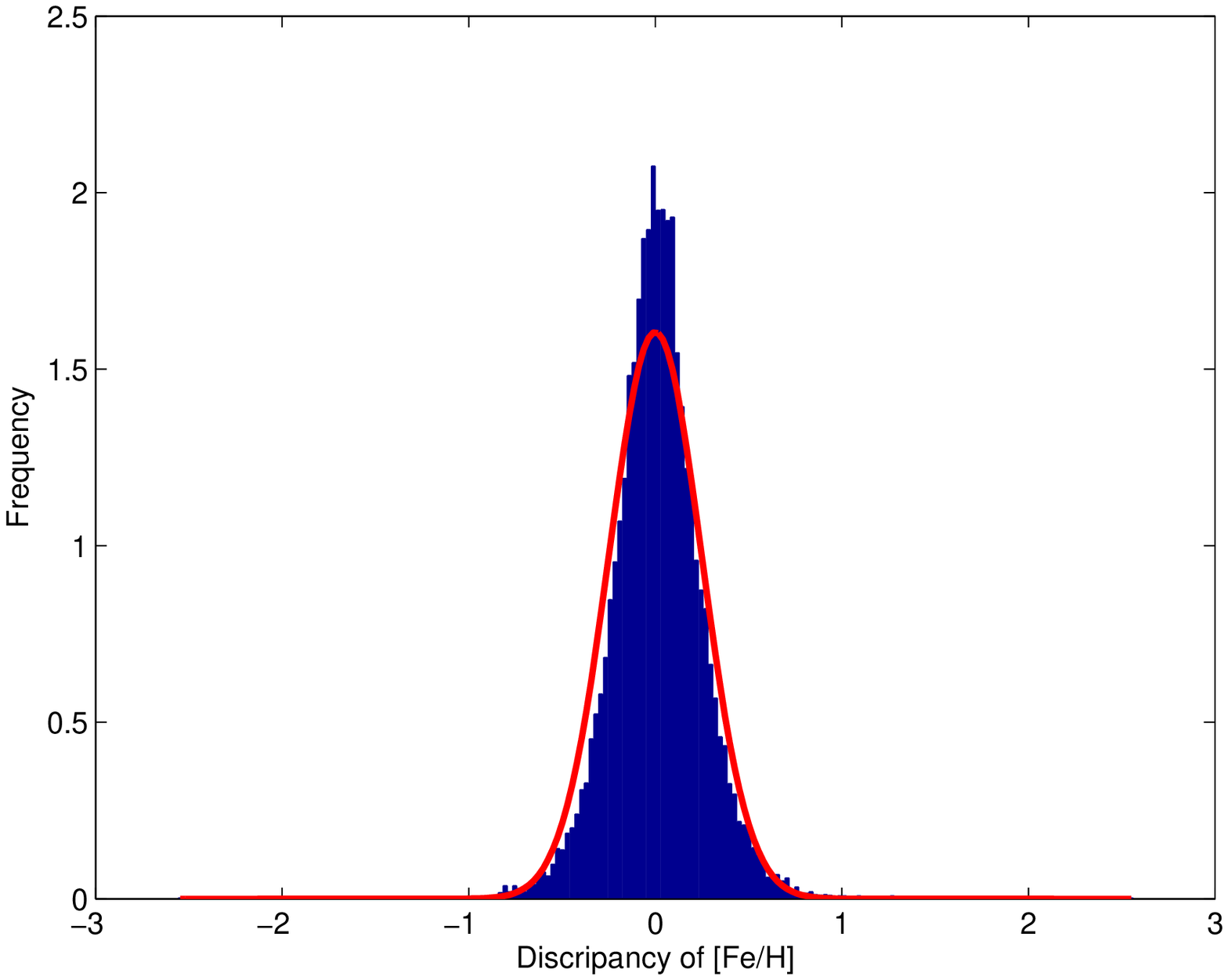}}
    \setlength{\abovecaptionskip}{-2pt}
  \caption{Discrepancy and bias of the estimation. We compare our estimation of log$~T_{\texttt{eff}}$, log$~g$, and [Fe/H] with the corresponding reference value provided by SSPP of SLOAN on the test set. The horizontal axis is the difference between the reference parameter provided by SSPP of SLOAN and the estimation of our proposed method. The vertical axis is the estimated probability density of the difference on the test set, and the red curve is a fitting of the density by a Gaussian distribution with identical mean and variance. In this experiment, features are described using the LI method.}
  \label{Fig:Discripancy:im} 
\end{figure*}

The theoretical foundation of proposed feature description method LI in equations (\ref{Equ:featuredescription:inte:Teff}), (\ref{Equ:featuredescription:inte:Logg}), and (\ref{Equ:featuredescription:inte:FeH}) is the law of large numbers (LLN) in probability theory. A preprocessed spectral flux $\tilde{x}(\lambda^l)$ consists of a theoretical-spectral component and a noise term
\begin{equation}\label{Equ:flux:decomposition}
   \tilde{x}(\lambda^l) = \tilde{x}_{th}(\lambda^l) + \epsilon(\lambda^l),
\end{equation}
where $\tilde{x}_{th}(\lambda^l)$ is a theoretical flux without contamination from noise at log(wavelength) $\lambda^l$, and $\epsilon(\lambda^l)$ is noise at the corresponding position. Suppose $\{\epsilon(\lambda^l),~\lambda^l \in [3.581862, 3.963961]\}$ is a set of independent and identically distributed random variables drawn from distributions with zero mean and finite variances $\sigma ^2$. The LLN states that the average of noises $\frac{\sum_{j=-k}^{j=k}{\epsilon(\lambda^l+j \times 0.0001)}}{2k+1}$ converges in probability and almost surely to the expected value 0 as $k \rightarrow \infty$, where $\lambda^l \leq 3.963961 -  k \times 0.0001$ and $\lambda^l \geq 3.581862 + k \times 0.0001$. In other words, it says that the negative effect from noise diminishes toward zero with $k$ increasing. Similarly, information from the theoretical fluxes in observed spectra is also erased gradually with $k$ increasing in equations (\ref{Equ:featuredescription:inte:Teff}), (\ref{Equ:featuredescription:inte:Logg}), and (\ref{Equ:featuredescription:inte:FeH}). Therefore, the performance of parameter estimation increases at the beginning on the whole, and after the effect of erasing theoretical fluxes overpower the effect of diminishing noise, the performance will degrade (Fig. \ref{Fig:Eva_MAE:TR}). In this work, we obtain the optimal $k$ based on the performance of the proposed scheme on validation set. Optimized $k$ are 6, 2 and 8 respectively for $T_{\texttt{eff}}$, log$~g$, and [Fe/H]. Final results are presented in Table \ref{Tab:Accuracy:R:Optimized}, Fig. \ref{Fig:consistency:im} and Fig. \ref{Fig:Discripancy:im}. It is shown that accuracy of the estimation based on the LI description is improved.

In the proposed LI approach, all of the detected features share a common smoothing parameter $k$.  In reality, yet, the detected features may be different from each other on scale. Therefore, two more deliberated schemes are to estimate an independent smoothing scale based on a validation set for every feature, or to determine the scales adaptively in detecting features, for example, the fused lasso method can detect the supporting interval for every feature \citep{Journal:Tibshirani:2005,Journal:Ye:2011}.

\begin{table}\scriptsize
\centering
\caption{Accuracy/Consistency on test set with features described by local integral near the detected typical positions. Integral radii are 6 for log $T_\texttt{eff}$, 2 for log$~g$ and 8 for [Fe/H] respectively.}
\begin{tabular}{ c c  c  c }
   \hline \hline
   evaluation method & log~Teff       & log~g     & [Fe/H]  \\  \hline
   MAE               & 0.007458       & 0.189557  & 0.182060\\
   SD                & 0.011189       &  0.270496 & 0.248504\\  \hline
\end{tabular}\label{Tab:Accuracy:R:Optimized}
\end{table}

\section{Feature Evaluation and Refinement}\label{Sec:Feature:eva}
In this section we investigate the compactness of the detected features. By compactness, we mean to study whether there is any redundancy in the set of detected features and how to detect and refine the features if any redundancy exists.

First, we introduce a measure to evaluate the significance/necessity of a feature. For ease of description, we take the evaluation of the features in Table \ref{Tab:LASSO:Features:Teff} for $T_{\texttt{eff}}$ as an example. The full set of the features can be denoted by
 \begin{equation}\label{Equ:Features:set:Teff}
 F_{Teff} = \{ T_j,~ 1\leq j \leq N_{Teff} \},
 \end{equation}
 and $\bar{F}^{i}_{Teff}$  represents a subset of features by deleting $T_i$ from $ F_{Teff}$:
 \begin{equation}\label{Equ:Features:set:Teffs}
 \begin{split}
    \bar{F}^{i}_{Teff} =& F_{Teff} - \{ T_i \}\\
= &\{ T_j,~ 1\leq j \leq N_{Teff}, j \neq i \}, \\
 \end{split}
 \end{equation}
where $N_{Teff}$ represents the number of detected features for $T_{\texttt{eff}}$ in Table \ref{Tab:LASSO:Features:Teff}, and $i = 1, \cdots, N_{Teff}$. In this work, $N_{Teff}$  is 10. We propose to evaluate the significance of $T_i$ by
\begin{equation}\label{Equ:significance}
      S(T_i) = MAE(\bar{F}^{i}_{Teff}) - MAE( F_{Teff}),
\end{equation}
where $MAE(\bar{F}^{i}_{Teff})$ and $MAE( F_{Teff})$ represent the mean absolute error of the estimation of atmospheric parameter $T_{\texttt{eff}}$ based on features $\bar{F}^{i}_{Teff}$ and $F_{Teff}$ respectively. If a feature $T_i$ is completely redundant, in theory the estimation performance should be unaffected after deleting it, and $S(T_i)$ should be zero. On the other hand, if feature $T_i$ is essential for estimating atmospheric parameter $T_{\texttt{eff}}$, then the accuracy should noticeably deteriorate after deleting it. Therefore, the proposed measure $S$ expresses the necessity of the detected features to parameter estimation. Evaluation results of the features in Table \ref{Tab:LASSO:Features:Teff} are presented in the second column of Table \ref{Tab:LASSO:Features:Teff:compactness}. The features of log$~g$ and [Fe/H] (Table \ref{Tab:LASSO:Features:Logg} and Table \ref{Tab:LASSO:Features:FeH}) can be evaluated similarly, and corresponding results are presented in the second column of Table \ref{Tab:LASSO:Features:Logg:compactness} and Table \ref{Tab:LASSO:Features:FeH:compactness} respectively.

\begin{table}\scriptsize
\centering
\caption{Compactness of the detected features in Table \ref{Tab:LASSO:Features:Teff} for estimating $T_{\texttt{eff}}$ from stellar spectra. S$_{va}$ and S$^{r}_{va}$ are the evaluate values of significance measure and relative significance measure respectively on validation set. MAE are the mean absolute errors on test set based on features $\bar{F}^i_{Teff}$. Items are sorted decreasingly based on S$^{r}_{va}$.}
\begin{tabular}{  c c c c }
  \hline \hline
label  &  S$_{va}$     &    S$^{r}_{va}$&    MAE      \\ \hline
T7        &   0.001587    &     0.2128     &    0.009211 \\
T8        &   0.000985    &     0.1321     &    0.008380 \\
T5        &   0.000638    &     0.0855     &    0.008139 \\
T6        &   0.000588    &     0.0788     &    0.007980 \\
T1        &   0.000082    &     0.0110     &    0.007570 \\
T4        &   0.000057    &     0.0076     &    0.007498 \\
T9        &   0.000019    &     0.0025     &    0.007521 \\
T10       &   0.000018    &     0.0024     &    0.007517 \\
T2        &   0.000010    &     0.0013     &    0.007466 \\
T3        &   -0.000003   &    -0.0004     &    0.007461 \\ \hline
\end{tabular}\label{Tab:LASSO:Features:Teff:compactness}
\end{table}

\begin{table}\scriptsize
\centering
\caption{Compactness of the detected features in Table \ref{Tab:LASSO:Features:Logg} for estimating log$~g$ from stellar spectra. S$_{va}$ and S$^{r}_{va}$ are the evaluate values of significance measure and relative significance measure respectively on validation set. MAE are the mean absolute errors on test set based on features $\bar{L}^i_{log~g}$. Items are sorted decreasingly based on S$^{r}_{va}$.}
\begin{tabular}{  c c c c }
  \hline \hline
label   &  S$_{va}$   &    S$^{r}_{va}$&   MAE     \\ \hline
L16        &  0.003656   &      0.0193    &   0.193511\\
L18        &  0.003182   &      0.0168    &   0.192749\\
L9         &  0.002700   &      0.0142    &   0.192719\\
L15        &  0.001455   &      0.0077    &   0.190796\\
L14        &  0.001067   &      0.0056    &   0.190436\\
L17        &  0.000948   &      0.0050    &   0.191023\\
L13        &  0.000884   &      0.0047    &   0.190412\\
L8         &  0.000752   &      0.0040    &   0.189936\\
L6         &  0.000675   &      0.0036    &   0.189539\\
L10        &  0.000577   &      0.0030    &   0.189894\\
L11        &  0.000560   &      0.0030    &   0.189815\\
L12        &  0.000548   &      0.0029    &   0.190194\\
L19        &  0.000530   &      0.0028    &   0.189892\\
L5         &  0.000314   &      0.0017    &   0.189607\\
L3         &  0.000200   &      0.0011    &   0.189695\\
L2         &  0.000013   &      0.0001    &   0.189978\\
L4         &  -0.000071  &     -0.0004    &   0.189228\\
L7         &  -0.000091  &     -0.0005    &   0.189375\\
L1         &  -0.000892  &     -0.0047    &   0.189230\\  \hline
\end{tabular}\label{Tab:LASSO:Features:Logg:compactness}
\end{table}

\begin{table}\scriptsize
\setlength{\abovecaptionskip}{-100pt}
\setlength{\belowcaptionskip}{-100pt}
\centering
\caption{Compactness of the detected features in Table \ref{Tab:LASSO:Features:FeH} for estimating [Fe/H] from stellar spectra. S$_{va}$ and S$^{r}_{va}$ are the evaluate values of significance measure and relative significance measure respectively on validation set. MAE are the mean absolute errors on test set based on features $\bar{F}^i_{[Fe/H]}$. Items are sorted decreasingly based on S$^{r}_{va}$.}
\begin{tabular}{  c c c c }
  \hline \hline
label &  S$_{va}$      &    S$^{r}_{va}$&   MAE     \\ \hline
F3       &  0.008238      &       0.0452   &   0.191239\\
F8       &  0.004414      &       0.0242   &   0.186801\\
F10      &  0.002754      &       0.0151   &   0.184995\\
F11      &  0.002444      &       0.0134   &   0.186255\\
F7       &  0.001979      &       0.0109   &   0.184598\\
F12      &  0.001651      &       0.0091   &   0.183202\\
F6       &  0.001152      &       0.0063   &   0.183398\\
F9       &  0.001136      &       0.0062   &   0.183192\\
F5       &  0.000922      &       0.0051   &   0.183404\\
F2       &  0.000799      &       0.0044   &   0.182874\\
F4       &  0.000710      &       0.0039   &   0.183193\\
F1       &  0.000485      &       0.0027   &   0.182613\\
F13      &  -0.000157     &      -0.0009   &   0.182116\\
F14      &  -0.000078     &      -0.0004   &   0.182141\\ \hline
\end{tabular}\label{Tab:LASSO:Features:FeH:compactness}
\end{table}

The magnitude of MAE varies from problem to problem; for example, in Table \ref{Tab:Accuracy:R0} and Table \ref{Tab:Accuracy:R:Optimized}, the MAE of $T_{\texttt{eff}}$ is noticeably less than that of log$~g$ and [Fe/H]. This magnitude can determine the potential value of significance evaluation. Therefore, we introduce the following relative evaluation scheme
\begin{equation}\label{Equ:significance_r}
      S^{r}(T_i) = \frac{MAE(\bar{F}^{i}_{Teff}) - MAE( F_{Teff})}{MAE(\bar{F}^{i}_{Teff})}.
\end{equation}
Similarily, $F_{log~g}$, $\bar{F}^{i}_{log~g}$, S$(L_{i})$, S$^r(L_{i})$, $F_{[Fe/H]}$, $\bar{F}^{i}_{[Fe/H]}$, S$(F_{i})$, S$^r(F_{i})$ can be defined for the features of $log~g$ and $[Fe/H]$.

 Corresponding results are presented in the third column of Table \ref{Tab:LASSO:Features:Teff:compactness}, Table \ref{Tab:LASSO:Features:Logg:compactness}, and Table \ref{Tab:LASSO:Features:FeH:compactness} respectively. For convenience, we name the evaluation schemes $S$ in equation (\ref{Equ:significance}) and $S^r$ in equation (\ref{Equ:significance_r}) as Significance(S) measure and Relative Significance (RS) measure respectively. The RS measure can be regarded as a standardized variant of the S measure.

The above evaluating results show that:
\begin{enumerate}
\item[1)~]{In the detected features for $T_\texttt{eff}$ and [Fe/H], no sufficient evidence shows the existence of redundancy (TABLE \ref{Tab:LASSO:Features:Teff:compactness}, TABLE \ref{Tab:LASSO:Features:FeH:compactness}).}
\item[2)~]{The evaluating results in TABLE \ref{Tab:LASSO:Features:Logg:compactness} show that there exist three redundant features --- L1, L4 and L7 --- in the detected features for log$~g$ due to over learning with high probability.}
\item[3)~]{Available evidence shows that T3, F13 and F14 are non-significant with high probability (TABLE \ref{Tab:LASSO:Features:Teff:compactness}, TABLE \ref{Tab:LASSO:Features:FeH:compactness}).}
\end{enumerate}
Overall, although the compactness of the detected features is excellent, there remains some redundancy and non-significant features. Fortunately, magnitude of the relative significance $S^{r}$ of the redundant and non-significant features is evidently smaller than that of others. Therefore, they can be detected by checking whether the relative evaluation value $S^{r}$ of a feature is smaller than a preset threshold, for example 0.001.

In theory, the significance evaluation of every feature should be non-negative. However, there exist both theoretical-spectral components and noise components in observed data (equation \ref{Equ:flux:decomposition}). The effectiveness of a redundant or non-significant feature is usually relatively low and can be overpowered by the effect of noise with a certain probability. Therefore, sometimes we can find that some detected features have negative significance evaluation, as in the case of the L4 in Table \ref{Tab:LASSO:Features:Logg:compactness}.

\section{On configuration of the proposed scheme and evaluation on spectra with ground-truth}\label{Sec:Config_evaOnSyn}
\subsection{Linearity v.s. nonlinearity}\label{Sec_sub:linearity_nonlinearity}

LASSO is a method based on a linear model,  detects features according to the degree of linear correlations between a response and predictors. It is intuitive to choose a linear method for estimating the atmospheric parameters from the detected features.

On the other hand, it is also possible that there exist some non-linear relationships between a response and its predictors with high linear correlation. For example, suppose $x=(x_1,x_2)$ are two predictors, $y$ is a response, if some observed samples of $(x_1,x_2,y)$ are as following: $(1,1,2)$, $(2, 2, 4)$, $(3, -3, 5)$ and $(4, 4 , 8)$; it is evident that there exist some non-linear relationships between the two predictors and the response, even though the linear correlations between $x_1$ and  $y$, and $x_2$ and $y$ are as high as  0.9968 and 0.4722 respectively based on the observations. The experimental results in Table \ref{Tab:Accuracy:R:Optimized} and Table \ref{Tab:effectiveness:Linear} indicates the existence of non-linear relationships between the detected features and the physical parameters.

\begin{table}\scriptsize
\setlength{\abovecaptionskip}{-100pt}
\setlength{\belowcaptionskip}{-100pt}
\centering
\caption{Performance (MAE) of two linear methods. Experimental configurations are same as the experiments in Table \ref{Tab:Accuracy:R:Optimized}. OLS (Ordinary Least Squares): linear least squares regression, SVR(linear): Support Vector machine Regresion with a linear kernel.  }
\begin{tabular}{  c c c c }
\hline \hline
evaluation method   &  T$_{eff}$     &       log$~g$    &   [Fe/H] \\ \hline
OLS        &  0.036510      &       0.301661   &   0.360890\\
SVR(linear)&  0.034152      &       0.253363   &   0.323512\\
 \hline
\end{tabular}\label{Tab:effectiveness:Linear}
\end{table}

\subsection{On choosing of estimation method}\label{Sec_sub:choose_regressors}

To capture the non-linearity in atmospheric parameter estimation, we investivate four typical non-linear regression methods: FNN (Feedforward neural network, implemented by the neural network toolbox in Matlab 2011b), GAM(Generalized Additive Models (smooth splines)\citep{Book:Hastie:1990}, implemented by the R package gam), MARS (Multivariate Adaptive Regression Splines \citep{Journal:Freidman:1991}, implemented by the R package mda), RF( Random Forest \citep{Journal:Breiman:2001,Journal:Liaw:2002}). Parameters of the estimation methods are choosed based on validation set.

Related evaluation results are presented in Table \ref{Tab:effectiveness:nonlinear} and Table \ref{Tab:Accuracy:R:Optimized}. It is shown that SVR is more applicable to this estimation problem.

\begin{table}\scriptsize
\setlength{\abovecaptionskip}{-100pt}
\setlength{\belowcaptionskip}{-100pt}
\centering
\caption{Performance (MAE) of four nonlinear methods. Experimental configurations are same as the experiments in Table \ref{Tab:Accuracy:R:Optimized}.}
\begin{tabular}{  c c c c }
\hline \hline
evaluation method  &  T$_{eff}$     &    log$~g$       &   [Fe/H]     \\ \hline
FNN       &  0.008980      &   0.186014       &   0.179565\\
GAM       &  0.008139      &   0.245167       &   0.245111\\
MARS      &  0.011335      &   0.243147       &   0.242703\\
RF        &  0.009478      &   0.228717       &   0.204248\\
 \hline
\end{tabular}\label{Tab:effectiveness:nonlinear}
\end{table}

\subsection{Evaluation on spectra with ground-truth}\label{Sec_sub:eva_Syn}

The proposed scheme is also evaluated on 18~969 synthetic spectra. The synthetic spectra are calculated from the SPECTRUM (v2.76) package \citep{Con:Gray:1994} with the ¡°New Grids of ATLAS9 Model Atmospheres¡± \citep{Journal:Castelli:2003} as the stellar atmosphere model. In generating the synthetic spectra, 830~828 atomic and molecular lines are used (contained in two files luke.lst and luke.nir.lst), and the used atomic and molecular data comes form file stdatom.dat, which includes solar atomic abundances from \citet{Journal:Grevesse:1998}. The SPECTRUM package and the three data files can all be downloaded from the website.\footnote{http://stellar.phys.appstate.edu/spectrum/download.html.}

Our grids of the synthetic stellar spectra span the parameter ranges [4000, 9750] K in T$_{eff}$ (45 values, step size 100K between 4000K and 7500K and 250 K between 7750K and 9750K), [1, 5] dex in log$~g$ (17 values, step size 0.25 dex steps), and [-3.6, 0.3] dex in [Fe/H] (27 values, step size 0.2 between -3.6 dex and -1, 0.1 between -1 dex and 0.3 dex).

The synthetic stellar spectra are also partitioned into three subsets: training set, validation set, and test set. Sizes of the three subsets are 8~500, 1~969 and 8~500 respectively. The training set are used for detecting features and computing the estimation model. Validation set and test set are used for optimization the parameters in SVR and evaluating the performance of the learned model respectively.

The detected features from synthetic training set are presented in Table \ref{Tab:LASSO:Features:Teff:syn}, Table \ref{Tab:LASSO:Features:Logg:syn} and Table \ref{Tab:LASSO:Features:FeH:syn}. In this experiment, we adjusted the threshold $t$ by hand to detect approximately same amount of features as the corresponding experiments on SDSS spectra. Numbers of the detected features are 9 for estimating T$_{eff}$, 19 for log$~g$ and 15 for [Fe/H]. Based on these features, the estimation results are presented Table \ref{Tab:effectiveness:Synthetic}. In \citet{Journal:Fiorentin:2007}, the best consistency on synthetic spectra are obtained based on 100 principal components, and the MAE are 0.0030 dex for log$T_{eff}$, 0.0251 for log$~g$, 0.0269 for [Fe/H](Table 1 in \citet{Journal:Fiorentin:2007}). Therefore, apart from much less complexity in computing spectral features, the proposed scheme in this work is also more accurate than the scheme based on PCA.

\begin{table}\scriptsize
\setlength{\abovecaptionskip}{-100pt}
\setlength{\belowcaptionskip}{-100pt}
\centering
\caption{Detected typical positions for estimating $T_{\texttt{eff}}$ from synthetic stellar spectra. TPW $\lambda^w$: Typical position in wavelength ({\AA}), TPL $\lambda^l$: Typical position in log(wavelength).}
\begin{tabular}{p{0.35cm}p{1.5cm} p{0.9cm}p{0.35cm}p{1.5cm} p{0.9cm}}
  \hline \hline
index         &TPW $\lambda^w$  &TPL $\lambda^l$& index         &TPW $\lambda^w$   &TPL $\lambda^l$        \\ \hline
~~1           &   3933.3446            &3.5948         & ~~2           &4036.1008                &3.6060\\
~~3           &   4221.4534            &3.6255         & ~~4           &4475.7106                &3.6509\\
~~5           &   4501.5492            &3.6534         & ~~6           &5753.8959                &3.7600\\
~~7           &   6496.2391            &3.8127         & ~~8           &6545.7890                &3.8160\\
~~9           &   6547.2964            &3.8161         &               &                         &      \\
\hline
\end{tabular}\label{Tab:LASSO:Features:Teff:syn}
\end{table}

\begin{table}\scriptsize
\setlength{\abovecaptionskip}{-100pt}
\setlength{\belowcaptionskip}{-100pt}
\centering
\caption{Detected typical positions for estimating log$~g$ from synthetic spectra. TPW $\lambda^w$: Typical position in wavelength  ({\AA}), TPL $\lambda^l$: Typical position in log(wavelength).}
\begin{tabular}{p{0.35cm}p{1.5cm} p{0.9cm}p{0.35cm}p{1.5cm} p{0.9cm}}
  \hline \hline
index         &   TPW $\lambda^w$           &TPL $\lambda^l$        & index         &TPW $\lambda^w$        &TPL $\lambda^l$        \\ \hline
~~~1           &   3835.8534            &3.5839         &~~~2           &3889.2154                &3.5899\\
~~~3           &   3933.3446            &3.5948         &~~~4           &3969.7394                &3.5988\\
~~~5           &   4101.6821            &3.6130         &~~~6           &4856.9317                &3.6864\\
~~~7           &   4858.0502            &3.6865         &~~~8           &5183.9643                &3.7147\\
~~~9           &   5240.3706            &3.7194         &~~~10          &5276.6951                &3.7224\\
~~~11          &   5316.9429            &3.7257         &~~~12          &5321.8423                &3.7261\\
~~~13          &   5323.0678            &3.7262         &~~~14          &5336.5674                &3.7273\\
~~~15          &   5368.6118            &3.7299         &~~~16          &5589.3589                &3.7474\\
~~~17          &   5657.9877            &3.7527         &~~~18          &5891.9900                &3.7703\\
~~~19          &   8467.6325            &3.9278         &               &                &\\
 \hline
\end{tabular}\label{Tab:LASSO:Features:Logg:syn}
\end{table}

\begin{table}\scriptsize
\setlength{\abovecaptionskip}{-100pt}
\setlength{\belowcaptionskip}{-100pt}
\centering
\caption{Detected typical positions for estimating [Fe/H] from synthetic spectra. TPW $\lambda^w$: Typical position in wavelength  ({\AA}), TPL $\lambda^l$: Typical position in log(wavelength).}
\begin{tabular}{p{0.35cm}p{1.5cm} p{0.9cm}p{0.35cm}p{1.5cm} p{0.9cm}}
  \hline \hline
index         &   TPW $\lambda^w$           &TPL $\lambda^l$        & index         &TPW $\lambda^w$        &TPL $\lambda^l$        \\ \hline
~~~1          &   3933.3446            &3.5948         &~~~2          &4340.7223               &3.6376\\
~~~3          &   4871.4921            &3.6877         &~~~4          &   5176.8073            &3.7141\\
~~~5          &   5183.9643            &3.7147         &~~~6          &   5275.4802            &3.7223\\
~~~7          &   5279.1257            &3.7226         &~~~8          &   5287.6415            &3.7233\\
~~~9          &   5304.7143            &3.7247         &~~10          &   5316.9429            &3.7257\\
~~11          &   5475.9818            &3.7385         &~~12          &   5527.9232            &3.7426\\
~~13          &   5588.0721            &3.7473         &~~14          &   5615.1582            &3.7494\\
~~15          &   8542.0479            &3.9316         &              &                        &      \\
 \hline
\end{tabular}\label{Tab:LASSO:Features:FeH:syn}
\end{table}

\begin{table}\scriptsize
\setlength{\abovecaptionskip}{-100pt}
\setlength{\belowcaptionskip}{-100pt}
\centering
\caption{Performance on synthetic spectra based on SVR and features in Table \ref{Tab:LASSO:Features:Teff:syn}, Table \ref{Tab:LASSO:Features:Logg:syn} and Table \ref{Tab:LASSO:Features:FeH:syn}. Feature are described by the LI method with integration radii k = 6, 2, 8 repectively for T$_{eff}$, log$~g$ and [Fe/H]. MAE are the mean absolute errors on synthetic test set.}
\begin{tabular}{  c c c c }
  \hline \hline
evaluation method &  T$_{eff}$      &    log$~g$&   [Fe/H]     \\ \hline
MAE       &  0.000801      &       0.017881   &   0.013142\\
SD        &  0.001277      &       0.071147   &   0.036305\\ \hline
\end{tabular}\label{Tab:effectiveness:Synthetic}
\end{table}

\subsection{LASSO for spectral feature selection: feasibility, potential risks and alternatives}
Feature selection is to choose a subset of variables that collectively have a good predictive power. According to the utilized evaluation metric on the predictive power, feature selection algorithms can be divided into three categories: filters, wrappers and embedded methods \citep{Journal:Guyon:2003}.

Wrappers measure the effectiveness of a subset of variables by the accuracy of a learning machine of interest (a regression model or a classifier). Every subset should be used to train a model of the selected learning machine. The amount of possible combination of variables increases exponentially with the number of observed variables. Therefore, wrappers are computationally intensive in case of a larger number of observed variables.
Filters perform feature selection by a measure independently of the learning machine of interest. This kind of methods is usually less computationally intensive than wrappers, but the selected features are not tuned to a specific learning machine of interest.
Embedded methods select features in learning the model of interest and the features selected by this kind of methods are optimal to a specific learning machine. Due to the computational feasibility problem, the optional models in embedded methods are limited, for example, a linear model.

In this work, we investigated the feasibility of exploring the possible subsets of spectral features for estimating atmospheric parameters by LASSO. LASSO is a feature selection method based on a linear model. However, experiments shows that there exist some non-linearity in the dependence of atmospheric parameters on observed spectral fluxes (Section \ref{Sec_sub:linearity_nonlinearity}). Therefore, the LASSO played a role of filters in this work and there exists the risk of missing high relevant features to the estimating model of interest (SVR in this work). To reduce the possibility of this risk, an optional scheme is to firstly select a larger subset of features by the large value $t$ in inequality (\ref{Equ:LASSO_constraint}), and then refine the features by a more computational embedded method on the selected subset, for example the recursive forward selection \citep{Journal:Liu:2006} and backward elimination \citep{Journal:Guyon:2003}. For the specific learning machine SVR, features can also be detected by a built-in SVM feature selection algorithms \citep{Journal:Becker:2009,Con:Weston:2000}.

\section{Related Research}\label{Sec:RelatedResearches}
To highlight the characteristics of the proposed scheme, related research is reviewed and analyzed in this section.

Due to the rapid development of spectrum-obtaining capability and the driven by demand, many attempts have been made to estimate the atmospheric parameters directly from spectra in literature. In automatically estimating physical parameters from a stellar spectrum, a key procedure is feature extraction, which determines the applicable range of the corresponding system, accuracy, efficiency, physical interpretability, and robustness to noise and distortion from calibration error. Therefore, we roughly classify related researches into three categories based on the feature-extracting methods used in them: line index method, template matching method, and the statistical index scheme.

\subsection{Line index method}
This kind of method is used to estimate atmospheric parameters by representing a stellar spectrum with a description of typical lines, which is directly related to our knowledge about the stellar spectrum and astrophysics. A prominent characteristic of the line index method is physical interpretability. Therefore, this is a favorite method in spectrum analysis.

For example, \citet{Journal:Muirhead:2012} investigated the estimation problem of effective temperature $T_\texttt{eff}$ and metallicity [M/H] for late-K and M-type planet-candidate host stars from the K-band spectra released by the Kepler Mission based on three spectral indices: the equivalent widths of NaI (2.210 $\mu$m) and CaI (2.260 $\mu$m) lines, and an index describing the change in flux between three 0.02$\mu$m wide bands centered at 2.245, 2.370, and 2.080 $\mu$m respectively. \citet{Journal:Rojas-Ayala:2012} further proposed a revised relationship that estimates metallicities [Fe/H] and [M/H] of M dwarfs based on the three spectral indices. \citet{Journal:Mishenina:2008} proposed a method to estimate effective temperature by line depth ratio, two methods to estimate surface gravity log$~g$ based on the ionization balance of iron and fitting of the wings of the CaI 6162.17$\AA$ line. The fundamental parameters of 66 B-type stars are determined by the equivalent widths and/or line profile shapes of continuum-normalized hydrogen, helium, and silicon line profiles in \citep{Journal:Lefever:2010}. \citet{Journal:Posbic:2012} developed a software to determine radial velocity $V_r$, effective temperature $T_\texttt{eff}$, surface gravity log$~g$, metallicity [Fe/H], and individual abundances by a scheme relying on line-by-line modeling. \citet{Journal:Lee:2008:a} and \citet{Journal:Luo:2008} each took a line index method as a component in developing their atmospheric parameter estimation systems for stellar spectra from SDSS and LAMOST/Guoshoujing Telescope respectively.

Despite the advantage of physical interpretability, the performance of this kind method depends on the reliability of detecting spectral lines and accuracy of their description, which are usually sensitive to noise and calibration distortion in application \citep{Journal:Han:2011,Journal:Han:2013}.

\subsection{Template matching method}
Suppose $S_l = \{(x^i, y_i), ~ i = 1 \cdots N\}$ is a library of templates and x is a stellar spectrum whose physical parameters $y(x)$ need to be estimated, where $x^i$ is a template spectrum and $y_i$ is the corresponding physical parameter. If $x^{i_0}$ is the most similar template to $x$, then a basic implementation of the template matching method is to assign $\hat{y}(x) = y_{i_0}$. The fundamental idea of this method is simple and intuitive: give the estimated value with the parameter of the most similar template.

Therefore, it is also widely investigated in atmospheric parameter estimation. Its basic steps are:
\begin{itemize}
\item{Construct a library of templates $S_l$;}
\item{Find $k$ most similar template spectra in $S_l$ for a spectrum $x$ whose physical parameter $y(x)$ needs to be estimated, where $k$ is a preset positive integer;}
\item{Estimate $y(x)$ by fusing the parameters of k most similar template spectra.}
\end{itemize}
 Key problems in this method are: 1) construction of the template library, which acts as a source or carrier of professional knowledge needed to parameterize stellar spectra and is closely related to the accuracy of estimation and applicable range of the corresponding system; 2) similarity measure between two spectra, which embodies our understanding to the problem to be tackled and also is related to accuracy; 3) scheme to organize the spectral template library and find the $k$ most similar template(s), as this scheme determines the efficiency of the system.

Due to the importance of the construction of the template library, this method has attracted considerable attention. \citet{Con:Gray:1994} investigated the construction of synthetic stellar spectra based on Kurucz models \citep{Con:Kurucz:1992} and developed a publicly available program, SPECTRUM. Based on the SPECTRUM and New grids of the ATLAS9 Model Atmosphere \citep{Journal:Serven:2005}, \citet{Journal:Du:2012} synthesized a comprehensive set of 2~890 near-infrared spectrum library with resolution wavelength sampling similar to the SDSS and LAMOST, and parameter ranges from 3~500 to 7~500K for effective temperature $T_\texttt{eff}$, from 0.5 to 5.0 dex for surface gravity log$~g$, and from -4.0 to 0.5 dex for [Fe/H]. \citet{Journal:Heiter:2002} presented several sets of grids of model stellar atmospheres computed by modified versions of the ATLAS9 code with parameter range from 4~000 to 10~000 K for $T_\texttt{eff}$, from 2.0 to 5.0 dex for log$~g$, and from -2.0 to 1.0 dex for metallicity [M/H]. \citet{Journal:Gustafsson:2008} developed and used a program MARCS, to construct late-type model atmospheres and presented a gird of about $10^4$ model atmospheres for stars with parameter range from 2~500K to 8~000K in $T_\texttt{eff}$, from -1 to 5 dex in log$~g$ and from -5 to +1 in [Me/H].

Based on the utilized evaluation scheme of similarity between spectra, the template matching method can be implemented in forms of the nearest neighbor method \citep{Journal:Zwitter:2005}, the k-nearest neighbor method \citep{Journal:Liu:2013}, the chi-square minimization method \citep{Journal:Jofre:2010,Journal:Prieto:2006}, the correlation coefficient method \citep{Journal:Liu:2012}, etc. For example, \citet{Journal:Katz:1998}, \citet{Journal:Soubiran:2000}, and \citet{Journal:Soubiran:2003} provided a software, TGMET, based on a reduced chi-square minimization scheme and investigated the problem to estimate physical parameters $T_\texttt{eff}$, log$~g$, and [Fe/H]. \citet{Journal:Shkedy:2007} developed a method using a hierarchical Bayesian principle to estimate fundamental stellar parameters and their associated uncertainties from the infrared 2.38-2.60 $\mu$m Infrared Space Observatory (ISO)-Short Wavelength Spectrometer (SWS) spectral data; in this method, both systematic and statistical measurement errors were taken into account. \citet{Journal:Liu:2013} comprehensively compared the chi-square minimization method, k-nearest neighbor method, and correlation coefficient method in estimating atmospheric parameters from stellar spectra. \citet{Journal:Koleva:2009} developed a full-spectrum fitting package, ULySS, and explored its application in parameterizing stellar spectra. Based on the ULySS, Wu et al. re-estimated the physical parameters for the CFLIB spectral database \citep{Journal:Wu:2011:a}, explored new Metal-poor Star Candidates from Guo Shoujing Telescope (LAMOST) Commissioning Observations \citep{Journal:Wu:2010}, and constructed a set of stellar spectral templates to estimate physical parameters from LAMOST stellar spectra \citep{Journal:Wu:2011:b}.

Apart from the advantages of simplicity and intuitiveness, the template matching method is essentially a global method that is sensitive to the accumulation of noise, distortion, and calibration error. With this method, it is also difficult to analyze and evaluate the effectiveness of local features of spectra, or to resolve the physical interpretation of a phenomenon.

\subsection{Statistical index scheme}

To estimate atmospheric parameter from stellar spectra, the statistical index scheme is dedicated to establishing a function mapping from spectral space\footnote{A stellar spectrum is regarded as a vector in a high dimensional space.} to the space of physical parameters by treating all of the fluxes of a spectrum equally and as a whole. Common methods of this kind include Principal Component Analysis (PCA)\citep{Book:Jolliffe:2002}, Wavelet\citep{Book:Mallat:2008}, Neural Network\citep{Book:Bishop:1995}, etc.

For example, \citet{Journal:Fiorentin:2007} first projected a spectrum into a 50-dimensional PCA space and then estimated physical parameters by learning a mapping from the PCA space to atmospheric parameter space with a nonlinear feedforward neural network. \citet{Journal:Zhang:2006} estimated atmospheric parameters by establishing a mapping from PCA space to parameter space by using a non-parametric estimator with variable window-width. \citet{Journal:Manteiga:2010} parameterized stellar spectra by extracting features based on Fourier analysis and Wavelet decomposition, and constructing a mapping from a feature space to the parameter space by feedforward networks with three layers. After extracting features by Haar wavelet, Lu et al. investigated the atmospheric parameterization problem by capturing the mapping to parameter space based on the Support Vector Regression (SVR) \citep{Journal:Lu:2013} and the non-parameter regression method \citep{Journal:Lu:2012}.

On works based on Neural Network, \citet{Journal:Bailer-Jones:2000} investigated the estimating precision of stellar parameters $T_\texttt{eff}$, log$~g$, and [M/H] by a feedforward non-linear network with two hidden layers on synthetic spectra with different resolution and signal-to-noise ratio. \citet{Journal:Snider:2001} explored the application of back-propagation neural networks with one and two hidden layers in estimating atmospheric parameters from medium-resolution spectra of F- and G-type stars. By a back propagation neural network, \citet{Journal:Giridhar:2006} studied the parameterization of a set of stellar spectra from the 2.3 m Vainu Bappu Telescope at Kavalur observatory, India. \citet{Journal:Willemsen:2005} researched parameterization of stellar spectra obtained at the VLT at ESO/Paranal (Chile) in visitor mode by using a feedforward neural network, which is trained on synthetic spectra using the model atmospheres from \citet{Journal:Castelli:1997} in combination with SPECTRUM \citep{Con:Gray:1994}. In the aforementioned works based on Neural Networks, there is no explicit or separated procedure for extracting spectral features. Actually, the data stream moving layer by layer from input to output is an iterative feature extraction procedure.

A prominent characteristic of this kind of method is that the parameterizing model of stellar spectra can be explored without need for prior of physical atmospheric model generating spectra; in machine leaning and artificial intelligence, methods with this characteristic are called black-box approaches. Therefore, the statistical index scheme is relatively easy to use. Furthermore, results of a method of this kind are obtained statistically from a lot of spectra, which usually result in good overall performance. Meanwhile, the existence of noise, distortion, and calibration error usually leads to incompleteness of the theoretical atmosphere model in reality, and we are obtaining stellar spectra on an unprecedented scale. Therefore, the statistical index scheme has a good potential usage in spectral parameterization and knowledge mining from a large set of spectra. Its limitations are the difficulty of resolving physical interpretations from the results of the statistical index scheme. This work investigated a novel statistical index scheme for stellar spectrum parameterization with good interpretability physically.

\section{Conclusion}\label{Sec:Conclusion}

We propose a nonlinear scheme to automatically estimate three primary atmospheric physical parameters, $T_\texttt{eff}$, log$~g$, and [Fe/H], from SDSS stellar spectra. This scheme is invoked by two sets of pre-parameterized stellar spectra, which act as two sources/carriers of professional knowledge needed to parameterize stellar spectra, and are called a training set and a validation set respectively in pattern recognition and machine learning. Therefore, the proposed scheme is flexible and can be updated conveniently by replacing the knowledge carriers with two new ones to meet developing needs.

The proposed model consists of the following five procedures: 1) Statistically detect typical wavelength positions of features from stellar spectra; 2) Compute the description of spectral features based on local information near the detected typical positions; 3) Refine features by evaluating compactness of the extracted features; 4) Learn a parameterizing model by SVR algorithm based on training data; 5) Estimate physical parameters by the learned model and description of spectra. Procedures 1), 2), 3) and 4) are for constructing a stellar parameterizing model, procedures 2) and 5) are used in parameterizing a new spectrum; procedure 3) is optional depending on our specific requirement for sparseness and accuracy.

One prominent characteristic of the proposed scheme is sparseness and locality. In this work, for example, every observed spectrum consists of 3~821 fluxes; our method detects 10, 19, and 14 typical wavelength positions to estimate $T_\texttt{eff}$, log$~g$, and [Fe/H] respectively. Then, a stellar spectra can be described by a vector of 10, 19 and 14 components respectively for $T_\texttt{eff}$, log$~g$, and [Fe/H]; this is a dramatic reduction of data compared to the original components number of 3~821 and the typical results with 50 components of related research \citep{Journal:Fiorentin:2007}. Therefore, the features detected by this method are very sparse, which is closely related to computing efficiency of the processing system and physical interpretability of related results.

Another typical characteristic is locality. We propose two methods to describe features. One is to use the observed fluxes at the detected typical positions. That is to say, we can just pick up 10, 19, and 14 fluxes at the detected wavelength positions from 3~821 fluxes as features to estimate $T_\texttt{eff}$, log$~g$, and [Fe/H] respectively. The second method is to accumulate the nearest 13, 5, or 17 fluxes at every detected position respectively for $T_\texttt{eff}$, log$~g$, and [Fe/H]. Based on the second method, to compute the features for $T_\texttt{eff}$, the needed computation is just 120 plus operation. On the contrary, if we extract 10 features by the traditional Principal Component Analysis (PCA) method, the computations are approximately 38~210 product operations and 38~200 plus operations for nearly every flux in a spectrum. Therefore, the proposed scheme is relatively very efficient. Furthermore, because the proposed method only uses the fluxes near the detected positions, it is more immune and robust to aggregation of noise and distortion from calibration error. For convenience, we name the aforementioned describing methods Point Description (PD) and Local Integration (LI) respectively.

Accuracies/Consistencies of our proposed scheme with respect to the pre-estimation by SSPP of SDSS are 0.007458 dex for log$~T_\texttt{eff}$ (101.609921 K for $T_\texttt{eff}$), 0.189557 dex for log$~g$, and 0.182060 for [Fe/H] if features are described by the LI method, where the accuracy is evaluated by mean absolute error (MAE). If features are described by the PD method, the accuracies are 0.009092 dex for log$~T_\texttt{eff}$ (124.545075 K for $T_\texttt{eff}$), 0.198928 dex for log$~g$, and 0.206814 dex for [Fe/H].
 In similar scenario, \citet{Journal:Fiorentin:2007} investigated the stellar parameter estimation problem and obtained accuracies 0.0126 dex for log$~T_\texttt{eff}$, 0.3644 dex for log$~g$ dex and 0.1949 dex for [Fe/H] on a test set (19~000 stellar spectra from SDSS);
 \citet{Journal:Jofre:2010} applied MA$\chi$ method to a sample of 17~274 metal-poor dwarf stars from SDSS/SEGUE and estimated the metallicity with averaged accuracies of 0.24 dex, the temperature with 130 K and log$~g$ with 0.5 dex;
 \citet{Journal:Xin:2013} proposed a scheme to parameterizing stellar spectra based on line index and artificial neural network, where the accuracies are 147.8123 K for  log$~T_\texttt{eff}$, 0.24757 dex for log$~g$ dex and 0.19942 dex for [Fe/H] on 9043 spectra from SDSS.
Therefore, compared to the results of related works in similar scenario, the performance of this scheme is excellent.

We also investigate the compactness of the detected features and introduce two concepts: Significance measure $S$ (in equation (\ref{Equ:significance})) and Relative Significance measure $S^r$ (in equation (\ref{Equ:significance_r})) for this purpose. By compactness, we mean to study whether there is redundancy in the detected features and, if any redundancy exist, how much. Research shows that we can refine the features by the measures $S$ and  $S^r$ on a validation set if there exists a demand for a more compact feature set.

\acknowledgments
The authors would like to extend their thanks to Dr. Jiannan Zhang, Juanjuan Ren, Yinbi Li, Yihan Song, Bing Du, Yue Wu, and Wei Du for their support and discussion. This work is supported by the National Natural Science Foundation of China (grant No: 61273248, 61075033), the Natural Science Foundation of Guangdong Province (S2011010003348), and the Open Project Program of the National Laboratory of Pattern Recognition (NLPR) (201001060).

\end{document}